\documentclass[showpacs,aps,prd,reprint,superscriptaddress,nofootinbib,longbibliography]{revtex4-1}
\usepackage[colorlinks=true, pdfstartview=FitV, linkcolor=magenta,citecolor=blue, urlcolor=magenta,
bookmarks=true, bookmarksnumbered=true, breaklinks]{hyperref}
\usepackage[dvipdfmx]{graphicx}
\usepackage{amsmath,amssymb,bm,color,longtable,mathrsfs,amsfonts,slashed}

\newcommand{\Slash}[1]{{\ooalign{\hfil/\hfil\crcr$#1$}}}

\begin{document}
\begin{flushright}
\end{flushright}

\title{Chiral separation effect catalyzed by heavy impurities}

\author{Daiki~Suenaga}
\email[]{suenaga@mail.ccnu.edu.cn}
\affiliation{Research Center for Nuclear Physics,
Osaka University, Ibaraki, 567-0048, Japan }

\author{Yasufumi Araki}
\email[]{araki.yasufumi@jaea.go.jp}
\affiliation{Advanced Science Research Center, Japan Atomic Energy Agency (JAEA), Tokai 319-1195, Japan}

\author{Kei~Suzuki}
\email[]{{k.suzuki.2010@th.phys.titech.ac.jp}}
\affiliation{Advanced Science Research Center, Japan Atomic Energy Agency (JAEA), Tokai 319-1195, Japan}

\author{Shigehiro~Yasui}
\email[]{yasuis@keio.jp}
\affiliation{Research and Education Center for Natural Sciences,\\ Keio University, Hiyoshi 4-1-1, Yokohama, Kanagawa 223-8521, Japan}
\affiliation{RIKEN iTHEMS, Wako, Saitama 351-0198, Japan}

\date{\today}

\begin{abstract}
We investigate the influence of Kondo effect, namely, the nonperturbative effect induced by heavy impurities, on the chiral separation effect (CSE) in quark matter. We employ a simple effective model incorporating the Kondo condensate made of a light quark and a heavy quark, and compute the response function of axial current to the magnetic field in static limit and dynamical limit. As a result, we find that the Kondo effect catalyzes the CSE in both the limits, and particularly the CSE in dynamical limit can be enhanced by a factor of approximately three. Our findings clearly show that the presence of heavy impurities in quark matter can play an important role in the transport phenomena of light quarks induced by a magnetic field.

\end{abstract}


\maketitle

\section{Introduction}
\label{sec:Introduction}
Investigation of currents induced by an external magnetic field is one of the most important subjects 
in association with heavy-ion collision (HIC) and neutron star (NS) physics. For example, it was proposed that a magnetic field applied in chirality imbalanced matter gives rise to a vector current, which is called the chiral magnetic effect (CME)~\cite{Kharzeev:2004ey,Kharzeev:2007tn,Kharzeev:2007jp,Fukushima:2008xe} (see also
Ref.~\cite{Vilenkin:1980fu,Nielsen:1983rb} for earlier works). Another example is the chiral separation effect (CSE)~\cite{Son:2004tq,Metlitski:2005pr,Newman:2005as} which refers to an axial current induced by a magnetic field in ordinary baryonic or quark matter. The CME has been mostly studied with respect to high-energy HICs in which a chirality imbalanced environment may be created by metastable states related to the $\theta$-vacua~\cite{Kharzeev:1998kz,Kharzeev:1999cz,Kharzeev:2001ev}. The CSE is expected to play an important role in low-energy HICs in which finite baryon density can be created, as well as in NSs.

In quark matter, in addition to light quarks ($u,d$) with the Fermi surface, heavy quarks ($c$) may exist as impurities.\footnote{In low-energy HICs, heavy quarks can be created by hard processes of gluons, and they do not participate in the equilibration of light quarks.} In such a situation, light quarks near the Fermi surface can be correlated with the heavy quarks in a nonperturbative way due to the non-Abelian interaction with $SU(N_{c})$ symmetry ($N_c$ is the number of colors) of quantum chromodynamics (QCD). As a result, a condensate made of a light quark and a heavy quark is created, which is the so-called {\it Kondo condensate}. The emergence of Kondo condensate and phenomena induced by the condensate are called {\it Kondo effect}.\footnote{The Kondo effect driven in QCD is sometimes called {\it QCD Kondo effect.}} While the Kondo effect was originally found in metal including impurities in the context of condensed-matter physics~\cite{Kondo:1964,Hewson,Yosida,Yamada,coleman_2015}, it has been theoretically shown to emerge in quark matter~\cite{Yasui:2013xr,Hattori:2015hka} (See Refs.~\cite{Ozaki:2015sya,Yasui:2016svc,Yasui:2016yet,Kanazawa:2016ihl,Kimura:2016zyv,Yasui:2017bey,Suzuki:2017gde,Yasui:2017izi,Kimura:2018vxj,Macias:2019vbl,Hattori:2019zig,Suenaga:2019car,Suenaga:2019jqu,Kanazawa:2020xje,Araki:2020fox,Araki:2020rok,Kimura:2020uhq} for related discussions).\footnote{The Kondo effect in high-energy physics emerges not only in quark matter with heavy quarks but also in nuclear matter with heavy hadrons~\cite{Yasui:2013xr,Yasui:2016hlz,Yasui:2016ngy,Yasui:2019ogk}. While the former is caused by the QCD interaction, the latter is caused by spin and/or isospin ones.}

The Kondo effect is known to change the transport phenomena characteristically. For example, this effect tends to suppress the electric conductivity of charged particles as we lower the temperature. Similarly, we expect that the Kondo effect also influences the transport phenomena induced by a magnetic field. Based on this expectation, in this paper we discuss modification of CSE due to the Kondo effect in quark matter, by focusing on the response function of the axial current to a magnetic field. In the ordinary quark matter where the Kondo condensate is absent, 
the magnitude of CSE was found to be slightly modified by perturbative corrections in quantum electrodynamics (QED)~\cite{Gorbar:2013upa}. On the other hand, in this paper, we investigate modification of the CSE, not from perturbative effects of QED but from nonperturbative corrections of QCD driven by the Kondo effect. Hence we expect a significant modification of the CSE, and in fact this is the case as we will see later. 

To focus on the impact of Kondo effect on the CSE, we employ a simple effective model which contains light quarks with a finite chemical potential and heavy quarks defined within the heavy quark effective theory (HQET)~\cite{Eichten:1989zv,Georgi:1990um,Neubert:1993mb,Manohar:2000dt}, in the presence of Kondo condensate. 
For evaluation of the response function of CSE, we make use of the linear response theory~\cite{Kapusta:2006pm}. This method yields the CSE in two types of limits which are the so-called {\it static limit} and {\it dynamical limit}. 
The former (latter) describes an induced current under an external magnetic field whose time dependence is slower (faster) than the equilibration of the spatial part of the system, and in general these two limits give different values of induced current~\cite{Kharzeev:2009pj,Hou:2011ze,Son:2012zy,Landsteiner:2013aba,Kharzeev:2016sut}. Thus, we investigate the response function of CSE in both the static limit and dynamical limit.

This paper is organized as follows. In Sec.~\ref{sec:Formalism} we present a model employed in our analysis and derive the Green's function of fermions modified by the Kondo condensate. In Sec.~\ref{sec:Analysis} we explain briefly the linear response theory and show our strategy to evaluate the CSE. In Sec.~\ref{sec:Results}, we present numerical results of response function of the CSE in both static limit and dynamical limit. In Sec.~\ref{sec:GaugeInvariance} we discuss gauge invariance of our computation, and in Sec.~\ref{sec:Conclusion} we conclude the present work.

\section{Model}
\label{sec:Formalism}

Here, we introduce an effective Lagrangian toward the investigation of CSE in the presence of Kondo condensate. In the present work we are particularly interested in the response of axial current to an external magnetic field, so that we do not derive the Kondo condensate in a concrete model, such as the Nambu--Jona-Lasinio--type (NJL-type) model which is powerful for the study of phase structure as done in the literatures~\cite{Yasui:2016svc,Yasui:2017izi}. Alternatively, we simply assume the presence of the Kondo condensate in a reasonable form, for the clarity of discussion.

The kinetic terms of massless light quarks at finite density and heavy quarks put as impurities are given by
\begin{eqnarray} 
{\cal L}_{\rm kin} = \bar{\psi}(i\Slash{D}+\mu\gamma_0)\psi + \Psi_v^\dagger iD_0\Psi_v \ . \label{LKinetic}
\end{eqnarray}
 In this Lagrangian, while the light quark field $\psi$ is given as an ordinary Dirac fermion, the heavy quark field $\Psi_v$ is given within the HQET~\cite{Eichten:1989zv,Georgi:1990um,Neubert:1993mb,Manohar:2000dt}, i.e., $\Psi_v$ is related to the ordinary Dirac field $\Psi$ as $\Psi_v = \frac{1+\gamma_0}{2}{\rm e}^{iM_Qt}\Psi$ ($M_Q$ is the mass of $\Psi$ and $t$ describes the time of the system) in the rest frame of the heavy quark, which allows us to take into account only the particle component of the heavy quark together with the energy and momentum of order $\Lambda_{\rm QCD}$. $\mu$ is a chemical potential related to the density of light quarks. 

In Eq.~(\ref{LKinetic}), we have introduced covariant derivatives for the quarks to incorporate a magnetic field as
\begin{eqnarray}
D_\mu\psi &=& (\partial_\mu+ie_q A_\mu)\psi \ ,  \nonumber\\
D_0\Psi_v &=& (\partial_0+ie_QA_0)\Psi_v\  \label{CovDerivative}
\end{eqnarray}
[$A^\mu=(A_0,{\bm A})$)]. In Eq.~(\ref{CovDerivative}), $e_q$ and $e_Q$ are the electric charges of light and heavy quarks, respectively. We note that the magnetic coupling between the heavy quark and the gauge field is absent in the Lagrangian~(\ref{LKinetic}) because of the heavy-quark spin-symmetry of the HQET~\cite{Eichten:1989zv,Georgi:1990um,Neubert:1993mb,Manohar:2000dt}. This is explicitly shown by the lack of spatial derivative of $\Psi_v$ in Eq.~(\ref{LKinetic}) in the rest frame. In other words, the heavy quark do not couple with the magnetic field directly.

Field-theoretically, the Kondo effect can be described by a condensate formed by a light quark and a heavy quark, referred to as the Kondo condensate, which is analogous to the diquark condensate for color superconductivity~\cite{Alford:1997zt,Rapp:1997zu,Alford:1998mk,Alford:2007xm} or the chiral condensate for spontaneous chiral-symmetry breaking~\cite{Nambu:1961tp}. One of the most concise effective terms incorporating the Kondo condensate may be given as
\begin{eqnarray}
{\cal L}_{\Delta} &=& \Delta_S(\bar{\psi}\Psi_v)  + {\bm \Delta}_V\cdot (\bar{\psi}{\bm \gamma}\Psi_v) + ({\rm c.c.}) \ . \label{LKondo}
\end{eqnarray}
Here we have included the vector condensate ${\bm \Delta}_V$ in addition to the scalar one $\Delta_S$ as naturally suggested by the chiral partner structure of HQET~\cite{Bardeen:1993ae,Nowak:1992um}. Namely, when the chiral symmetry is significantly restored at finite density, the masses of scalar and vector heavy-light modes tend to degenerate with the heavy-quark spin symmetry~\cite{Suenaga:2014sga,Harada:2016uca,Suenaga:2017deu}, which suggests that those modes play comparable roles in quark matter. We note that pseudoscalar and axial-vector condensates could also be present in Eq.~(\ref{LKondo}), but we have not included them due to the assumption of parity invariance of the ground state. According to Refs.~\cite{Yasui:2016svc,Yasui:2017izi} based on the NJL-type analysis, the condensates in momentum space may take the ansatz of
\begin{eqnarray}
&& \Delta_S = \Delta\ , \nonumber\\
&& {\bm \Delta}_V = \Delta \hat{\bm p}\ , \label{Hedgehog}
\end{eqnarray}
($\hat{\bm p} = {\bm p}/|{\bm p}|$) where $\Delta$ is a constant providing an order-parameter in the ground state. 
In particular, the second line in Eq.~(\ref{Hedgehog}) is called the {\it hedgehog ansatz}. 

In what follows, we assume that the light and heavy quarks carry the identical electric charge as
\begin{eqnarray}
e\equiv e_q=e_Q\ , \label{ChargeE}
\end{eqnarray}
which allows us to avoid complexity caused by the spontaneous breakdown of the electromagnetic $U(1)_{\rm EM}$ symmetry and the appearance of the so-called Nambu-Goldstone (NG) mode.
For example, if we regard the heavy quark as the charm quark, then the light quark is identified as the up quark,
leading to $e\equiv e_q=e_Q=+\frac{2}{3}\hat{e}$, where $\hat{e}$ is the elementary charge.
For the bottom quark and down quark, $e\equiv e_q=e_Q=-\frac{1}{3}\hat{e}$.

\begin{figure}[htbp]
\centering
\includegraphics[clip, width=1.0\columnwidth]{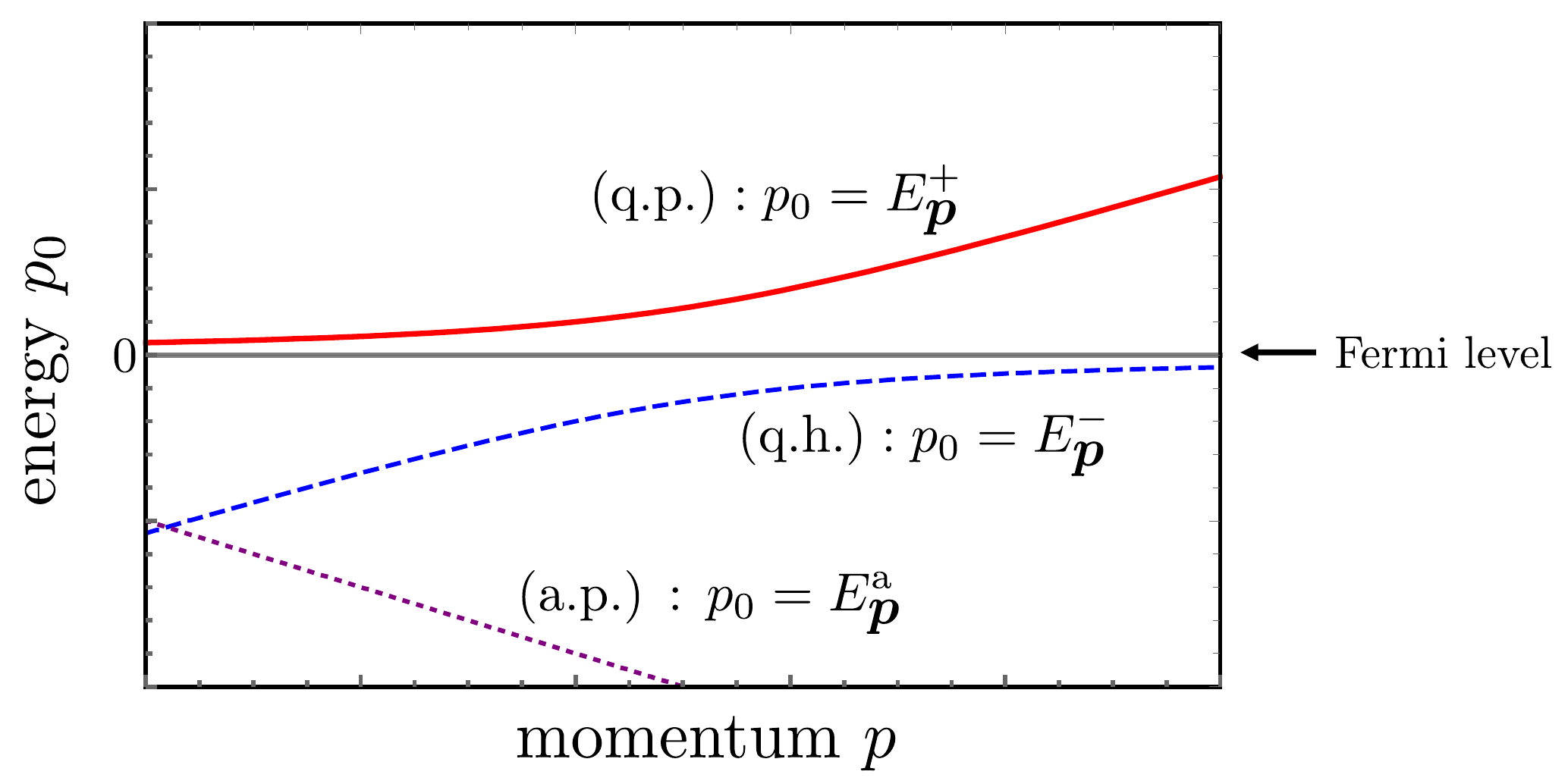}
\caption{A schematic behavior of the dispersions in Eq.~(\ref{Dispersions}) ($p=|{\bm p}|$). Solid red, dashed blue, and dotted purple curves correspond to $p_0=E_{\bm p}^{+}$ (q.p.), $p_0=E_{\bm p}^{-}$ (q.h.), and $p_0=E_{\bm p}^{\rm a}$ (a.p.), respectively. The gray horizontal line indicates $p_0=0$ corresponding to the Fermi level.}
\label{fig:Dispersions}
\end{figure}

Next, we derive a Green's function of the fermions incorporating $\Delta$ in the absence of the external gauge field ${\bm A}$. The effective Largangian employed in this study is given by the sum of Eqs.~(\ref{LKinetic}) and~(\ref{LKondo}):
\begin{eqnarray}
{\cal L} = {\cal L}_{\rm kin} + {\cal L}_{\Delta}\ . \label{LTotal}
\end{eqnarray}
Thus, by reading quadratic terms of the light and heavy quarks from the Lagrangian~(\ref{LTotal}) without ${\bm A}$ in momentum space, we can easily get the Green's function of the fermions as
\begin{widetext}
\begin{eqnarray}
\tilde{\cal G}_0(p_0,{\bm p}) &=& \frac{i}{(p_0-E^{\rm a}_{\bm p})(p_0-E_{\bm p}^+)(p_0-E_{\bm p}^-)} \nonumber\\
&& \times \left(
\begin{array}{ccc}
p_0(p_0+\mu)-|\Delta|^2 & -\left(|{\bm p}|p_0+|\Delta|^2\right)\hat{\bm p}\cdot{\bm \sigma} & -\Delta^* (p_0+|{\bm p}|+\mu) \\
\left(|{\bm p}|p_0+|\Delta|^2\right)\hat{\bm p}\cdot{\bm \sigma}  & -p_0(p_0+\mu)+|\Delta|^2 &  -\Delta^*(p_0+|{\bm p}|+\mu) \hat{\bm p}\cdot{\bm \sigma} \\
 -\Delta (p_0+|{\bm p}|+\mu) &\Delta (p_0+|{\bm p}|+\mu) \hat{\bm p}\cdot{\bm \sigma} &(p_0-|{\bm p}|+\mu)(p_0+|{\bm p}|+\mu) \\
\end{array}
\right) \ ,\label{G0Tilde}
\end{eqnarray}
\end{widetext}
in the Dirac representation, with the dispersion relations for the fermions 
\begin{eqnarray}
 E_{\bm p}^{+} &=& \frac{1}{2}\left(|{\bm p}|-\mu + \sqrt{(|{\bm p}|-\mu)^2+8|\Delta|^2}\right)\ , \nonumber\\
E_{\bm p}^{-} &=& \frac{1}{2}\left(|{\bm p}|-\mu - \sqrt{(|{\bm p}|-\mu)^2+8|\Delta|^2}\right)\ , \nonumber\\
E^{\rm a}_{\bm p} &=& -|{\bm p}|-\mu\ .  \label{Dispersions}
\end{eqnarray}
We note that the inverse of the Green's function is a $6\times6$ matrix because the light quark field $\psi$ is a four-component Dirac spinor while the heavy quark field $\Psi_v$ is essentially two-component one carrying only its particle component.
Throughout this paper, the modes carrying the dispersion of $E_{\bm p}^{+}$, $E_{\bm p}^{-}$ and $E_{\bm p}^{\rm a}$ are denoted by ``q.p.'' (Kondo quasiparticle), ``q.h.'' (Kondo quasihole) and ``a.p.'' (light antiparticle), respectively. For later use, we plot a schematic behavior of these dispersions in Fig.~\ref{fig:Dispersions}. In this figure, solid red, dashed blue, and dotted purple curves correspond to $p_0=E_{\bm p}^{+}$ (q.p.), $p_0=E_{\bm p}^{-}$ (q.h.), and $p_0=E_{\bm p}^{\rm a}$ (a.p.), respectively. The gray horizontal line indicates $p_0=0$ corresponding to the Fermi level. We find that $E_{\bm p}^{+}$ is always positive while $E_{\bm p}^{-}$ and $E_{\bm p}^{\rm a}$ are always negative as long as $\Delta$ is finite.

Equation~(\ref{G0Tilde}) is the Green's function found in the previous works~\cite{Yasui:2016svc,Yasui:2017izi}. In the present paper, we rewrite Eq.~(\ref{G0Tilde}) into a more convenient form which shows its physical meaning manifestly. Namely, we transform Eq.~(\ref{G0Tilde}) into the following form:
\begin{eqnarray}
\tilde{\cal G}_0(p_0,{\bm p}) = \left(
\begin{array}{cc}
\tilde{\cal G}_0^{\bar{\psi}\psi} (p_0,{\bm p}) & \tilde{\cal G}_0^{\bar{\psi}\Psi_v}  (p_0,{\bm p}) \\
\tilde{\cal G}_0^{{\Psi}_v^\dagger\psi} (p_0,{\bm p}) & \tilde{\cal G}_0^{{\Psi}_v^\dagger\Psi_v} (p_0,{\bm p}) \\  
\end{array}
\right) \ ,\label{GZeroAnother}
\end{eqnarray}
where the elements are given as
\begin{eqnarray}
\tilde{\cal G}_0^{\bar{\psi}\psi} (p_0,{\bm p})  &=& i\left[\frac{U_{+}({\bm p})}{p_0-E_{\bm p}^{+}}+ \frac{U_{-}({\bm p})}{p_0-E_{\bm p}^{-}}\right]\Lambda_{\rm p} \nonumber\\
&&  + i\frac{U_{\rm a}({\bm p})}{p_0-{E}_{\bm p}^{\rm a}}\Lambda_{\rm a} \, ,\nonumber\\
\tilde{\cal G}_0^{\bar{\psi}\Psi_v}(p_0,{\bm p}) &=& i\left[\frac{V^*_{+}({\bm p})}{p_0-E_{\bm p}^{+}}+ \frac{V^*_{-}({\bm p})}{p_0-E_{\bm p}^{-}}\right]\Lambda_{{\rm p}{\rm H}}\ , \nonumber\\
\tilde{\cal G}_0^{{\Psi}_v^\dagger\psi}(p_0,{\bm p}) &=& i\left[\frac{V_{+}({\bm p})}{p_0-E_{\bm p}^{+}}+ \frac{V_{-}({\bm p})}{p_0-E_{\bm p}^{-}}\right]\Lambda_{{\rm H}{\rm p}}\ , \nonumber\\
\tilde{\cal G}_0^{{\Psi}_v^\dagger\Psi_v}(p_0,{\bm p}) &=& i\left[\frac{W_{+}({\bm p})}{p_0-E_{\bm p}^{+}}+ \frac{W_{-}({\bm p})}{p_0-E_{\bm p}^{-}}\right]{\bm 1}\ . \label{GZeroElements}
\end{eqnarray}
In these expressions, $\tilde{\cal G}_0^{\bar{\psi}\psi} $, $\tilde{\cal G}_0^{\bar{\psi}\Psi_v} $, $\tilde{\cal G}_0^{{\Psi}_v^\dagger\psi}$, and $\tilde{\cal G}_0^{{\Psi}_v^\dagger\Psi_v}$ are $4\times4$, $4\times2$, $2\times4$, and $2\times2$ matrices, respectively. Also,
\begin{eqnarray}
\Lambda_{\rm p} \equiv \frac{1+\hat{\bm p}\cdot{\bm \alpha}}{2}\gamma_0\ , \ \ \Lambda_{\rm a} \equiv \frac{1-\hat{\bm p}\cdot{\bm \alpha}}{2}\gamma_0 \label{ProjectionPA}
\end{eqnarray}
(${\bm \alpha} = \gamma_0{\bm \gamma}$) are the projection operators for positive- and negative-energy states of the light quark, respectively. Similarly
\begin{eqnarray}
\Lambda_{{\rm p}{\rm H}} \equiv \left(
\begin{array}{c}
1  \\
\hat{\bm p}\cdot{\bm \sigma} \\
\end{array}
\right)\ , \ \ \Lambda_{{\rm H}{\rm p}} \equiv \left(
\begin{array}{cc}
1  & -\hat{\bm p}\cdot{\bm \sigma} \\
\end{array}
\right)
\end{eqnarray}
are the operators mixing positive-energy states of the light quark and the heavy quark.\footnote{One can understand these operators easily by
\begin{eqnarray}
\Lambda_{\rm p}\Lambda_{\rm H} = \left(
\begin{array}{cc}
1 & 0 \\
\hat{\bm p}\cdot{\bm \sigma} & 0 \\
\end{array}
\right)\ , \ \ \Lambda_{\rm H}\Lambda_{\rm p} = \left(
\begin{array}{cc}
1 & -\hat{\bm p}\cdot{\bm \sigma} \\
0 & 0 \\
\end{array}
\right)
\end{eqnarray}
with $\Lambda_{\rm H} = (1+\gamma_0)/2$.}
${\bm 1}$ is a $2\times2$ unit matrix.
The weight factors $U({\bm p})$'s, $V({\bm p})$'s, and $W({\bm p})$'s in Eq.~(\ref{GZeroElements}) are 
\begin{eqnarray}
&&U_{+}({\bm p})= \frac{2 (|\Delta|^2+|{\bm p}|E_{\bm p}^{+})}{(E_{\bm p}^{-}-E_{\bm p}^{+})(E_{\bm p}^{\rm a}-E_{\bm p}^{+})} \ , \nonumber\\
&& U_{-}({\bm p}) = \frac{2 (|\Delta|^2+|{\bm p}|E_{\bm p}^{-})}{(E_{\bm p}^{+}-E_{\bm p}^{-})(E_{\bm p}^{\rm a}-E_{\bm p}^{-})} \ , \nonumber\\
&& U_{\rm a}({\bm p}) = 1\ , \nonumber\\
&& V_{+}({\bm p}) = \frac{\Delta}{E_{\bm p}^{-}-E_{\bm p}^{+}} \ , \ \ V_{-}({\bm p}) = \frac{\Delta}{E_{\bm p}^{+}-E_{\bm p}^{-}} \ , \nonumber\\
 && W_{+}({\bm p}) = \frac{E_{\bm p}^{+}-|{\bm p}|+\mu}{E_{\bm p}^{+}-E_{\bm p}^{-}} \ , \ \ W_{-}({\bm p}) = \frac{E_{\bm p}^{-}-|{\bm p}|+\mu}{E_{\bm p}^{-}-E_{\bm p}^{+}}\ . \nonumber\\ \label{WeightFactors}
\end{eqnarray}
Equations~(\ref{GZeroElements}) and~(\ref{WeightFactors}) clearly show that only positive-energy components of the light quark correlate with the heavy quark. Namely $U_{\rm a}({\bm p})$ in Eq.~(\ref{WeightFactors}) is not changed from unity. In what follows we assume $\Delta$ to be real.

\section{Analysis}
\label{sec:Analysis}

\subsection{Linear response theory}
\label{sec:LRT}

In Sec.~\ref{sec:Formalism} we introduced an effective Lagrangian with an appropriate ansatz describing the Kondo condensate, and derived the Green's function of fermions. Here, we show our strategy to investigate the CSE in the presence of Kondo condensate based on them. In the following calculation, we set $A_0=0$ since the magnetic field is not generated by $A_0$.

The axial current is defined by $j_5^i = \bar{\psi}\gamma^i\gamma_5\psi$ ($i=1,2,3$) including only the light-quark degrees of freedom. Its expectation value can be given by
\begin{eqnarray}
\langle j_5^i(t,{\bm x})\rangle_\beta = \frac{1}{\Omega}\lim_{q_0,{\bm q}\to0}\langle \tilde{j}_5^i(q_0,{\bm q})\rangle_\beta\ , \label{J5Definition}
\end{eqnarray}
where $\langle\cdots\rangle_\beta$ stands for a thermodynamic expectation value and $\Omega=\beta V$ ($\beta=1/T$ is the inverse of temperature, and $V$ is the three-dimensional volume), and $\tilde{j}_5^i(q_0,{\bm q})$ is the Fourier transformation of $j_5^i(t,{\bm x})$. Thus, first we need to evaluate the expectation value of axial current in momentum space. In the present study, we are particularly interested in quark matter under a magnetic field with magnitude smaller than that of the Fermi energy: $\sqrt{eB}\ll \mu$. Therefore we can evaluate the CSE within the framework of the linear response theory~\cite{Kapusta:2006pm}.\footnote{The axial current can be calculated directly, but the computation is rather complicated in general. See Appendix~\ref{sec:AnotherStatic} for another derivation of the static limit of the axial current in case of $e=e_{q}=e_{Q}$.}
The linear response theory tells us that the response function to an external field takes the form of the {\it retarded} one. 
The retarded function could be given by an analytic continuation of the one computed within the imaginary-time formalism. Thus $\langle \tilde{j}_5^i(q_0,{\bm q})\rangle_\beta$ in the right-hand side (RHS) of Eq.~(\ref{J5Definition}) is provided by
\begin{eqnarray}
\langle \tilde{j}_5^i(q_0,{\bm q})\rangle_\beta = \langle \tilde{j}_5^i(i\bar{\omega}_n,{\bm q})\rangle_\beta|_{i\bar{\omega}_n\to q_0+i\eta}\ , \label{ZeroLimitJ5}
\end{eqnarray}
with $\eta$ an infinitesimal positive number, in which $\langle \tilde{j}_5^i(i\bar{\omega}_n,{\bm q})\rangle_\beta$ is given by
\begin{eqnarray}
&&\langle \tilde{j}_5^i(i\bar{\omega}_n,{\bm q})\rangle_\beta  \nonumber\\
&\approx&   eN_cT\sum_m\int\frac{d^3p}{(2\pi)^3}{\rm tr}\Big[\Gamma^i\Gamma_5\tilde{\cal G}_0(i{\omega}_m+i\bar{\omega}_n,{\bm p}+{\bm q})\nonumber\\
&& \times \tilde{A}^j(i\bar{\omega}_n,{\bm q})\Gamma^j\tilde{\cal G}_0(i{\omega}_m,{\bm p})\Big]\ . \label{J5General}
\end{eqnarray}
Here $6\times6$ gamma matrices $\Gamma^i$ and $\Gamma_5$ are defined by
\begin{eqnarray}
&& \Gamma^i = \left(
\begin{array}{cc}
\gamma^i & 0 \\
0 & 0 \\
\end{array}
\right) = \left(
\begin{array}{ccc}
0 & \sigma^i & 0 \\
-\sigma^i & 0 & 0 \\
0 & 0 & 0 \\
\end{array}
\right)\  , \nonumber\\
&& \Gamma_5 = \left(
\begin{array}{cc}
\gamma_5 & 0 \\
0 & 0 \\
\end{array}
\right)  =\left(
\begin{array}{ccc}
0 & 1 & 0 \\
1 & 0 & 0 \\
0 & 0 & 0 \\
\end{array}
\right)\ , \label{J5Imaginary}
\end{eqnarray}
respectively, which act on only light quark components. 
Note that $\bar{\omega}_n=2\pi n/\beta$ is the Matsubara frequency for bosons which shows the periodic boundary condition for the imaginary-time direction, while ${\omega}_n=(2n+1)\pi/\beta$ is for fermions which shows the anti-periodic boundary condition. $N_c$ is the number of colors, and we set $N_c=3$ throughout this paper. 
$\tilde{\cal G}_0(i{\omega}_m,{\bm p})$ in Eq.~(\ref{J5General}) can be obtained by simply replacing $p_{0}$ as $p_0\to i{\omega}_m$ in Eq.~(\ref{GZeroAnother}). 

Before moving on to the detailed calculation of Eq.~(\ref{J5General}), we mention the zero-momentum limit in Eq.~(\ref{J5Definition}). In medium, due to the lack of Lorentz invariance, it is possible to take the following two types of zero momentum limits:
\begin{itemize}
\item {\it static\ limit}: \ \ \ \ \ \ \ \  $\langle j_5^i\rangle^{\rm sta}_\beta \equiv \frac{1}{\Omega}\lim_{{\bm q}\to0}\langle \tilde{j}_5^i(0,{\bm q})\rangle_\beta$ ,
\item {\it dynamical\ limit}: \ $\langle j_5^i\rangle^{\rm dyn}_\beta \equiv \frac{1}{\Omega}\lim_{q_0\to0}\langle \tilde{j}_5^i(q_0,{\bm 0})\rangle_\beta$ . 
\end{itemize}
The static limit (dynamical limit) describes the response to a magnetic field whose time dependence is slower (faster) than the equilibration of the spatial part of the system. These two limits yield different results in general~\cite{Kharzeev:2009pj,Hou:2011ze,Son:2012zy,Landsteiner:2013aba,Kharzeev:2016sut}. Therefore in the following analysis, we will investigate the CSE in both the limits.

\subsection{Calculation of $\langle \tilde{j}_5^i(i\bar{\omega}_n,$\mbox{\boldmath $q$}$)\rangle_\beta $}
\label{sec:CalculationJ5}

In Sec.~\ref{sec:LRT}, we showed our strategy to evaluate the axial current $\langle j_5^i(t,{\bm x})\rangle_\beta$ within the framework of the linear response theory and presented a rather general expression of the axial current $\langle \tilde{j}_5^i(i\bar{\omega}_n,{\bm q})\rangle_\beta$ in momentum space in Eq.~(\ref{J5General}) in the imaginary-time formalism.  In this subsection, we further proceed the analytical evaluation of $\langle \tilde{j}_5^i(i\bar{\omega}_n,{\bm q})\rangle_\beta$.

By inserting Eq.~(\ref{GZeroAnother}) into Eq.~(\ref{J5General}), the trace with respect to the Dirac indices in Eq.~(\ref{J5Imaginary}) is reduced to the one for the light quark as
\begin{eqnarray}
\langle \tilde{j}_5^i(i\bar{\omega}_n,{\bm q})\rangle_\beta  &=&   eN_cT\sum_m\int\frac{d^3p}{(2\pi)^3}{\rm tr}\Big[\gamma^i\gamma_5\tilde{\cal G}^{\bar{\psi}\psi}_0(i{\omega}_m',{\bm p}')\nonumber\\
&& \times \tilde{A}^j(i\bar{\omega}_n,{\bm q})\gamma^j\tilde{\cal G}^{\bar{\psi}\psi}_0(i{\omega}_m,{\bm p})\Big] \label{J54By4}
\end{eqnarray}
(${\bm p}'={\bm p}+{\bm q}$ and $i{\omega}_m'=i{\omega}_m+i\bar{\omega}_n$). This reduction could be done because the axial current is defined with respect to only the light-quark degrees of freedom, and the magnetic field only couples with them now. Recalling the first line in Eq.~(\ref{GZeroElements}), the trace calculation is straightforwardly carried out, which yields
 \begin{eqnarray}
&& \langle \tilde{j}_5^i(i\bar{\omega}_n,{\bm q})\rangle_\beta  \nonumber\\
&=&   -ieN_c\epsilon^{ijk}\tilde{A}^k(i\bar{\omega}_n,{\bm q}) T\sum_m\int\frac{d^3p}{(2\pi)^3} \nonumber\\
&& \sum_{\zeta,\zeta'=+, -, {\rm a}}(\epsilon_{\zeta'}\hat{p}'^j-\epsilon_\zeta\hat{p}^j)\frac{U_{\zeta'}({\bm p}')U_\zeta({\bm p})}{(i{\omega}_m'-E_{{\bm p}'}^{\zeta'})(i{\omega}_m-E_{\bm p}^\zeta)}\ . \nonumber\\
\label{J5Zeta}
\end{eqnarray} 
In Eq.~(\ref{J5Zeta}), the ``sign function'' $\epsilon_\zeta$ is defined by 
\begin{eqnarray}
\epsilon_{+} = \epsilon_{-} = + 1\ ,
\end{eqnarray}
while
\begin{eqnarray}
\epsilon_{\rm a} = -1\ ,
\end{eqnarray}
with the subscripts ``$+$'', ``$-$'', and ``a'' corresponding to the ones in Eq.~(\ref{Dispersions}).
In deriving Eq.~(\ref{J5Zeta}), we have used an identity
\begin{eqnarray}
{\rm tr}[\gamma^i\gamma_5\Lambda'_{\xi'}\gamma_0\gamma^j\Lambda_\xi\gamma_0] = -i\epsilon^{ijk}(\epsilon_{\xi'}\hat{p}'^k-\epsilon_\xi\hat{p}^k)\ , \label{TraceFormula}
\end{eqnarray}
with the help of the formula ${\rm tr}[\gamma_5\gamma^0\gamma^i\gamma^j\gamma^k] =-4i\epsilon^{ijk}$, where $\xi,\xi'={\rm p}, {\rm a}$. In Eq.~(\ref{TraceFormula}), the ``sign function'' is
\begin{eqnarray}
\epsilon_{\rm p}=+1\ , \epsilon_{\rm a}=-1\ ,
\end{eqnarray}
with the subscripts ``p'' and ``a'' corresponding to the ones in Eq.~(\ref{ProjectionPA}).

\begin{figure*}[t!]
\centering
\includegraphics*[clip, width=2.0\columnwidth]{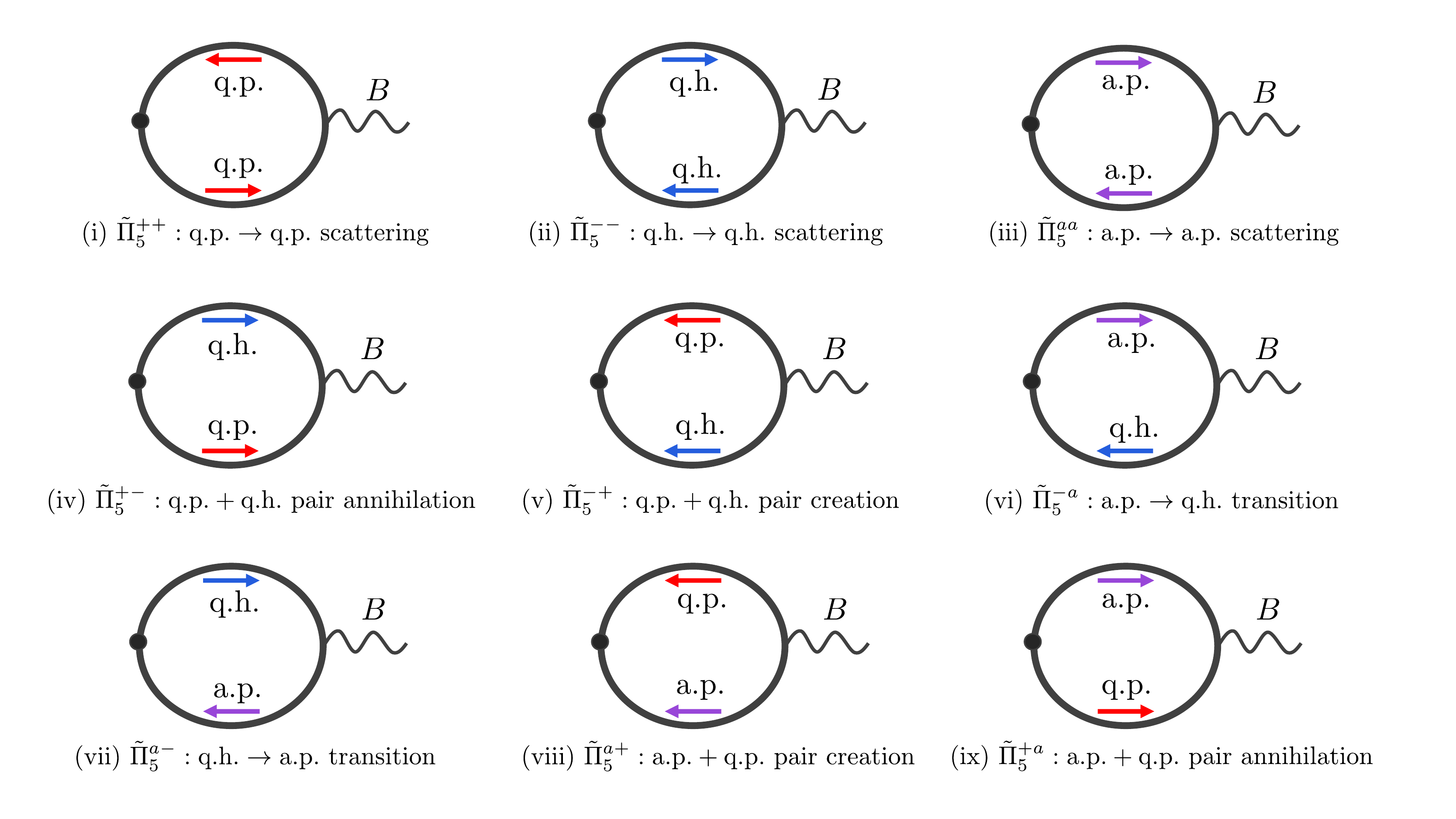}
\caption{Diagrammatic interpretation of each term in Eq.~(\ref{Pi5Each}). The arrows are put such that the corresponding modes carry their positive energies.}
\label{fig:Diagrams}
\end{figure*}

Equation~(\ref{J5Zeta}) includes an infinite sum with respect to the Matsubara frequency. This summation can be done by transforming the infinite series into a contour integral in the complex plane. Namely, we make use of the Abel-Plana formula~\cite{Kapusta:2006pm} 
\begin{eqnarray}
&& T\sum_m F(p_0=i{\omega}_m)  \nonumber\\
&=& \frac{1}{2\pi i} \int_{-i\infty}^{i\infty} dp_0 \frac{1}{2}\Big[F(p_0)+F(-p_0) \Big] \nonumber\\
&&-\frac{1}{2\pi i} \int_{-i\infty+\eta}^{i\infty+\eta} dp_0\Big[F(p_0)+F(-p_0)\Big]f_F(p_0) \ , \label{ABFermionFiniteT}
\end{eqnarray}
where $F(p_0)$ is an arbitrary function which is analytic on the imaginary axis, and $f_F(p_0) = \frac{1}{{\rm e}^{\beta p_0}+1}$ is the Fermi-Dirac distribution function. The calculation can be carried out without difficulty by means of the Cauchy's residue theorem, yielding
\begin{eqnarray}
\langle \tilde{j}_5^i(i\bar{\omega}_n,{\bm q})\rangle_\beta = ie\epsilon^{ijk}\tilde{A}^k(i\bar{\omega}_n,{\bm q})\tilde{\Pi}_5^j(i\bar{\omega}_n,{\bm q})\ ,\label{J5Convenient}
\end{eqnarray}
in which we have defined
\begin{widetext}
\begin{eqnarray}
\tilde{\Pi}_5^j(i\bar{\omega}_n,{\bm q}) \equiv  -N_c \int\frac{d^3p}{(2\pi)^3}\sum_{\zeta,\zeta'={+}, {-},{\rm a}}(\epsilon_{\zeta'}\hat{p}'^j-\epsilon_\zeta\hat{p}^j)\frac{U_{\zeta'}({\bm p}') U_{\zeta}({\bm p})}{i\bar{\omega}_n-E_{{\bm p}'}^{\zeta'}+E_{\bm p}^{\zeta}}\left[f_F(E_{\bm p}^\zeta)-f_F(E_{{\bm p}'}^{\zeta'})\right]\ .  \label{Pi5JDef}
\end{eqnarray}
\end{widetext}
In Eq.~(\ref{J5Convenient}), $\tilde{\Pi}_5^j(i\bar{\omega}_n,{\bm q})$ must be proportional to $q^j$ due to its Lorentz structure. Thus, by defining $\tilde{\Pi}_5(i\bar{\omega}_n,{\bm q})$ in terms of 
\begin{eqnarray}
\tilde{\Pi}_5^j(i\bar{\omega}_n,{\bm q}) \equiv \tilde{\Pi}_5(i\bar{\omega}_n,{\bm q})q^j\ , \label{Pi5JRelation}
\end{eqnarray}
now Eq.~(\ref{J5Convenient}) reads
\begin{eqnarray}
\langle \tilde{j}_5^i(i\bar{\omega}_n,{\bm q})\rangle_\beta 
&=& ie\epsilon^{ijk}q^j\tilde{A}^k(i\bar{\omega}_n,{\bm q})\tilde{\Pi}_5(i\bar{\omega}_n,{\bm q}) \nonumber\\
&=& e\tilde{B}(i\bar{\omega}_n,{\bm q}) \tilde{\Pi}_5(i\bar{\omega}_n,{\bm q})\ . \label{J5Magnetic}
\end{eqnarray}
In the second line of Eq.~(\ref{J5Magnetic}), we have made use of $\tilde{B}^i(i\bar{\omega}_n,{\bm q}) = i\epsilon^{ijk}q^j\tilde{A}^k(i\bar{\omega}_n,{\bm q})$ [$\tilde{B}^i(q_0,{\bm q})$ is defined by the Fourier transformation of $B^i(t,{\bm x})$]. We note that $\tilde{\Pi}_5(i\bar{\omega}_n,{\bm q})$ can be ``solved'' as  
\begin{eqnarray}
\tilde{\Pi}_5(i\bar{\omega}_n,{\bm q}) = \frac{1}{|{\bm q}|^2} \tilde{\Pi}_5^j(i\bar{\omega}_n,{\bm q}) q^j \label{Pi5Pi5J}
\end{eqnarray}
from Eq.~(\ref{Pi5JRelation}). 

Equation~(\ref{J5Magnetic}) shows that $\tilde{\Pi}_5(i\bar{\omega}_n,{\bm q})$ is the response function of axial current to a magnetic field. Concretely, from Eqs.~(\ref{Pi5JDef}) and~(\ref{Pi5Pi5J}), the response function is expressed as
\begin{eqnarray}
\tilde{\Pi}_5(i\bar{\omega}_n,{\bm q}) = \sum_{\zeta,\zeta'=+,-,{\rm a}}\tilde{\Pi}^{\zeta'\zeta}_5(i\bar{\omega}_n,{\bm q})  \label{Pi5TildeDef} 
\end{eqnarray}
with
\begin{widetext}
\begin{eqnarray}
\tilde{\Pi}^{\zeta'\zeta}_5(i\bar{\omega}_n,{\bm q}) \equiv -\frac{N_c}{|{\bm q}|^2}\int\frac{d^3p}{(2\pi)^3}\Big[\epsilon_{\zeta'}(\hat{\bm p}'\cdot{\bm q})-\epsilon_\zeta(\hat{\bm p}\cdot{\bm q})\Big]\frac{U_{\zeta'}({\bm p}') U_{\zeta}({\bm p})}{i\bar{\omega}_n-E_{{\bm p}'}^{\zeta'}+E_{\bm p}^{\zeta}}\left(f_F(E_{\bm p}^\zeta)-f_F(E_{{\bm p}'}^{\zeta'})\right)\ . \label{Pi5Each}
\end{eqnarray} 
\end{widetext}


After performing the analytic continuation to the real-time formalism and taking zero limit of the external momentum ($q_0,{\bm q}$) properly, static and dynamical limits for $\langle j_5^i\rangle_\beta$ can be evaluated from Eq.~(\ref{J5Magnetic}) as
\begin{eqnarray}
&&\langle j_5^i(t,{\bm x})\rangle_\beta^{\rm sta} = eB\tilde{\Pi}_5^{\rm sta} , \nonumber\\
&& \langle j_5^i(t,{\bm x})\rangle_\beta^{\rm dyn} = eB\tilde{\Pi}_5^{\rm dyn}\ ,  
\end{eqnarray}
respectively from Eqs.~(\ref{J5Definition}) and~(\ref{ZeroLimitJ5}), with the definitions of
\begin{eqnarray}
&&\tilde{\Pi}_5^{\rm sta} \equiv \lim_{{\bm q}\to{\bm 0}}\tilde{\Pi}_5(0,{\bm q})\ , \nonumber\\
&&\tilde{\Pi}_5^{\rm dyn} \equiv \lim_{q_0\to 0}\tilde{\Pi}_5(q_0,{\bm 0})\ . \label{Pi5Tot}
\end{eqnarray}
We also used $\tilde{B}(q_0=0,{\bm q}={\bm 0}) = \Omega B$. By carrying out the angular integral in Eq.~(\ref{Pi5TildeDef}) in two limits, the response function can be analytically obtained apart from the $|{\bm p}|$ integral. 
We note that a rather detailed discussion on the response function in static and dynamical limits is given in Appendix~\ref{sec:GeneralD}. 

At the end of this subsection, we explain the details of the nine contributions in Eq.~(\ref{Pi5Each}) obtained by setting $\zeta,\zeta'=+,-,{\rm a}$. Diagrammatically, these contributions can be understood as in Fig.~\ref{fig:Diagrams}. In this figure, the arrows are put in such a way that the corresponding modes carry their positive energies. We note that the momentum integrals of the loops in Fig.~\ref{fig:Diagrams} do not include any ultraviolet divergences. According to Eqs.~(\ref{PiDifference}) and~(\ref{PiNoDifference}) in Appendix~\ref{sec:GeneralD}, each contribution in Eq.~(\ref{Pi5Each}) in static limit and dynamical limit after the analytic continuation is found to satisfy
\begin{eqnarray}
\lim_{{\bm q}\to{\bm 0}}\tilde{\Pi}_5^{\zeta\zeta}(0,{\bm q}) &=& -N_c\int\frac{d^3p}{(2\pi)^3}{\cal I}_\zeta({\bm p})\frac{\partial f_F(E_{\bm p}^\zeta)}{\partial|{\bm p}|} \ ,\nonumber\\ 
\lim_{q_0\to0}\tilde{\Pi}_5^{\zeta\zeta}(q_0,{\bm 0}) &=& 0\ ,  \label{Pi5Summary1}
\end{eqnarray}
[${\cal I}_\zeta({\bm p})$ is a function of ${\bm p}$~\footnote{The expression of ${\cal I}_\zeta({\bm p})$ may be intricate, so that we do not show it explicitly here.}] for $\tilde{\Pi}_5^{++}$, $\tilde{\Pi}_5^{--}$ and $\tilde{\Pi}_5^{\rm {aa}}$, while
\begin{eqnarray}
\lim_{{\bm q}\to{\bm 0}}\tilde{\Pi}_5^{\zeta'\zeta}(0,{\bm q}) = \lim_{q_0\to0}\tilde{\Pi}_5^{\zeta'\zeta}(q_0,{\bm 0}) \label{Pi5Summary2}
\end{eqnarray}
for the remaining ones $(\zeta\neq\zeta')$. In other words, only the diagrams of (i), (ii), and (iii) in Fig.~\ref{fig:Diagrams}, which correspond to the scattering processes of the identical particle by the magnetic field: ${\rm q.p.}\to{\rm q.p.}$, ${\rm q.h.}\to{\rm q.h.}$, and ${\rm a.p.}\to {\rm a.p.}$ (or {\it intraband scattering processes} in the terminology of condensed matter physics), can generate the difference between the results in static limit and dynamical limit. Such a difference between the static and dynamical limits was also found in Ref.~\cite{Araki:2020rok}. In this reference, we investigated the spin-orbital crossed susceptibility, namely the response of spin polarization to a magnetic field, in which a relativistic (linear) band and a nonrelativistic band coexist.


\begin{figure*}[htbp]
    \begin{minipage}[t]{0.66\columnwidth}
        \begin{center}
            \includegraphics[clip, width=1.0\columnwidth]{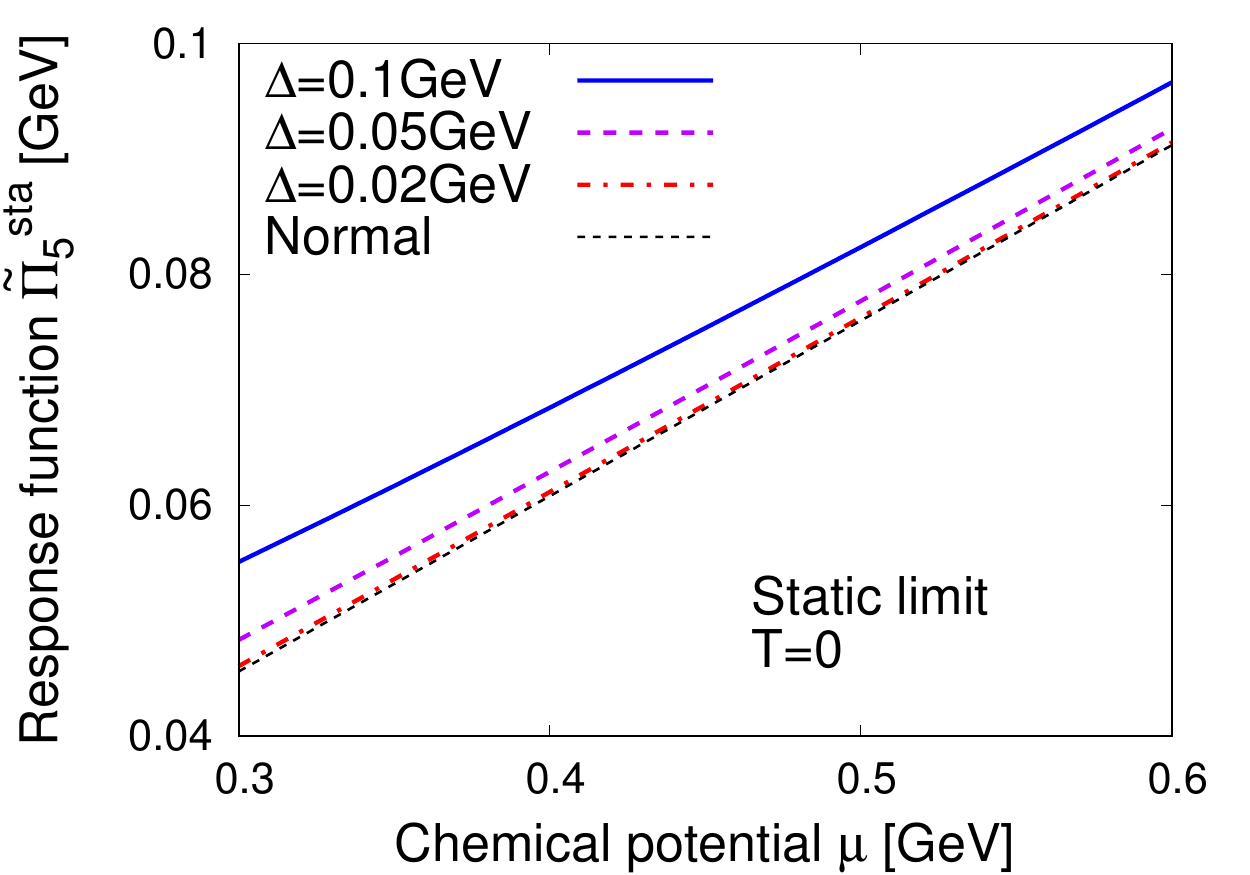}
        \end{center}
    \end{minipage}%
    \begin{minipage}[t]{0.66\columnwidth}
        \begin{center}
            \includegraphics[clip, width=1.0\columnwidth]{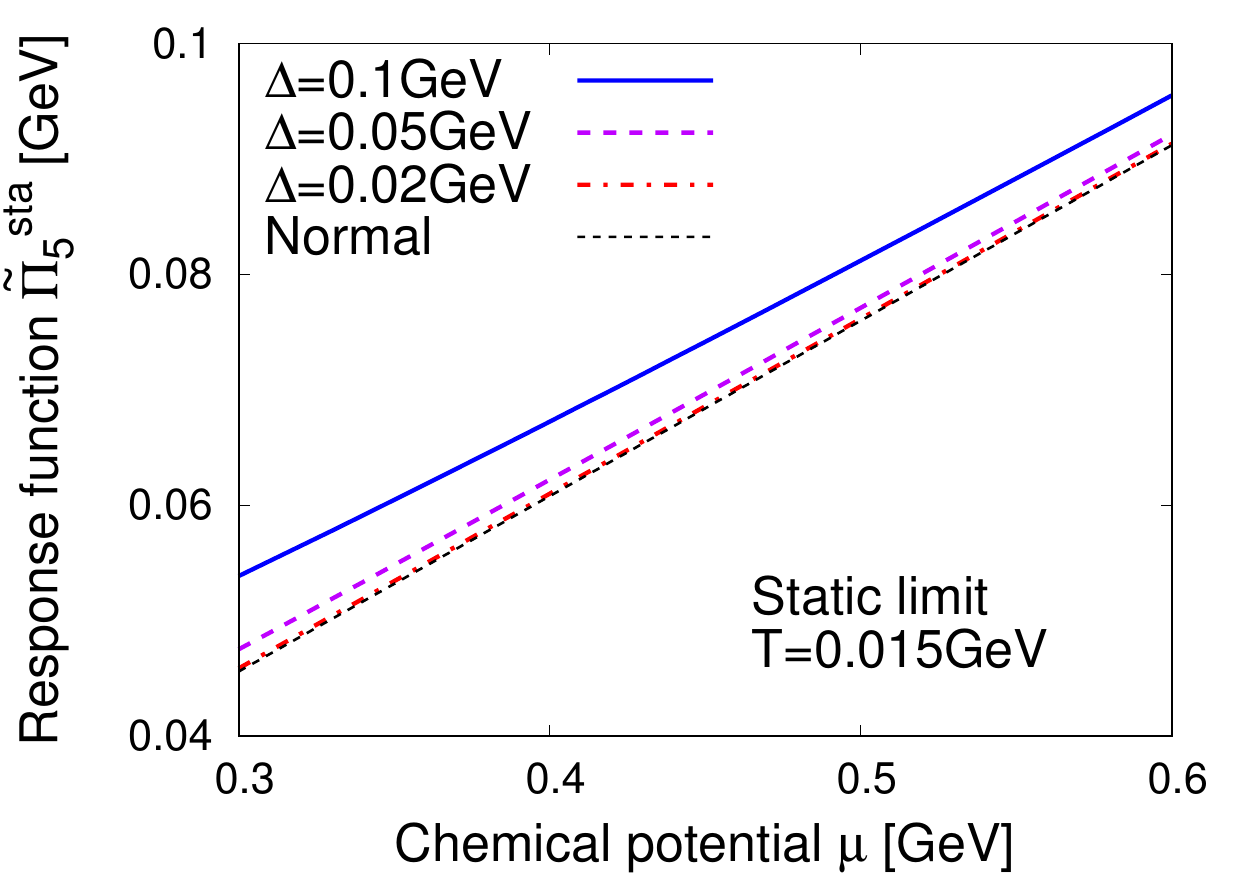}
        \end{center}
    \end{minipage}%
    \begin{minipage}[t]{0.66\columnwidth}
        \begin{center}
            \includegraphics[clip, width=1.0\columnwidth]{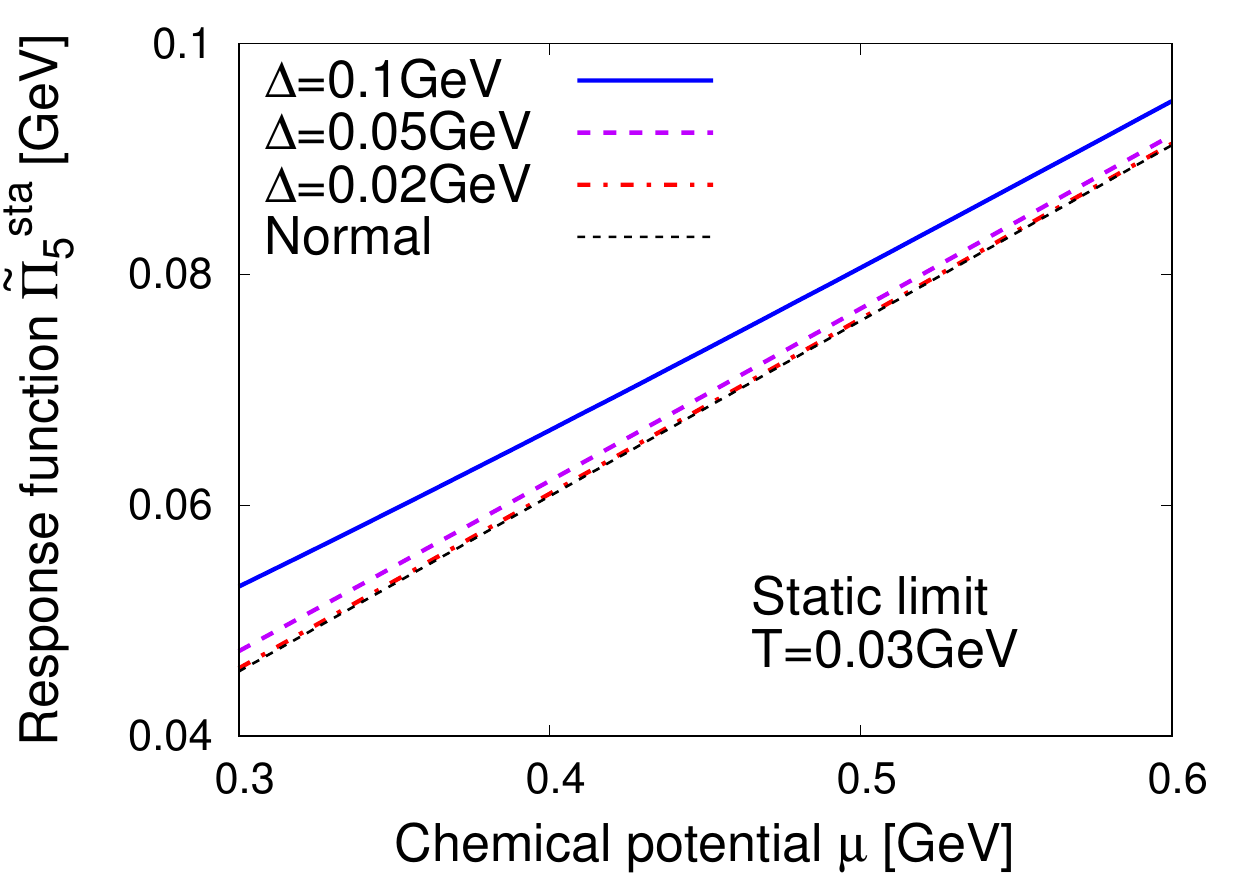}
        \end{center}
    \end{minipage}
\caption{The $\mu$ dependence of response function in static limit ($\tilde{\Pi}_5^{\rm sta}$) with $\Delta=0.02$ GeV (red dashed-dotted), $\Delta=0.05$ GeV (purple dashed), and $\Delta=0.1$ GeV (blue solid). The black dashed curve indicates the response function in the normal matter given in Eq.~(\ref{Pi5NormS}).}
\label{fig:StaticMuDep}
\end{figure*}
\begin{figure*}[htbp]
    \begin{minipage}[t]{0.66\columnwidth}
        \begin{center}
            \includegraphics[clip, width=1.0\columnwidth]{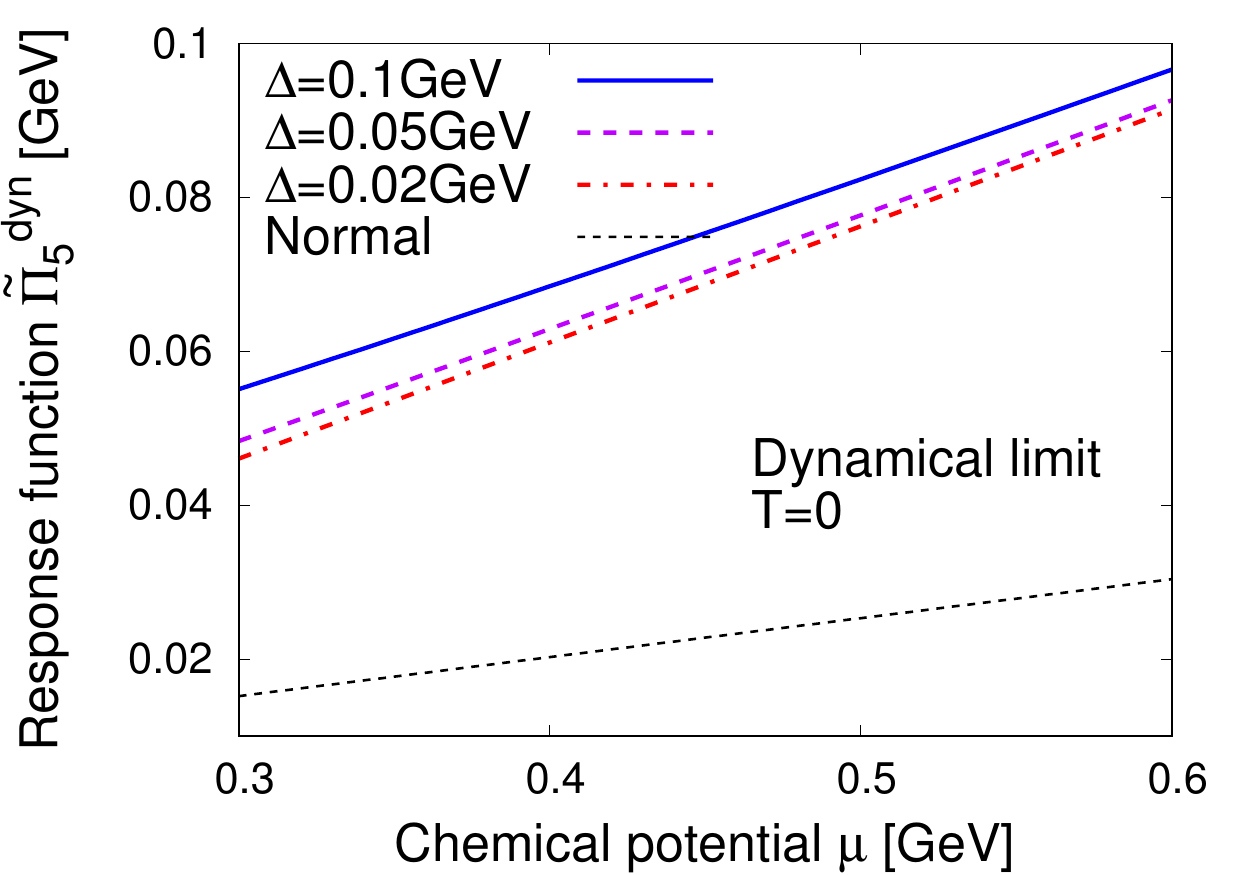}
        \end{center}
    \end{minipage}%
    \begin{minipage}[t]{0.66\columnwidth}
        \begin{center}
            \includegraphics[clip, width=1.0\columnwidth]{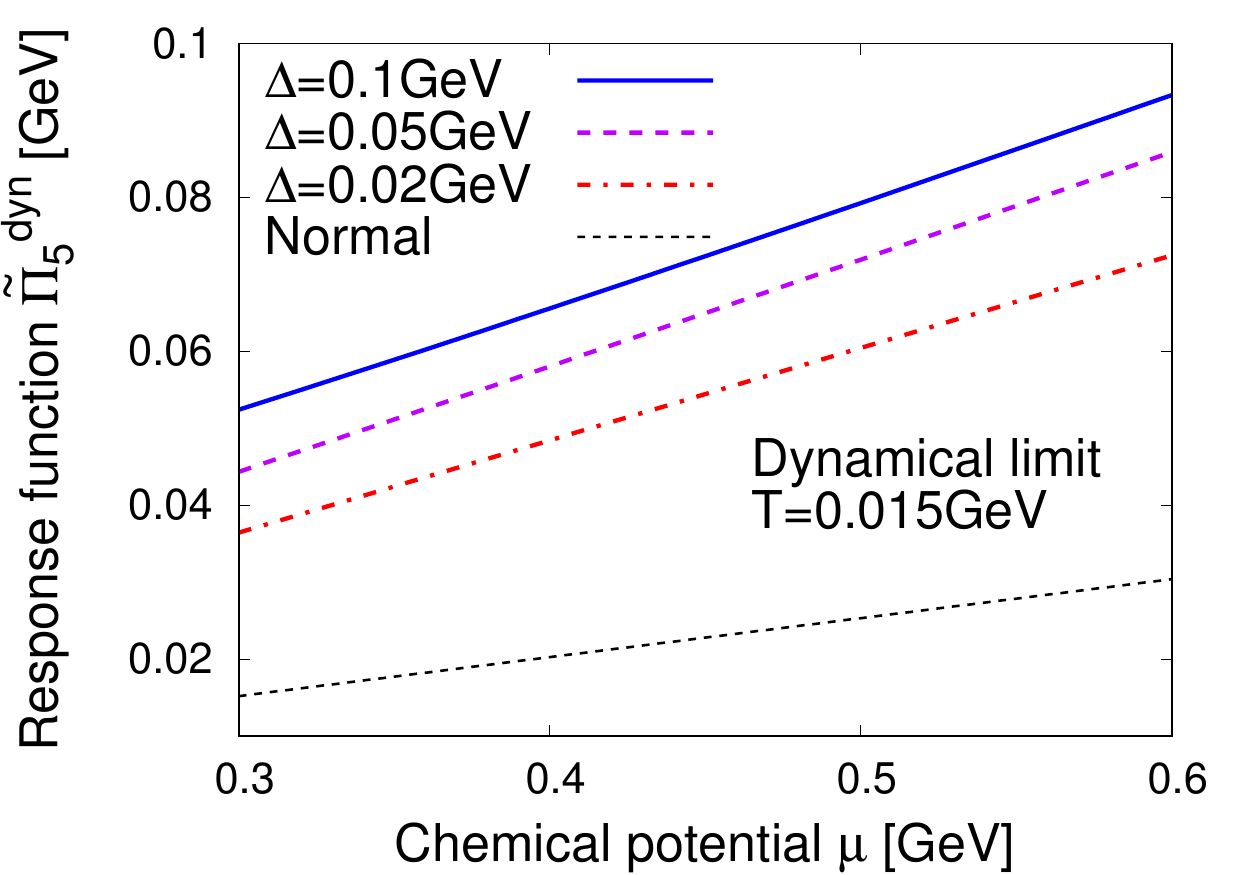}
        \end{center}
    \end{minipage}%
    \begin{minipage}[t]{0.66\columnwidth}
        \begin{center}
            \includegraphics[clip, width=1.0\columnwidth]{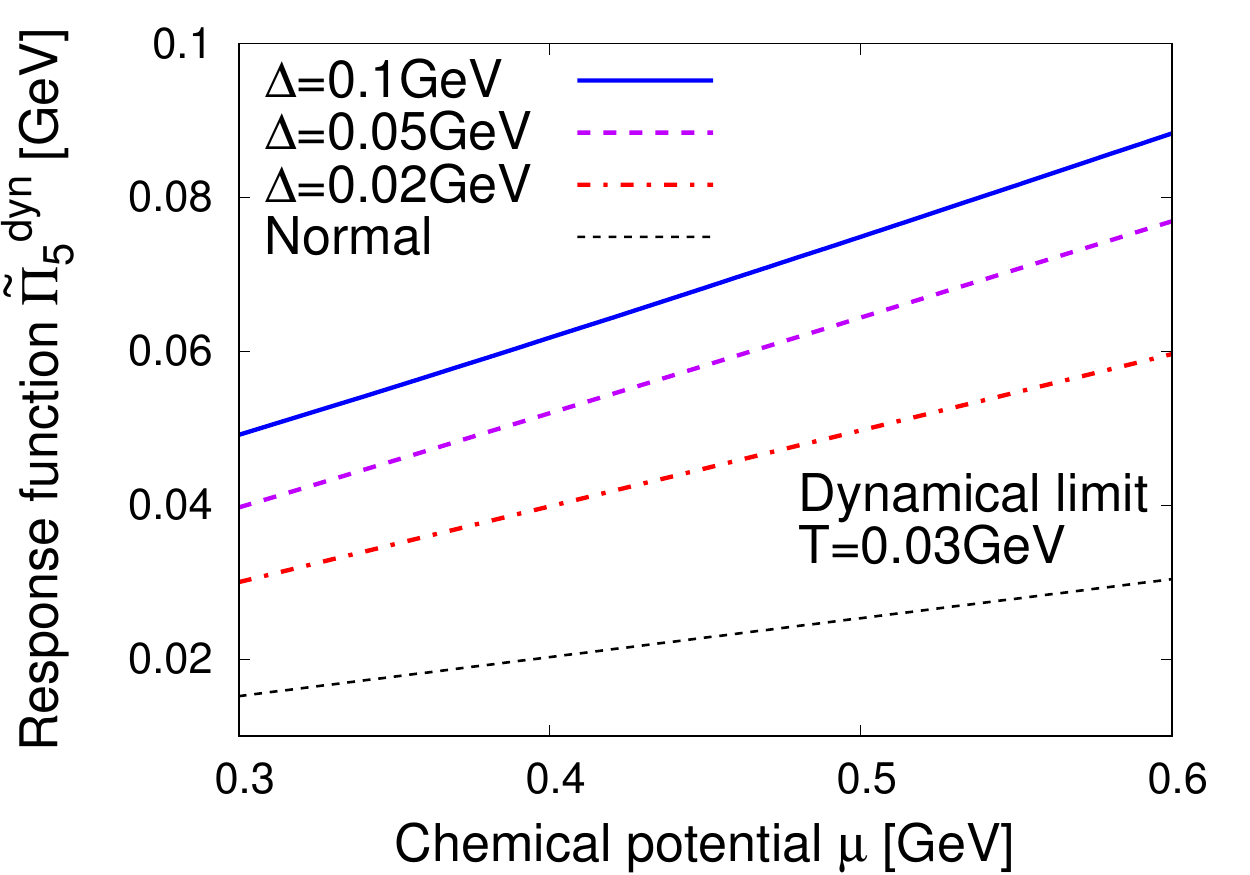}
        \end{center}
    \end{minipage}
\caption{The $\mu$ dependence of response function in dynamical limit ($\tilde{\Pi}_5^{\rm dyn}$) with $\Delta=0.02$ GeV (red dashed-dotted), $\Delta=0.05$ GeV (purple dashed), and $\Delta=0.1$ GeV (blue solid). The black dashed curve indicates the response function in the normal matter given in Eq.~(\ref{Pi5NormD}).}
\label{fig:DynamicalMuDep}
\end{figure*}

\begin{figure}[htbp]
    \begin{minipage}[t]{1.0\columnwidth}
        \begin{center}
            \includegraphics[clip, width=1.0\columnwidth]{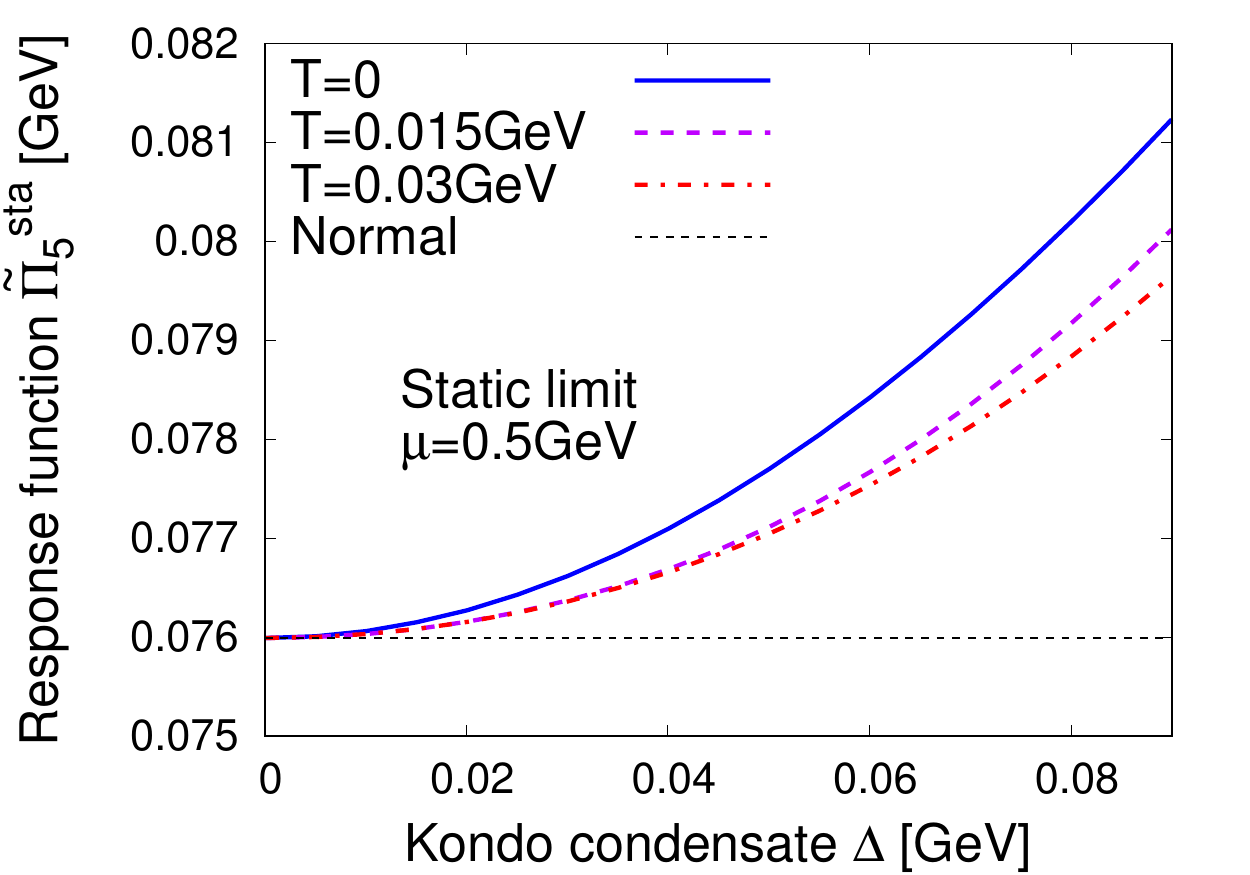}
            \includegraphics[clip, width=1.0\columnwidth]{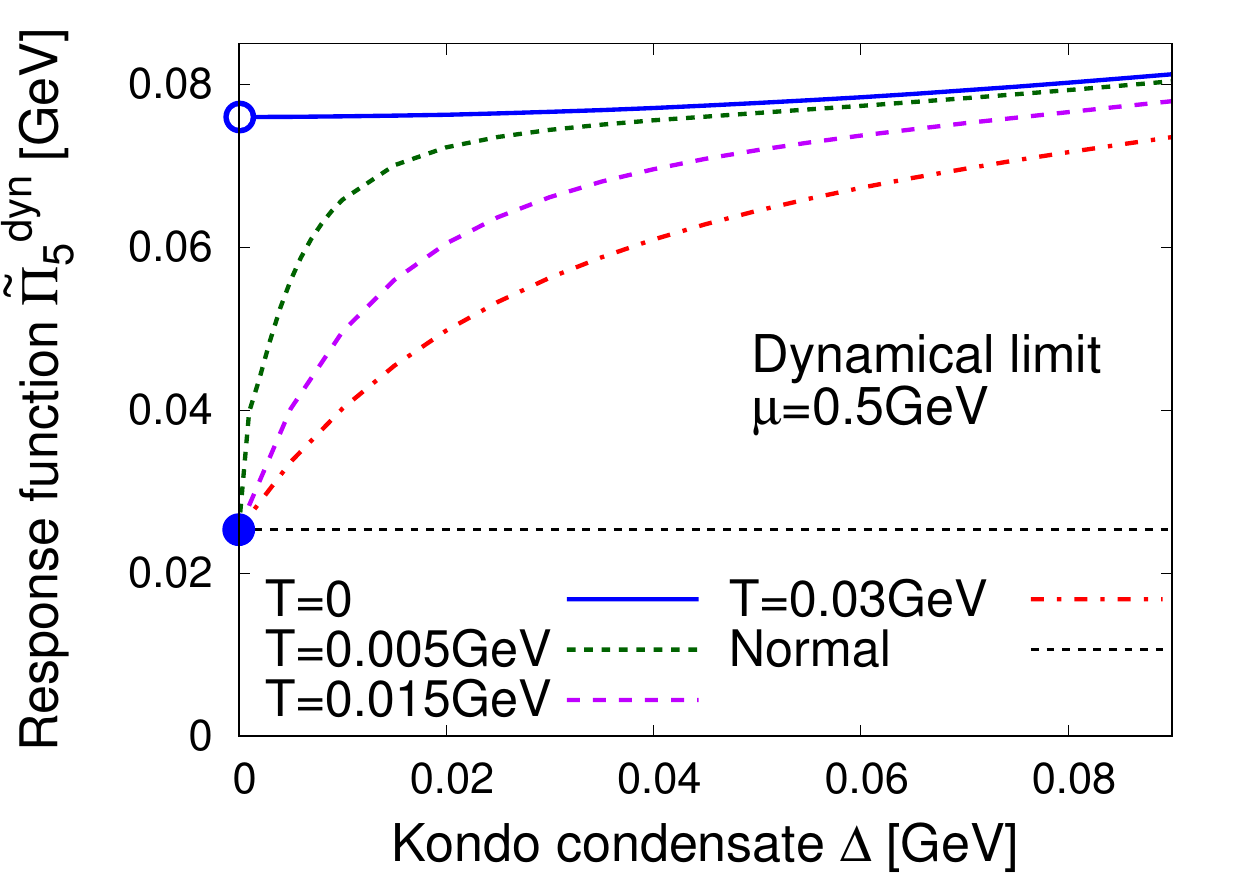}
        \end{center}
    \end{minipage}
\caption{The gap dependence of response function in static limit ($\tilde{\Pi}_5^{\rm sta}$) (upper panel) and in dynamical limit ($\tilde{\Pi}_5^{\rm dyn}$) (lower panel) at $\mu=0.5$ GeV and at several temperatures. The black dashed line indicates the response function in the normal matter given in Eq.~(\ref{Pi5NormS}).
Blue filled or open circle in dynamical limit mean the discontinuous behavior at $T=0$ (see the main text).
}
\label{fig:DeltaDep}
\end{figure}

\begin{figure}[htbp]
    \begin{minipage}[t]{1.0\columnwidth}
        \begin{center}
            \includegraphics[clip, width=1.0\columnwidth]{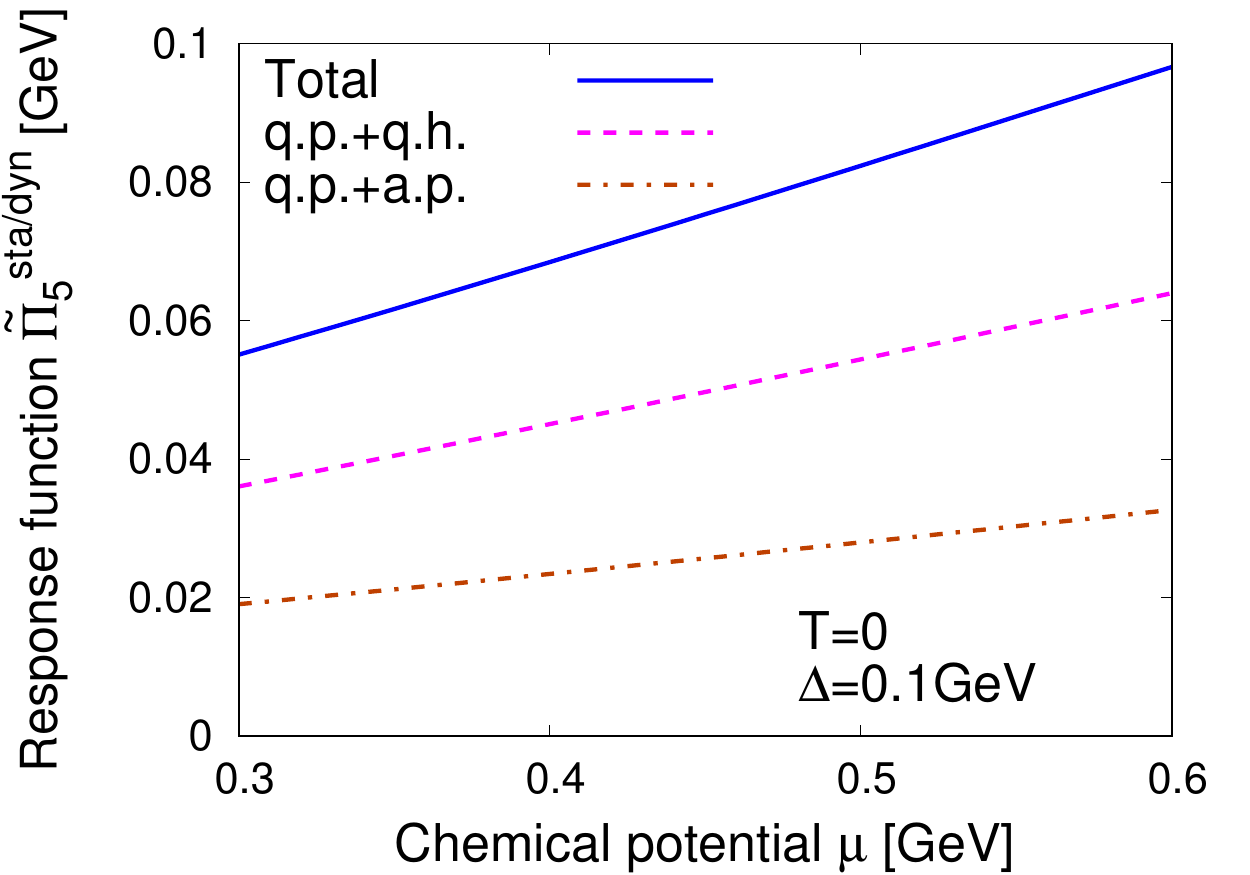}
            \includegraphics[clip, width=1.0\columnwidth]{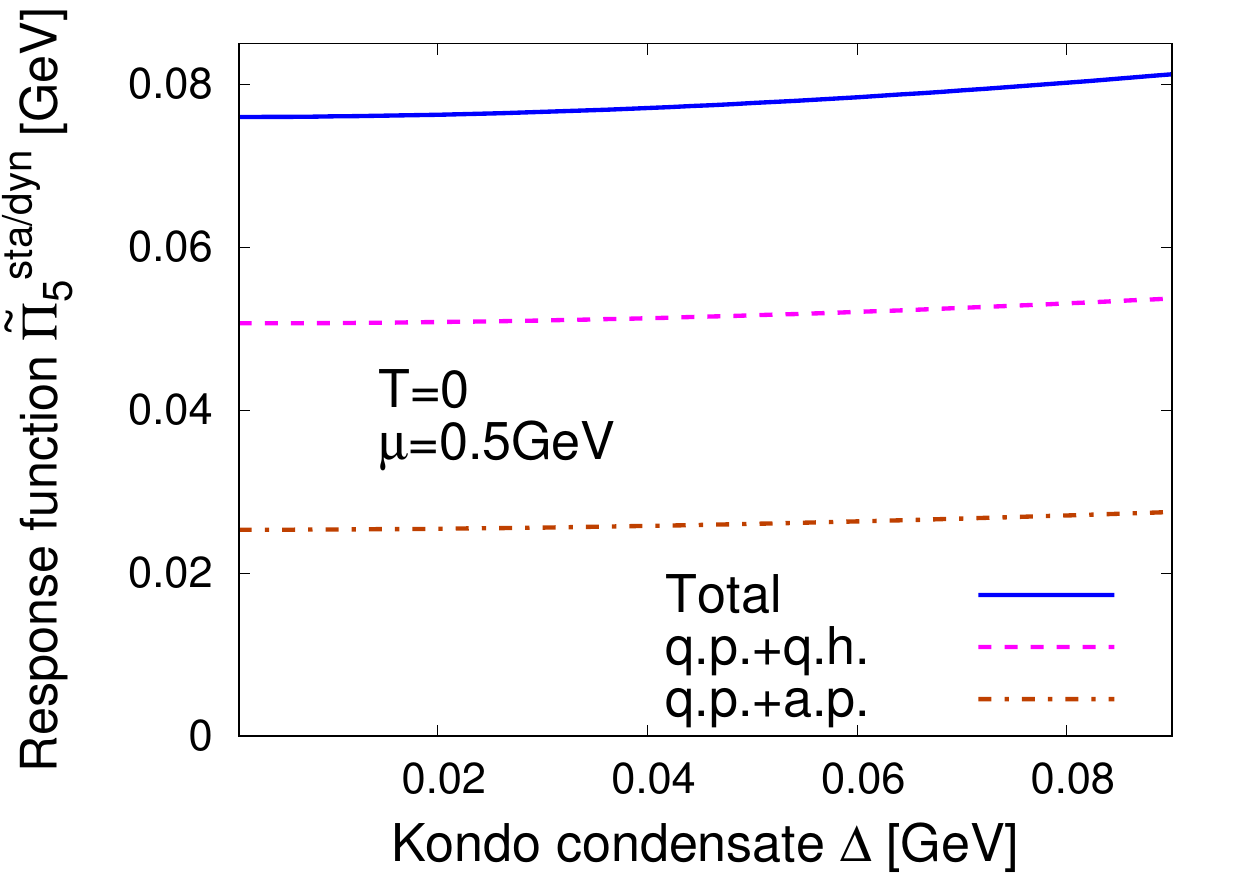}
        \end{center}
    \end{minipage}
\caption{The $\mu$ dependence at $T=0$ and $\Delta=0.1$ GeV of each process contributing to the response function (upper panel), and the $\Delta$ dependence at $T=0$ and $\mu=0.5$ GeV of them (lower panel). We compare the total value of response function (blue solid), the sum of pair creation and pair annihilation processes of ${\rm q.p.} +{\rm q.h.}$ (magenta dashed), and the sum of pair creation and pair annihilation processes of ${\rm q.p.} +{\rm a.p.}$ (brown dashed-dotted).} 
\label{fig:Components}
\end{figure}

\section{Numerical results}
\label{sec:Results}
In Sec.~\ref{sec:Analysis} we showed the detailed procedure to compute the axial current induced by a magnetic field in the presence of Kondo condensate in static and dynamical limits. In this section, we present numerical results of Eq.~(\ref{Pi5Tot}) and examine the impact of Kondo effect on CSE. For the sake of clarity, here we do not solve the gap equation to determine the magnitude, and its density dependence of $\Delta$ and simply regard it as a constant parameter. 

\subsection{Density dependence of the CSE}
\label{sec:Result1}

First, we show the density dependence of the response function of CSE obtained in the present work.

In Fig.~\ref{fig:StaticMuDep} and Fig~\ref{fig:DynamicalMuDep}, the $\mu$ dependences of response function $\tilde{\Pi}_5^{\rm sta}$ in static limit and $\tilde{\Pi}_5^{\rm dyn}$ in dynamical limit with $\Delta=0.02$ GeV (red), $\Delta=0.05$ GeV (purple), and $\Delta=0.1$ GeV (blue) are plotted, respectively. The black dashed line indicates the response function in the normal matter ($\Delta=0$), i.e., 
\begin{eqnarray}
\tilde{\Pi}^{\rm sta}_5|_{\Delta=0} &=& \frac{N_c}{2\pi^2}\mu
 \ , \label{Pi5NormS}\\
\tilde{\Pi}^{\rm dyn}_5|_{\Delta=0} &=& \frac{N_c}{6\pi^2}\mu 
\ , \label{Pi5NormD}
\end{eqnarray}
at any temperatures. We have put this line so as to compare the results with and without Kondo effect more clearly. We have shown the results in the range of $0.3\, {\rm GeV}<\mu<0.6\, {\rm GeV}$ in which the Kondo condensate is expected to appear significantly within  reasonable model parameters, as suggested by the previous works~\cite{Yasui:2016svc,Yasui:2017izi}.

Figures~\ref{fig:StaticMuDep} and~\ref{fig:DynamicalMuDep} show that the Kondo effect enhances the CSE in both static limit and dynamical limit. In particular, the response function in dynamical limit can be enhanced by a factor of approximately three. Besides, those figures show that the results in static limit and dynamical limit at $T=0$ coincide with each other: $\tilde{\Pi}^{\rm sta}_5 = \tilde{\Pi}^{\rm dyn}_5$. Note that the magnitude of $\Delta$ was found to be $\Delta\approx 0.085$ GeV at $\mu=0.5$ GeV in the previous work based on the NJL-type analysis~\cite{Yasui:2016svc,Yasui:2017izi}. Therefore, we can conclude that the influence of Kondo effect on CSE with a reasonable value of $\Delta$ is significantly large, showing that the heavy impurities in quark matter is expected to play an important role in the transport phenomena induced by a magnetic field. The large enhancement of the response function in dynamical limit is consistent with the finding in Ref.~\cite{Araki:2020rok}.\footnote{In Ref.~\cite{Araki:2020rok} we have seen that the response function to a magnetic field in dynamical limit gets enhanced for general dispersions with a condensate. The large enhancement found in the present study can be understood as its extreme case in the heavy-quark limit, i.e., the heavy-quark dispersion is flat in momentum space.}

\subsection{$\Delta$ dependence of the CSE}
\label{sec:Result2}

Next, we study the $\Delta$ dependence of the CSE to examine the influence of Kondo condensate $\Delta$ on the CSE in more detail.

In Fig.~\ref{fig:DeltaDep} we display the $\Delta$ dependence of the response function in static limit (upper panel) and dynamical limit (lower panel) at $\mu=0.5$ GeV and at several temperatures. The black dashed line indicates the response function in the normal matter given in Eq.~(\ref{Pi5NormS}). 
The lower panel tells us that, $\tilde{\Pi}_5^{\rm dyn}$ for small $\Delta$ with $\Delta\neq0$ is enhanced largely due to an effect generated by the Kondo condensate, although it converges on the result of normal matter in dynamical limit provided in Eq.~(\ref{Pi5NormD}) at $\Delta=0$. Such a sudden and large enhancement around $\Delta=0$ stems from the fact that the effect of $\Delta$ is nonperturbative around the Fermi level, which can be understood as follows. The CSE is mostly generated by the response of fermions in the vicinity of the Fermi level ($|{\bm p}|\approx\mu$) in the momentum integrals of the loops. However, the factor of $\sqrt{(|{\bm p}|-\mu)^2+8\Delta^2}$ appearing in $E_{\bm p}^{\pm}$ prohibits us from expanding $E_{\bm p}^\pm$ with respect to $\Delta$ around $|{\bm p}|=\mu$ now, because the magnitudes of $(|{\bm p}|-\mu)^2$ and $8\Delta^2$ are comparable order in this case. As a result, the Taylor expansion of the response function with respect to $\Delta$ should fail, and hence the modification of $\tilde{\Pi}_5^{\rm dyn}$ at small $\Delta$ is significant.

The signature of the nonperturbative effect by $\Delta$ can be seen prominently at $T=0$ in the lower panel of Fig.~\ref{fig:DeltaDep}. Namely, at $T=0$, $\tilde{\Pi}_5^{\rm dyn}$ at $\Delta\neq0$ varies continuously, but only at $\Delta=0$ it jumps to the result of normal matter in dynamical limit discontinuously. Here, we provide some mathematical explanations for the origin of the discontinuity. As shown in Eqs.~(\ref{Pi5Summary1}) and~(\ref{Pi5Summary2}), the difference between the results in static and dynamical limits originates from the scattering process of the identical particles, i.e., from the diagrams (i), (ii), and (iii) in Fig.~\ref{fig:Diagrams}. Below we focus on the diagram (i). At zero temperature, from Eq.~(\ref{Pi5Summary1}), the response functions from the diagram (i) in static and dynamical limits are of the forms
\begin{eqnarray}
&&\lim_{{\bm q}\to{\bm 0}}\tilde{\Pi}_5^{++}(0,{\bm q})\overset{T=0}{=} -N_c\int\frac{d^3p}{(2\pi)^3}{\cal I}_+({\bm p}) \frac{E_{\bm p}^{+}}{E_{\bm p}^{+}-E_{\bm p}^{-}}\delta(E_{\bm p}^{+}) \ , \nonumber\\
&&\lim_{q_0\to0}\tilde{\Pi}_5^{++}(q_0,{\bm 0}) \overset{T=0}{=} 0 \ , \label{limitDeltaCSE}
\end{eqnarray}
respectively. 
These results coincide with each other, even when $\Delta$ is infinitesimally small, since $E_{\bm p}^{+}$ and $E_{\bm p}^{+}-E_{\bm p}^{-}$ are always positive in this case, resulting in 
\begin{eqnarray}
\lim_{{\bm q}\to{\bm 0}}\tilde{\Pi}_5^{++}(0,{\bm q})\overset{T=0}{=} 0\ . \label{limitDeltaCSE2}
\end{eqnarray}
Hence, $\tilde{\Pi}^{\rm sta}_5 = \tilde{\Pi}^{\rm dyn}_5$ holds for $\Delta\neq0$. On the other hand, when $\Delta$ is exactly zero, the upper equation of Eq.~(\ref{limitDeltaCSE}) is no longer zero 
because $E_{\bm p}^+|_{\Delta=0} = \theta(|{\bm p}|-\mu)(|{\bm p}|-\mu)$ and $E_{\bm p}^-|_{\Delta=0} = \theta(\mu-|{\bm p}|)(|{\bm p}|-\mu)$ can reach zero, yielding $\tilde{\Pi}^{\rm sta}_5 \neq \tilde{\Pi}^{\rm dyn}_5$. As a result, the discontinuous behavior of the blue curve of lower panel in Fig.~\ref{fig:DeltaDep} at $\Delta\to0$ emerges. 
The similar explanation can be also applied to (ii) and (iii). The discontinuity at $\Delta=0$, however, should not be taken too seriously, because in realistic measurements, this discontinuity would be smeared by the finite relaxation time.


In association with the sudden and large enhancement at $T\neq0$ (and the discontinuity at $T=0$) in dynamical limit, in Ref.~\cite{Satow:2014lva}, it was found that, in the normal matter the damping rate of free quark enhances the CSE in dynamical limit to coincide with the one in static limit.
Thus, by focusing on the fact that nonzero $\Delta$ causes the sudden and large enhancement of the response function in dynamical limit resulting in the similar behavior to static limit, as shown in Fig.~\ref{fig:DeltaDep}, 
we speculate that there would be an analogy between the inclusion of damping rate in the normal matter and the presence of Kondo condensate $\Delta$. 
A detailed comparison between the present analysis and the one in Ref.~\cite{Satow:2014lva} is left for future study.

\subsection{Separate contributions to the CSE}
\label{sec:Result3}

Finally, we give a physical interpretation of how the each diagram in Fig.~\ref{fig:Diagrams} contributes to the response function. 

Again for the sake of clarity, we restrict ourselves to vanishing temperature. At $T=0$, the Pauli principle forces the intermediate states of the loops to excite above the Fermi level, i.e., at least one of the modes inside the loop must be ${\rm q.p.}$ because only this mode lies above the Fermi level as shown in Fig.~\ref{fig:Dispersions}. In fact we can confirm that only the diagrams (iv) and (v) for ${\rm q.p.}+{\rm q.h.}$ and (viii) and (ix) for ${\rm q.p.}+{\rm a.p.}$
in Fig.~\ref{fig:Diagrams} contribute to the CSE through the pair creation and pair annihilation processes. 
We note that the contribution from the diagram (i) vanishes at zero temperature as explained around Eqs.~(\ref{limitDeltaCSE}) and~(\ref{limitDeltaCSE2}) in detail.

In Fig.~\ref{fig:Components} we show the $\mu$ dependence of each process contributing to the response function at $T=0$ and $\Delta=0.1$ GeV (upper panel), and the $\Delta$ dependence of them at $T=0$ and $\mu=0.5$ GeV (lower panel). We note that the point for $\Delta=0$ is excluded in the lower panel to avoid the discontinuity explained in Sec.~\ref{sec:Result2}.
In Fig.~\ref{fig:Components}, the blue curve indicates the total value of response function, while the magenta curve corresponds to the sum of  ${\rm q.p.} +{\rm q.h.}$ pair creation and pair annihilation, and the brown one corresponds to the sum of ${\rm q.p.} +{\rm a.p.}$ pair creation and pair annihilation. Only these processes contribute to the response function due to the reason stated above. Note that at zero temperature the response functions in static and dynamical limits coincide: $\tilde{\Pi}_5^{\rm sta} = \tilde{\Pi}_5^{\rm dyn}$ as found in Sec.~\ref{sec:Result1}.
Figure~\ref{fig:Components} shows that the contribution from ${\rm q.p.}+{\rm q.h.}$ process and that of ${\rm q.p.}+{\rm a.p.}$ are approximately in the ratio of $2:1$ for any values of $\mu$ or $\Delta$. This ratio is easily understood because the former process includes two modes (both ${\rm q.p.}$ and ${\rm q.h.}$) lying in the vicinity of the Fermi level while the latter includes only one mode (only $ {\rm q.p.}$) there. 

At finite temperature, the above picture does not change significantly as long as the temperature is smaller than the chemical potential $\mu$. In fact this is the case in the present work so that the contributions from diagrams (i), (ii), (iii), (vi), and (vii) in Fig.~\ref{fig:Diagrams} remain small.

\section{Gauge invariance}
\label{sec:GaugeInvariance}
Here we discuss the gauge invariance in our calculations. If we assume a {\it charged} Kondo condensate with $e_q\neq e_Q$, then the condensate carries an electric charge and leads to the spontaneous breakdown of the electromagnetic $U(1)_{\rm EM}$ invariance.
In this case, we have to correct the vertex for the coupling between the quarks and the magnetic field in order to recover the gauge invariance, leading to the appearance of the NG mode. This treatment is an analogue of the derivation of NG mode appearing in the theory of superconductivity~\cite{schrieffer1999theory}, or that of pion in chiral symmetry breaking~\cite{cheng1994gauge}.
On the other hand, even when we assume the {\it charge-neutral} Kondo condensate as in the present study [recall Eq. (5)], the nontrivial structure of the hedgehog ansatz~(\ref{Hedgehog}) still requires to correct the vertex for the gauge invariance. In actuality, such a vertex correction is found not to affect our results considerably as it will be explained later. Rigorous arguments on the gauge invariance are, however, important and useful in the study of response to the external gauge field.



The underlying theory of the effective model~(\ref{LTotal}), i.e., QCD and QED, must possess the electromagnetic $U(1)_{\rm EM}$ gauge invariance. Thus the Ward-Takahashi identity together with the Dyson-Schwinger equation with respect to vector-axialvector current correlator tells us that the vertex $\Gamma^\mu$ between the gauge field and the quarks must obey the following identity~\cite{Nambu:1960tm}
\begin{eqnarray}
q_\mu\Gamma^\mu = -i{\cal G}^{-1}(p_0',{\bm p}')Q + Qi{\cal G}^{-1}(p_0,{\bm p}) \label{WTIdentity}
\end{eqnarray}
($p'=p+q$), where $\Gamma^\mu$ is the $6\times 6$ vertex matrix 
\begin{eqnarray}
\Gamma^\mu \equiv \left(
\begin{array}{cc}
\Gamma^\mu_{A\bar{\psi}\psi} & \Gamma^\mu_{A\bar{\psi}\Psi_v} \\
\Gamma^\mu_{A\Psi_v^\dagger\psi} & \Gamma^\mu_{A\Psi_v^\dagger\Psi_v} \\
\end{array}
\right) .
\end{eqnarray}
Here, for example, the subscript of $\Gamma^\mu_{A\bar{\psi}\psi}$ represents the type of gauge interaction such as $A^\mu\bar{\psi}\psi$. $Q$ is a charge matrix
\begin{eqnarray}
Q = \left(
\begin{array}{ccc}
e_q & 0 \\
0 & e_Q \\
\end{array}
\right)\ ,
\end{eqnarray}
where we have assumed the electric charges of light and heavy quarks are not identical for a general discussion. $i{\cal G}^{-1}(p_0,{\bm p})$ is the inverse of the full Green's function in general. Within the present work, it can be read by the Lagrangian~(\ref{LTotal}) [where we do not necessarily assume Eq.~(\ref{ChargeE})]:
\begin{eqnarray}
i{\cal G}^{-1}(p_0,{\bm p}) = \left(
\begin{array}{cc}
i{\cal G}_{\bar{\psi}\psi}^{-1}(p_0,{\bm p}) & i{\cal G}_{\bar{\psi}\Psi_v}^{-1}(p_0,{\bm p}) \\
i{\cal G}_{\Psi_v^\dagger\psi}^{-1}(p_0,{\bm p}) & i{\cal G}_{\Psi_v^\dagger\Psi_v}^{-1}(p_0,{\bm p}) \\
\end{array}
\right)
\end{eqnarray}
with
\begin{eqnarray}
i{\cal G}_{\bar{\psi}\psi}^{-1}(p_0,{\bm p}) &=&  \Slash{p}+\mu\gamma_0 \ , \nonumber\\
 i{\cal G}_{\bar{\psi}\Psi_v}^{-1}(p_0,{\bm p})  &=& \Delta\left(
\begin{array}{c}
1 \\
-\hat{\bm p}\cdot{\bm \sigma} \\
\end{array}
\right)\ , \nonumber\\
i{\cal G}_{\Psi_v^\dagger\psi}^{-1}(p_0,{\bm p}) &=& \Delta\left(
\begin{array}{cc}
 1& \hat{\bm p}\cdot{\bm \sigma} \\
\end{array}
\right)\ , \nonumber\\
i{\cal G}_{\Psi_v^\dagger\Psi_v}^{-1}(p_0,{\bm p})  &=& p_0 {\bm 1}\ .
\end{eqnarray}
Thus, each component of the Ward-Takahashi identity in Eq.~(\ref{WTIdentity}) reads
\begin{eqnarray}
q_\mu\Gamma^\mu_{A\bar{\psi}\psi} &=& -e_q\Slash{q}\ , \nonumber\\
q_\mu\Gamma^\mu_{A\bar{\psi}\Psi_v} &=& \Delta\left(
\begin{array}{c}
e_q-e_Q \\
e_Q\hat{\bm p}'\cdot{\bm \sigma}-e_q\hat{\bm p}\cdot{\bm \sigma} \\
\end{array}
\right) \ , \nonumber\\
q_\mu\Gamma^\mu_{A\Psi_v^\dagger\psi} &=&\Delta\left(
\begin{array}{cc}
e_Q-e_q & -e_q\hat{\bm p}'\cdot{\bm \sigma} + e_Q\hat{\bm p}\cdot{\bm \sigma} \\
\end{array}
\right)\ ,  \nonumber\\
q_\mu \Gamma^\mu_{A\Psi_v^\dagger\Psi_v} &=& -e_Qq_0 {\bm 1}\ . \label{WTICorrections}
\end{eqnarray}
For the diagonal components, $A^\mu\bar{\psi}\psi$ and $A^\mu\Psi_v^\dagger\Psi_v$, we can safely use their bare vertices:
\begin{eqnarray}
\Gamma^\mu_{A\bar{\psi}\psi} = -e_q\gamma^\mu\ , \ \ \Gamma^\mu_{A\Psi_v^\dagger\Psi_v} = -e_Q\delta^{\mu0} {\bm 1}\ .
\end{eqnarray}
On the other hand, for the off-diagonal components, $A^\mu\bar{\psi}\Psi_v$ and $A^\mu\Psi_v^\dagger\psi$, we cannot use their bare vertices, and they need to be corrected in such a way that they satisfy the identities in Eq.~(\ref{WTICorrections}).

When $e_q\neq e_Q$, the second and third identities in Eq.~(\ref{WTICorrections}) suggest an existence of collective mode (NG mode) coupling with the vertex, because the RHS in the limit of $q_0,{\bm q}\to0$ does not vanish in this case~\cite{Nambu:1960tm}. The appearance of NG mode can be also understood by the fact that $\Delta\sim\langle\Psi_v^\dagger\psi\rangle$ with $e_q\neq e_Q$ carries its nonzero electric charge leading to the spontaneous breakdown of $U(1)_{\rm EM}$ symmetry. This collective mode (NG mode) is analogous to the one in the theory of superconductivity~\cite{schrieffer1999theory}.

On the other hand, when $e \equiv e_q=e_Q$ as we have assumed in the present study, the NG mode related to the spontaneous breakdown of $U(1)_{\rm EM}$ symmetry does not appear. However, modifications of the vertices are still required, as seen by the RHS of second and third identities in Eq.~(\ref{WTICorrections}):
\begin{eqnarray}
q_\mu\Gamma^\mu_{A\bar{\psi}\Psi_v} &=& e\Delta\left(
\begin{array}{c}
0 \\
(\hat{\bm p}'-\hat{\bm p})\cdot{\bm \sigma} \\
\end{array}
\right) \ , \nonumber\\
q_\mu\Gamma^\mu_{A\Psi_v^\dagger\psi} &=&e\Delta\left(
\begin{array}{cc}
0 & -(\hat{\bm p}'-\hat{\bm p})\cdot{\bm \sigma} \\
\end{array}
\right)\ . \label{WTICorrections2}
\end{eqnarray}
Obviously the modifications originate from the nontrivial momentum alignment caused by hedgehog ansatz~(\ref{Hedgehog}). Equation~(\ref{WTICorrections2}) can be solved with respect to $\Gamma^\mu_{A\bar{\psi}\Psi_v}$ and $\Gamma^\mu_{A\Psi_v^\dagger\psi}$ by imposing some assumptions. For example, if we restrict ourselves to the vicinity of $q^\mu=0$, then the Ward-Takahashi identities for $\Gamma^\mu_{A\bar{\psi}\Psi_v}$ and $\Gamma^\mu_{A\Psi_v^\dagger\psi}$ are solved, and we obtain
\begin{eqnarray}
\Gamma^\mu_{A\bar{\psi}\Psi_v} &\approx& -e\Delta\delta^{i\mu}\left(
\begin{array}{c}
0 \\
 \frac{\sigma^i-(\hat{\bm p}\cdot{\bm \sigma})\hat{p}^i}{|{\bm p}|} \\
\end{array}
\right)\ , \nonumber\\
\Gamma^\mu_{A\Psi_v^\dagger\psi} &\approx&  e\Delta\delta^{i\mu}\left(
\begin{array}{cc}
0 & \frac{\sigma^i-(\hat{\bm p}\cdot{\bm \sigma})\hat{p}^i}{|{\bm p}|} \\
\end{array}
\right)\ . \label{VCorrections}
\end{eqnarray}

The vertex corrections in Eq.~(\ref{VCorrections}) can give rise to additional contributions to the axial current as
\begin{eqnarray}
&&\langle \delta\tilde{j}_5^i(i\bar{\omega}_n,{\bm q})\rangle_\beta  \nonumber\\
&\approx&  -N_cT\sum_m\int\frac{d^3p}{(2\pi)^3}{\rm tr}\Big[\Gamma^i\Gamma_5\tilde{\cal G}_0(i{\omega}_m+i\bar{\omega}_n,{\bm p}+{\bm q})\nonumber\\
&& \times \tilde{A}^j(i\bar{\omega}_n,{\bm q})
\left(
\begin{array}{cc}
0 & \Gamma_{A\bar{\psi}\Psi_v}^j \\
\Gamma_{A\Psi_v^\dagger\psi}^j & 0 \\
\end{array}
\right) \tilde{\cal G}_0(i{\omega}_m,{\bm p})\Big]\, , \label{J5corrections}
\end{eqnarray}
[compare to Eq.~(\ref{J5General})]. Here, we note that the RHS of Eq.~(\ref{J5corrections}) is proportional to $\Delta^2$,  because Eq.~(\ref{VCorrections}) is proportional to $\Delta$ and computation of the trace in Eq.~(\ref{J5corrections}) picks up another $\Delta$ in $V_{+}({\bm p})$ or $V_{-}({\bm p})$ in Eq.~(\ref{GZeroElements}). Namely the additional contributions from the vertex corrections may be small in a realistic situation where $\Delta$ is always smaller than $\mu$.

The above discussion suggests that, while the method in deriving the response functions in Sec.~\ref{sec:Results} seem to violate the $U(1)_{\rm EM}$ gauge invariance due to lack of the vertex corrections, such a violation is expected to be relatively small and gives only a subleading effect on our findings as long as we assume $e \equiv e_q=e_Q$.
On the other hand, when we take into account the difference between the electric charges of light and heavy quarks, the effects from the vertex correction would become rather important because of the appearance of NG mode~\cite{Nambu:1960tm}. Therefore, the study of CSE in the presence of Kondo condensate with other choices of electric charges would be of interest, and we leave such an examination for future publication.




\section{Conclusions}
\label{sec:Conclusion}

In this paper we studied the axial current induced by a magnetic field, i.e., the chiral separation effect (CSE) with the Kondo effect. To demonstrate the impact of Kondo effect on the CSE in a transparent way, we employed a simple effective Lagrangian incorporating a condensate between a light quark and a heavy quark, which is the so-called Kondo condensate.

When we evaluate the CSE by means of the linear response theory, we can define the axial current in the two types of limits: static limit and dynamical limit. The former (latter) describes an induced current under an external magnetic field whose time dependence is slower (faster) than the equilibration of the spatial part of the system, and in general they yield different results. Thus we studied the response function of the axial current to the magnetic field in those two limits. We found that the Kondo effect enhances the CSE in both the static limit and dynamical limit, and particularly the CSE in dynamical limit can be increased by a factor of approximately three in the range of the reasonable value of Kondo condensate. Those results clearly show that the Kondo effect arising from heavy impurities in quark matter, can play an important role in the transport phenomena of light quarks induced by a magnetic field.

The results obtained in this paper are expected to contribute to a better understanding of transport phenomena induced by a magnetic field in quark matter with heavy impurities. In condensed matter physics, our findings are related to the electron spin polarization (spin magnetization) in the response to a
magnetic field in Dirac/Weyl semimetals having a band structure including both relativistic and nonrelativistic degrees of freedom,
since the spin polarization is related to axial current by spin-momentum locking~\cite{Araki:2020rok}. 
In particular, the response function of the CSE to a magnetic field is equivalent to the so-called spin-orbital crossed susceptibility~\cite{Tserkovnyak,Koshino,Nakai,Suzuura,Ando,Ominato} which constitutes a part of the magnetic susceptibility, as pointed out in Ref.~\cite{Araki:2020rok}.

In what follows, we give discussions related to the present study. While we investigated the CSE with Kondo effect at vanishing frequency of magnetic field in this paper, our analysis can be applied to the one at finite frequency as in Refs.~\cite{Kharzeev:2009pj,Yee:2009vw,Landsteiner:2013aba,Satow:2014lva,Kharzeev:2016sut} for the CME and in Ref.~\cite{Landsteiner:2013aba} for the CSE. Such a study will be important for a more realistic situation such as (low-energy and peripheral) heavy-ion collisions where the strength of a magnetic field produced after a nucleus-nucleus collision evolves with time. 

Moreover, our analysis can be easily applied to the study of the chiral magnetic effect (CME)~\cite{Kharzeev:2004ey,Kharzeev:2007tn,Kharzeev:2007jp,Fukushima:2008xe} and the chiral vortical effect (CVE)~\cite{Vilenkin:1979ui,Vilenkin:1980zv,Erdmenger:2008rm,Banerjee:2008th,Son:2009tf,Landsteiner:2011cp,Stephanov:2012ki,Abramchuk:2018jhd} with the Kondo effect. The CME describes the vector current induced by a magnetic field in chirality imbalanced matter~\cite{Kharzeev:2004ey,Kharzeev:2007tn,Kharzeev:2007jp,Fukushima:2008xe}, whereas the CVE describes the vector or axial current induced by a magnetic field with rotation in ordinary and/or chirality-imbalanced matter~\cite{Vilenkin:1979ui,Vilenkin:1980zv,Erdmenger:2008rm,Banerjee:2008th,Son:2009tf,Landsteiner:2011cp,Stephanov:2012ki,Abramchuk:2018jhd}, respectively. Without the Kondo effect, while the CSE and CVE were found to be affected by a quark mass~\cite{Metlitski:2005pr,Gorbar:2013upa,Guo:2016dnm,Flachi:2017vlp,Lin:2018aon} or quantum corrections~\cite{Gorbar:2013upa}, the CME may be protected by the axial anomaly. Thus, the study of CVE and CME in addition to CSE toward the comprehensive understanding of chiral transport phenomena with Kondo effect, particularly from the viewpoint of axial anomaly, would be also of great interest.

\section*{Acknowledgements}

D. S. wishes to thank Keio University and Japan Atomic Energy Agency (JAEA) for their hospitalities during his stay there. Y. A. is supported by the Leading Initiative for Excellent Young Researchers (LEADER). K. S. is supported by Japan Society for the Promotion of Science (JSPS) KAKENHI (Grant Nos. JP17K14277
and JP20K14476). S. Y. is supported by JSPS KAKENHI (Grant No. JP17K05435), by the Ministry of Education, Culture, Sports, Science (MEXT)-Supported Program for the Strategic Research Foundation at Private Universities Topological Science (Grant No. S1511006) and by the Interdisciplinary Theoretical and Mathematical Sciences Program
(iTHEMS) at RIKEN.


\appendix

\section{Another derivation of CSE in static limit}
\label{sec:AnotherStatic}

Here, we show another derivation of CSE for $e=e_{q}=e_{Q}$ in the presence of Kondo condensate in static limit. It should be noted that the CSE calculated in this appendix does not include any vertex corrections.

\subsection{Green's function in a magnetic field}
\label{sec:GreenFunction}

To show the CSE in static limit in an alternative way, we go back to the original Lagrangian~(\ref{LTotal}):
\begin{eqnarray} 
{\cal L} &=& \bar{\psi}(i\Slash{D}+\mu\gamma_0)\psi + \Psi_v^\dagger iD_{0}\Psi_v \nonumber\\ 
&+& \Delta\left[\bar{\psi}(1-i{\bm \gamma}\cdot\hat{\bm \nabla})\Psi_v\right] + ({\rm c.c.}) \ , \label{LOriginal}
\end{eqnarray}
and derive the Green's function of quasiparticle in a magnetic field by a different treatment from the one done in main text. In Eq.~(\ref{LOriginal}), $D_\mu\psi=(\partial_\mu+ieA_\mu)\psi$ is a covariant derivative with respect to the magnetic field and the differential operator $-i\hat{\bm \nabla}\equiv (-i{\bm \nabla})/|-i{\bm \nabla}|$ corresponds to the hedgehog ansatz defined by the second line of Eq.~(\ref{Hedgehog}). The presence of a magnetic field breaks the translational invariance perpendicular to the direction of magnetic field, which suggests that it is not straightforward to obtain the Green's function in momentum space. For this reason, we first consider the Green's function in coordinate space. In what follows, we choose $z$ direction as the direction of magnetic field. 

The effective Lagrangian~(\ref{LOriginal}) can be written as
\begin{eqnarray}
{\cal L} = \big( \bar{\psi}, \Psi_v^\dagger\big) i{\cal G}^{-1}\left(
\begin{array}{c}
\psi \\
\Psi_v \\
\end{array}
\right)\ , \\ \nonumber
\end{eqnarray}
where
\begin{eqnarray}
i{\cal G}^{-1} = i{\cal G}_0^{-1} + {\cal V} \label{InvGAppendix}
\end{eqnarray}
is the inverse of the Green's function of the fermions (incorporating the gauge field) in coordinate space, with the free quasiparticle part
\begin{eqnarray}
i{\cal G}_0^{-1} =\left(
\begin{array}{ccc}
i{\partial}_0+\mu & i{\bm \nabla}\cdot{\bm \sigma} & \Delta^* \\
-i{\bm \nabla}{\bm \sigma} & -i{\partial}_0-\mu & \Delta^* i\hat{{\bm \nabla}}\cdot{\bm \sigma} \\
\Delta & -\Delta i \hat{\bm \nabla}\cdot{\bm \sigma} & i{\partial}_0  \\ 
\end{array}
\right) \label{G0Appendix}
 \end{eqnarray}
and the contributions from magnetic field
\begin{eqnarray} 
{\cal V} = \left(
 \begin{array}{ccc}
0 & e{\bm A}\cdot{\bm \sigma} & 0 \\
-e{\bm A}\cdot{\bm \sigma} & 0 & 0 \\
0 & 0 & 0 \\
\end{array}
\right) .\label{VAppendix}
\end{eqnarray}
Now the Landau gauge ${\bm A} = (-B y,0,0)$ is taken. In this paper we assume the magnetic field is so weak that the inverse of the Green's function~(\ref{InvGAppendix}) can be approximated up to the first order of $eB$ as
\begin{eqnarray}
{\cal G} = i \big(i{\cal G}_0^{-1}+{\cal V} \big)^{-1} \approx {\cal G}_0 +{\cal G}_1\  , \label{GApprox}
\end{eqnarray}
with ${\cal G}_1 \equiv {\cal G}_0 i{\cal V}{\cal G}_0$. In coordinate space, the free quasiparticle part ${\cal G}_0$ is straightforwardly expressed by means of the Fourier transformation thanks to the translational invariance as
\begin{eqnarray}
{\cal G}_0(r,r') = \int\frac{d^4p}{(2\pi)^4}\tilde{\cal G}_0(p_0,{\bm p}){\rm e}^{-ip\cdot(r-r')}\ , \label{G0R}
\end{eqnarray}
where the one in momentum space is of the form
\begin{widetext}
\begin{eqnarray}
\tilde{\cal G}_0(p_0,{\bm p}) &=& \frac{i}{(p_0-{E}^{\rm a}_{\bm p})(p_0-E_{\bm p}^{+})(p_0-E_{\bm p}^{-})} \nonumber\\
&& \times \left(
\begin{array}{ccc}
p_0(p_0+\mu)-|\Delta|^2 & -\left[|{\bm p}|p_0+|\Delta|^2\right]\hat{\bm p}\cdot{\bm \sigma} & -\Delta^* (p_0+|{\bm p}|+\mu) \\
\left[|{\bm p}|p_0+|\Delta|^2\right]\hat{\bm p}\cdot{\bm \sigma}  & -p_0(p_0+\mu)+|\Delta|^2 &  -\Delta^*(p_0+|{\bm p}|+\mu) \hat{\bm p}\cdot{\bm \sigma} \\
 -\Delta (p_0+|{\bm p}|+\mu) &\Delta (p_0+|{\bm p}|+\mu) \hat{\bm p}\cdot{\bm \sigma} &(p_0-|{\bm p}|+\mu)(p_0+|{\bm p}|+\mu) \\
\end{array}
\right) \ ,\label{G0Momentum}
\end{eqnarray}
\end{widetext}
with the dispersion relations of the quasiparticles being
\begin{eqnarray}
E_{\bm p}^{+} &=& \frac{1}{2}\left(|{\bm p}|-\mu + \sqrt{(|{\bm p}|-\mu)^2+8|\Delta|^2}\right)\ , \nonumber\\
E_{\bm p}^{-} &=& \frac{1}{2}\left(|{\bm p}|-\mu - \sqrt{(|{\bm p}|-\mu)^2+8|\Delta|^2}\right)\ , \nonumber\\
E^{\rm a}_{\bm p} &=& -|{\bm p}|-\mu\ .
\end{eqnarray}
These equations coincide with Eqs.~(\ref{G0Tilde}) and~(\ref{Dispersions}). In what follows we assume $\Delta$ to be real.

Unlike ${\cal G}_0$, the ${\cal G}_1$ part in Eq.~(\ref{GApprox}) is rather complicated due to the violation of the translational invariance. In fact, by defining a $6\times 6$ matrix
\begin{eqnarray}
{\mathscr V} = \left(
\begin{array}{ccc}
0 & -\sigma^1 & 0 \\
\sigma^1 & 0 & 0 \\
0 & 0 & 0 \\
\end{array}
\right) 
\end{eqnarray}
for convenience, the ${\cal G}_1$ part in Eq.~(\ref{GApprox}) can be evaluated as
\begin{widetext}
\begin{eqnarray}
{\cal G}_1(r,r') &=& ieB\int d^4r''\int\frac{d^4p}{(2\pi)^4}\frac{d^4p'}{(2\pi)^4}{\cal G}_0(p){\rm e}^{-ip\cdot (r-r'')} {\mathscr V} {\cal G}_0(p')y''{\rm e}^{-ip'\cdot (r''-r')} \nonumber\\
&=&  -eB\int d^4r''\int\frac{d^4p}{(2\pi)^4}\frac{d^4p'}{(2\pi)^4}{\cal G}_0(p) {\mathscr V}{\cal G}_0(p')\left[\partial_{p_y-p'_y}{\rm e}^{i(p-p')\cdot r''} \right] {\rm e}^{-ip\cdot r + ip'\cdot r'} \nonumber\\
&=&eB\int\frac{d^4p}{(2\pi)^4}\frac{d^4p'}{(2\pi)^4}\partial_{p_y-p'_y}\left[\frac{1}{{\cal G}_0(p)}{\mathscr V} \frac{1}{{\cal G}_0(p')}{\rm e}^{-ip\cdot r + ip'\cdot r'} \right] (2\pi)^4\delta^4(p-p') \nonumber\\
&=& \frac{eB}{2}\int\frac{d^4p}{(2\pi)^4} \Bigg\{\Big[\partial_{p_y}{\cal G}_0(p)\Big]  {\mathscr V}{\cal G}_0(p)-{\cal G}_0(p){\mathscr V}\Big[\partial_{p_y}{\cal G}_0(p)\Big] + i(y+y'){\cal G}_0(p){\mathscr V}{\cal G}_0\Bigg\} {\rm e}^{-ip\cdot (r- r')} \label{G1Evaluate}
\end{eqnarray}
\end{widetext}
with the help of Eq.~(\ref{G0R}). To obtain the third line in Eq.~(\ref{G1Evaluate}), we have made use of the integration by parts. The last term in the fourth line in Eq.~(\ref{G1Evaluate}), which is proportional to $y+y'$, obviously breaks the translational invariance, which stems from the so-called Schwinger phase~\cite{Schwinger:1951nm}. Here we ignore such a term~\cite{Gorbar:2013upa,Miransky:2015ava}, and Eq.~(\ref{G1Evaluate}) takes the form of
\begin{eqnarray}
{\cal G}_1(r,r') = \int\frac{d^4p}{(2\pi)^4}\tilde{\cal G}_1(p_0,{\bm p}){\rm e}^{-ip\cdot(r-r')}\ ,
\end{eqnarray}
where the counterpart in momentum space is given by
\begin{eqnarray}
\tilde{\cal G}_1(p_0,{\bm p}) &=& -\frac{eB}{2} \frac{{\cal I}}{|{\bm p}|^3(p_0-E^{\rm a}_{\bm p})^2(p_0-E_{\bm p}^{+})^2(p_0-E_{\bm p}^{-})^2} \nonumber\\
\label{G1Momentum}
\end{eqnarray}
with
\begin{eqnarray}
{\cal I} \equiv \left(
\begin{array}{cccccc}
{\cal I}_1 & {\cal I}_2 & {\cal I}_3 & 0 & {\cal I}_4 & {\cal I}_5 \\
-{\cal I}_2 & -{\cal I}_1 &0 & {\cal I}_3 & -{\cal I}_5 &{\cal I}_4^* \\
-{\cal I}_3  & 0 & -{\cal I}_1  & -{\cal I}_2 & -i{\cal I}_6 & {\cal I}_7 \\
0 & -{\cal I}_3 & {\cal I}_2 & {\cal I}_1 & {\cal I}_7^* & -i{\cal I}_6^* \\
-{\cal I}_4^* & {\cal I}_5 & i{\cal I}_6^*  & {\cal I}_7 & {\cal I}_8 & {\cal I}_9 \\
-{\cal I}_5 & -{\cal I}_4 & {\cal I}_7^*  & i{\cal I}_6 & -{\cal I}_9 & -{\cal I}_8 \\
\end{array}
\right) \ , \label{MatrixG1}
\end{eqnarray}
and the matrix elements are
\begin{eqnarray}
{\cal I}_1  = -2i\Big[\Delta^2-p_0(p_0+\mu)\Big]\Big[\Delta^2(p_x^2+p_z^2)+|{\bm p}|^3p_0\Big]\ , \nonumber\\
\end{eqnarray}
\begin{eqnarray}
{\cal I}_2 = -2p_yp_z\Delta^2\Big[\Delta^2-p_0(p_0+\mu)\Big]\ ,
\end{eqnarray}
\begin{eqnarray}
{\cal I}_3 = -2ip_z|{\bm p}|(\Delta^2+|{\bm p}|p_0)^2\ ,
\end{eqnarray}
\begin{eqnarray}
{\cal I}_4 &=& -\Delta  \Bigg\{|{\bm p}|p_y(2\Delta^2+p_0|{\bm p}|)(p_x-ip_y) \nonumber\\
&& +ip_0\Big[|{\bm p}|^2(p_0+|{\bm p}|+\mu)^2+ip_y(p_0+\mu)^2(p_x+ip_y)\Big] \nonumber\\
&& +2p_xp_y\Delta^2(p_0+\mu) \Bigg\} \ , 
\end{eqnarray}
\begin{eqnarray}
{\cal I}_5 = -p_yp_z\Delta \Big\{2\Delta^2|{\bm p}|+p_0\big[(p_0+\mu)^2+|{\bm p}|^2\big]\Big\}\ , \nonumber\\
\end{eqnarray}
\begin{eqnarray}
{\cal I}_6 &=&  2\Delta p_z\Big\{|{\bm p}|(p_0+|{\bm p}|+\mu)\big(\Delta^2+|{\bm p}|p_0\big) \nonumber\\
&&-ip_xp_y\big[\Delta^2-p_0(p_0+\mu)\big]\Big\}\ ,
\end{eqnarray}
\begin{eqnarray}
{\cal I}_7 =  2\Delta p_y\Big[p_z^2+ip_y(p_x-ip_y)\Big]\Big[\Delta^2-p_0(p_0+\mu)\Big]\ , \nonumber\\
\end{eqnarray}
\begin{eqnarray}
{\cal I}_8 =  2i\Delta^2(p_x^2+p_z^2)(p_0+|{\bm p}|+\mu)^2\ ,
\end{eqnarray}
\begin{eqnarray}
{\cal I}_9 =  2\Delta^2p_yp_z(p_0+|{\bm p}|+\mu)^2\ .
\end{eqnarray}

Summarizing the above derivation, the Green's function of the quasiparticle in a weak magnetic field in momentum space $\tilde{\cal G}(p_0,{\bm p})$ can be written as
\begin{eqnarray}
\tilde{\cal G}(p_0,{\bm p}) \approx \tilde{\cal G}_0(p_0,{\bm p})+ \tilde{\cal G}_1(p_0,{\bm p}) \label{GMomentum}
\end{eqnarray}
with $\tilde{\cal G}_0(p_0,{\bm p})$ and $\tilde{\cal G}_1(p_0,{\bm p})$ being given by Eqs.~(\ref{G0Momentum}) and~(\ref{G1Momentum}), respectively, apart from the Schwinger phase.

Note that when the Kondo condensate is absent, i.e., when we take $\Delta=0$ in Eq.~(\ref{GMomentum}), the Green's function is reduced to
\begin{eqnarray}
\tilde{\cal G}(p_0,{\bm p}) \to \left(
\begin{array}{cc}
\tilde{\cal G}_\psi(p_0,{\bm p}) & 0 \\
0 & \tilde{\cal G}_{\Psi_v}(p_0) \\
\end{array}
\right) \ ,
\end{eqnarray}
with
\begin{eqnarray}
\tilde{\cal G}_\psi(p_0,{\bm p}) &=& \frac{i(\Slash{p}+\mu\gamma^0)}{(p_0+\mu)^2-|{\bm p}|^2}  \nonumber\\
&& + eB\gamma^1\gamma^2\frac{(p_0+\mu)\gamma^0-p_z\gamma^3}{\big[(p_0+\mu)^2-|{\bm p}|^2\big]^2} \ ,   \label{GShovkovy}\\
\tilde{\cal G}_{\Psi_v}(p_0) &=& \frac{i}{p_0} \ .
\end{eqnarray}
Our result of Eq.~(\ref{GShovkovy}) is consistent with the one found in Appendix A of Ref.~\cite{Miransky:2015ava}.

\subsection{Calculation of the axial current}

In this subsection, we calculate the axial current by making use of the Green's function obtained in Appendix~\ref{sec:GreenFunction}.

From Eq.~(\ref{GMomentum}), by employing the imaginary-time formalism the axial current is now given by
\begin{eqnarray}
\langle j_5^i(t,{\bm x})\rangle = -i N_cT\sum_m\int\frac{d^3p}{(2\pi)^3}{\rm tr}\left[\Gamma^i\Gamma_5 \tilde{\cal G}(i{\omega}_m,{\bm p})\right] \ ,\nonumber\\
\end{eqnarray}
[${\omega}_m=(2m+1)\pi T$] with
\begin{eqnarray}
&& \Gamma^i = \left(
\begin{array}{cc}
\gamma^i & 0 \\
0 & 0 \\
\end{array}
\right) = \left(
\begin{array}{ccc}
0 & \sigma^i & 0 \\
-\sigma^i & 0 & 0 \\
0 & 0 & 0 \\
\end{array}
\right)\  , \nonumber\\
&& \Gamma_5 = \left(
\begin{array}{cc}
\gamma_5 & 0 \\
0 & 0 \\
\end{array}
\right)  =\left(
\begin{array}{ccc}
0 & 1 & 0 \\
1 & 0 & 0 \\
0 & 0 & 0 \\
\end{array}
\right)\ .
\end{eqnarray}
Here the nonzero component is given only for $i=3$ which is of the form
\begin{eqnarray}
\langle j_5^{i=3}\rangle &=& 4eN_cB T\sum_m\int\frac{d^3p}{(2\pi)^3} \nonumber\\
&\times&\frac{\Big[\Delta^2-i{\omega}_m(i{\omega}_m+\mu)\Big]\Big[\frac{2}{3}\Delta^2|{\bm p}|^2+i{\omega}_m|{\bm p}|^3\Big]}{|{\bm p}|^3(i{\omega}_m-E^{\rm a}_{\bm p})^2(i{\omega}_m-E_{\bm p}^{+})^2(i{\omega}_m-E_{\bm p}^{-})^2} \ .\nonumber\\\label{J5Third}
\end{eqnarray}
Therefore, the Abel-Plana formula~(\ref{ABFermionFiniteT}) enables us to carry out the Matsubara summation straightforwardly. The resultant expression is lengthy and not illuminating so that we do not show explicitly here, but it was found to be the same as the one obtained in static limit within the procedure employed in the main text.


\section{General discussions on each contribution of response function in Eq.~(\ref{Pi5Each})}
\label{sec:GeneralD}

In this appendix, we give general discussions on each contribution of response function in Eq.~(\ref{Pi5Each}) in more detail, and explain the difference between static limit and dynamical limit. For convenience, we again show Eq.~(\ref{Pi5Each}) here:
\begin{eqnarray}
\tilde{\Pi}^{\zeta'\zeta}_5(i\bar{\omega}_n,{\bm q}) &=& -\frac{N_c}{|{\bm q}|^2}\int\frac{d^3p}{(2\pi)^3}\Big[\epsilon_{\zeta'}(\hat{\bm p}'\cdot{\bm q})-\epsilon_\zeta(\hat{\bm p}\cdot{\bm q})\Big] \nonumber\\
&\times&\frac{U_{\zeta'}({\bm p}') U_{\zeta}({\bm p})}{i\bar{\omega}_n-E_{{\bm p}'}^{\zeta'}+E_{\bm p}^{\zeta}}\left(f_F(E_{\bm p}^\zeta)-f_F(E_{{\bm p}'}^{\zeta'})\right) \nonumber\\
\label{Pi5EachApp}
\end{eqnarray}  
with $\zeta,\zeta'=+,-,{\rm a}$.

With respect to Eq.~(\ref{Pi5EachApp}), in order to evaluate the response function properly, we need to take $q_0,{\bm q}\to0$ limit after the analytic continuation as seen from Eq.~(\ref{Pi5Tot}) together with Eq.~(\ref{Pi5TildeDef}). Here, such a limit suggests that the nine contributions in Eq.~(\ref{Pi5EachApp}) are mathematically classified into three cases of
\begin{description}
\item[(I)] \ \ \ \ \ $\zeta=\zeta'$ case ($\tilde{\Pi}_5^{++}$, $\tilde{\Pi}_5^{--}$, $\tilde{\Pi}_5^{\rm aa}$),
\item[(II)] \ \ \ \ $\zeta\neq\zeta'$ and $\epsilon_\zeta= \epsilon_{\zeta'}$ case ($\tilde{\Pi}_5^{+-}$, $\tilde{\Pi}_5^{-+}$),
\item[(III)] \ \ \ $\zeta\neq\zeta'$ and $\epsilon_\zeta \neq \epsilon_{\zeta'}$ case ($\tilde{\Pi}_5^{+{\rm a}}$, $\tilde{\Pi}_5^{{\rm a}+}$, $\tilde{\Pi}_5^{-{\rm a}}$, $\tilde{\Pi}_5^{{\rm a}-}$). 
\end{description}
Recall that the physical meaning of these processes is given in Fig.~\ref{fig:Diagrams}.

First, in the case (I), we find
\begin{eqnarray}
&&\epsilon_{\zeta}(\hat{\bm p}'\cdot{\bm q})-\epsilon_\zeta(\hat{\bm p}\cdot{\bm q})  = C_\zeta^{(2)}|{\bm q}|^2 + {\cal O}({\bm q}^3) \ ,\nonumber\\
&&q_0-E_{{\bm p}'}^{\zeta}+E_{\bm p}^{\zeta} = q_0+D_\zeta^{(1)}|{\bm q}| + {\cal O}({\bm q}^2)\ , \label{ExpandCaseI}
\end{eqnarray}
where $C_\zeta^{(2)}$ and $D_\zeta^{(1)}$ are coefficients independent of ${\bm q}$ but dependent on ${\bm p}$ (the number in the parenthesis corresponds to the order of ${\bm q}$). In this appendix, we do not show the explicit expressions of coefficients, because they are not important, and it is enough to consider only the order of expansion with respect to ${\bm q}$.

From Eq.~(\ref{ExpandCaseI}), the real part of $\tilde{\Pi}_5^{\zeta'\zeta}$ at small $(q_0,{\bm q})$ takes the form of
\begin{eqnarray}
\tilde{\Pi}_5^{\zeta\zeta}(q_0,{\bm q}) &\approx& -\frac{N_c}{|{\bm q}|^2}\int\frac{d^3p}{(2\pi)^3}C_\zeta^{(2)}|{\bm q}|^2 \frac{ U_\zeta({\bm p})U_\zeta({\bm p})}{q_0+D_\zeta^{(1)}|{\bm q}|}  \nonumber\\
&\times&\left[- (\hat{\bm p}\cdot{\bm q})\frac{\partial f_F(E_{\bm p}^\zeta)}{\partial|{\bm p}|}\right] \nonumber\\ 
&=& -N_c\int\frac{d^3p}{(2\pi)^3}\frac{N^{(1)}_\zeta|{\bm q}|}{q_0+D^{(1)}_\zeta|{\bm q}|}\frac{\partial f_F(E_{\bm p}^\zeta)}{\partial|{\bm p}|} \label{Pi5Singular}
\end{eqnarray}
by the Taylor expansion. In Eq.~(\ref{Pi5Singular}), we have defined $N_\zeta^{(1)} \equiv -(\hat{\bm p}\cdot\hat{\bm q})C_\zeta^{(2)}U_\zeta({\bm p})U_\zeta({\bm p})$ ($\hat{\bm q} \equiv {\bm q}/|{\bm q}|$). 
Therefore, Eq.~(\ref{Pi5Singular}) yields
\begin{eqnarray}
\lim_{{\bm q}\to{\bm 0}}\tilde{\Pi}_5^{\zeta\zeta}(0,{\bm q}) &=& -N_c\int\frac{d^3p}{(2\pi)^3}\frac{N^{(1)}_\zeta}{D^{(1)}_\zeta}\frac{\partial f_F(E_{\bm p}^\zeta)}{\partial|{\bm p}|} \nonumber\\ 
\lim_{q_0\to0}\tilde{\Pi}_5^{\zeta\zeta}(q_0,{\bm 0}) &=& 0\ , \label{ZetaZetaPi5}
\end{eqnarray}
which clearly shows that the results in static limit and dynamical limit differ due to a singularity caused by the factor $1/(q_0+D_\zeta^{(1)}|{\bm q}|)$ in the $q_0,{\bm q}\to0$ limit.

Next, in the case (II), we find
\begin{eqnarray}
&&\epsilon_{\zeta'}(\hat{\bm p}'\cdot{\bm q})-\epsilon_\zeta(\hat{\bm p}\cdot{\bm q})  = C^{(2)}_{\zeta'\zeta}|{\bm q}|^2  + {\cal O}({\bm q}^3) \ ,\nonumber\\
&&q_0-E_{{\bm p}'}^{\zeta'}+E_{\bm p}^{\zeta} = q_0+D_{\zeta'\zeta}^{(0)} + {\cal O}({\bm q}^1) \ ,
\end{eqnarray}
so that the same expansion as in Eq.~(\ref{Pi5Singular}) reads
\begin{eqnarray}
\tilde{\Pi}_5^{\zeta'\zeta}(q_0,{\bm q}) &\approx& -\frac{N_c}{|{\bm q}|^2}\int\frac{d^3p}{(2\pi)^3}C^{(2)}_{\zeta'\zeta}|{\bm q}|^2 \frac{U_{\zeta'}({\bm p})U_\zeta({\bm p})}{q_0+D_{\zeta'\zeta}^{(0)}} \nonumber\\
&\times&\left(f_F(E_{\bm p}^\zeta)-f_F(E_{{\bm p}}^{\zeta'})\right) \nonumber\\
&\equiv&-N_c\int\frac{d^3p}{(2\pi)^3}\frac{N^{(0)}_{\zeta'\zeta}}{q_0+D^{(0)}_{\zeta'\zeta}} \nonumber\\
&\times&\left(f_F(E_{\bm p}^\zeta)-f_F(E_{{\bm p}}^{\zeta'})\right)
\end{eqnarray}
[$N^{(0)}_{\zeta'\zeta} \equiv C_{\zeta'\zeta}^{(2)}U_{\zeta'}({\bm p})U_\zeta({\bm p})$]. Namely, in this case $D^{(0)}_{\zeta'\zeta}$ in the denominator acts as a regulator preventing us from getting the singularity, resulting in that there is no difference between static and dynamical limits:
\begin{eqnarray}
&&\lim_{{\bm q}\to{\bm 0}}\tilde{\Pi}_5^{\zeta'\zeta}(0,{\bm q}) = \lim_{q_0\to0}\tilde{\Pi}_5^{\zeta'\zeta}(q_0,{\bm 0}) \nonumber\\
&& = -N_c\int\frac{d^3p}{(2\pi)^3}\frac{C^{(0)}_{\zeta'\zeta}}{D^{(0)}_{\zeta'\zeta}} \left(f_F(E_{\bm p}^\zeta)-f_F(E_{{\bm p}}^{\zeta'})\right)\ .  \label{Pi5CaseII}
\end{eqnarray}

Finally, in the case (III), we get
\begin{eqnarray}
&& \epsilon_{\zeta'}(\hat{\bm p}'\cdot{\bm q})-\epsilon_\zeta(\hat{\bm p}\cdot{\bm q}) = C_{\zeta'\zeta}^{(1)}|{\bm q}| + C_{\zeta'\zeta}^{(2)}|{\bm q}^2| + {\cal O}({\bm q}^3) \ , \nonumber\\
&& q_0-E_{{\bm p}'}^{\zeta'}+E_{\bm p}^{\zeta} = q_0+D_{\zeta'\zeta}^{(0)} + {\cal O}({\bm q}^1) \ ,
\end{eqnarray}
so that the same expansion as in Eq.~(\ref{Pi5Singular}) leads to a slightly complicated form:
\begin{eqnarray}
\tilde{\Pi}_5^{\zeta'\zeta}(q_0,{\bm q}) &\approx& -\frac{N_c}{|{\bm q}|^2}\int\frac{d^3p}{(2\pi)^3}\left(C_{\zeta'\zeta}^{(1)}|{\bm q}| + C_{\zeta'\zeta}^{(2)}|{\bm q}^2| \right)\nonumber\\
&\times& \frac{\left[U_{\zeta'}({\bm p}) +(\hat{\bm p}\cdot{\bm q}) \frac{\partial U_{\zeta'}({\bm p})}{\partial|{\bm p}|}\right]U_\zeta({\bm p})}{q_0+D^{(0)}_{\zeta'\zeta}} \nonumber\\
&\times&\left[f_F(E_{\bm p}^\zeta)-f_F(E_{{\bm p}}^{\zeta'})-(\hat{\bm p}\cdot{\bm q})\frac{\partial f_F(E_{\bm p}^{\zeta'})}{\partial|{\bm p}|}\right] \nonumber\\
&\approx& -N_c\int\frac{d^3p}{(2\pi)^3}\frac{\frac{1}{|{\bm q}|}{N}_{\zeta'\zeta}^{(-1)} + {N}^{(0)}_{\zeta'\zeta}}{q_0+D^{(0)}_{\zeta'\zeta}} \nonumber\\
&\times&\left[f_F(E_{\bm p}^\zeta)-f_F(E_{{\bm p}}^{\zeta'})-(\hat{\bm p}\cdot{\bm q})\frac{\partial f_F(E_{\rm p}^{\zeta'})}{\partial|{\bm p}|}\right]\ . \nonumber\\ \label{Pi5Genaral3}
\end{eqnarray}
Here we have defined $N_{\zeta'\zeta}^{(-1)} \equiv C_{\zeta'\zeta}^{(1)}U_{\zeta'}({\bm p})U_\zeta({\bm p})$ and $N_{\zeta'\zeta}^{(0)} \equiv (\hat{\bm p}\cdot\hat{\bm q})C_{\zeta'\zeta}^{(1)}\frac{\partial U_{\zeta'}({\bm p})}{\partial|{\bm p}|}U_\zeta({\bm p}) + C_{\zeta'\zeta}^{(2)}U_{\zeta'}({\bm p})U_\zeta({\bm p})$. Equation~(\ref{Pi5Genaral3}) seemingly causes a divergence in the ${\bm q}\to0$ limit due to the $1/|{\bm q}|$ in the numerator. However, such a problematic contribution will be removed after the angular integral. Namely in both static limit and dynamical limit, Eq.~(\ref{Pi5Genaral3}) converges, leading to the same result as
\begin{eqnarray}
&&\lim_{{\bm q}\to{\bm 0}}\tilde{\Pi}_5^{\zeta'\zeta}(0,{\bm q}) = \lim_{q_0\to0}\tilde{\Pi}_5^{\zeta'\zeta}(q_0,{\bm 0}) \nonumber\\
&&= -N_c\int\frac{d^3p}{(2\pi)^3}\Bigg\{\frac{N^{(0)}_{\zeta'\zeta}}{D^{(0)}_{\zeta'\zeta}} \left(f_F(E_{\bm p}^\zeta)-f_F(E_{{\bm p}}^{\zeta'})\right) \nonumber\\
&& - (\hat{\bm p}\cdot\hat{\bm q})\frac{N_{\zeta'\zeta}^{(-1)}}{D^{(0)}_{\zeta'\zeta}}\frac{\partial f_F(E_{\rm p}^{\zeta'})}{\partial|{\bm p}|} \Bigg\}\ . \label{Pi5CaseIII}
\end{eqnarray}

Summarizing the above discussions, we have confirmed that 
\begin{eqnarray}
\lim_{{\bm q}\to{\bm 0}}\tilde{\Pi}_5^{\zeta\zeta}(0,{\bm q}) &=& -N_c\int\frac{d^3p}{(2\pi)^3}{\cal I}_\zeta({\bm p})\frac{\partial f_F(E_{\bm p}^\zeta)}{\partial|{\bm p}|} \nonumber\\ 
\lim_{q_0\to0}\tilde{\Pi}_5^{\zeta\zeta}(q_0,{\bm 0}) &=& 0 \label{PiDifference}
\end{eqnarray}
for the case (I) from Eq.~(\ref{ZetaZetaPi5}), while
\begin{eqnarray}
\lim_{{\bm q}\to{\bm 0}}\tilde{\Pi}_5^{\zeta'\zeta}(0,{\bm q}) = \lim_{q_0\to0}\tilde{\Pi}_5^{\zeta'\zeta}(q_0,{\bm 0}) \label{PiNoDifference}
\end{eqnarray}
for the case (II) and case (III) from Eqs.~(\ref{Pi5CaseII}) and~(\ref{Pi5CaseIII}), respectively. In Eq.~(\ref{PiDifference}), we have defined ${\cal I}_\zeta({\bm p}) \equiv N_\zeta^{(1)}/D_\zeta^{(1)}$ to simplify the expression. Equations~(\ref{PiDifference}) and~(\ref{PiNoDifference}) show that the difference between the response function in static limit and dynamical limit can be generated in only case (I) which corresponds to the scattering process of ${\rm q.p.}\to{\rm q.p.}$, ${\rm q.h.}\to{\rm q.h.}$ or ${\rm a.p.}\to {\rm a.p.}$. 

\bibliography{reference}

\begin{thebibliography}{84}%
\makeatletter
\providecommand \@ifxundefined [1]{%
 \@ifx{#1\undefined}
}%
\providecommand \@ifnum [1]{%
 \ifnum #1\expandafter \@firstoftwo
 \else \expandafter \@secondoftwo
 \fi
}%
\providecommand \@ifx [1]{%
 \ifx #1\expandafter \@firstoftwo
 \else \expandafter \@secondoftwo
 \fi
}%
\providecommand \natexlab [1]{#1}%
\providecommand \enquote  [1]{``#1''}%
\providecommand \bibnamefont  [1]{#1}%
\providecommand \bibfnamefont [1]{#1}%
\providecommand \citenamefont [1]{#1}%
\providecommand \href@noop [0]{\@secondoftwo}%
\providecommand \href [0]{\begingroup \@sanitize@url \@href}%
\providecommand \@href[1]{\@@startlink{#1}\@@href}%
\providecommand \@@href[1]{\endgroup#1\@@endlink}%
\providecommand \@sanitize@url [0]{\catcode `\\12\catcode `\$12\catcode
  `\&12\catcode `\#12\catcode `\^12\catcode `\_12\catcode `\%12\relax}%
\providecommand \@@startlink[1]{}%
\providecommand \@@endlink[0]{}%
\providecommand \url  [0]{\begingroup\@sanitize@url \@url }%
\providecommand \@url [1]{\endgroup\@href {#1}{\urlprefix }}%
\providecommand \urlprefix  [0]{URL }%
\providecommand \Eprint [0]{\href }%
\providecommand \doibase [0]{http://dx.doi.org/}%
\providecommand \selectlanguage [0]{\@gobble}%
\providecommand \bibinfo  [0]{\@secondoftwo}%
\providecommand \bibfield  [0]{\@secondoftwo}%
\providecommand \translation [1]{[#1]}%
\providecommand \BibitemOpen [0]{}%
\providecommand \bibitemStop [0]{}%
\providecommand \bibitemNoStop [0]{.\EOS\space}%
\providecommand \EOS [0]{\spacefactor3000\relax}%
\providecommand \BibitemShut  [1]{\csname bibitem#1\endcsname}%
\let\auto@bib@innerbib\@empty
\bibitem [{\citenamefont {Kharzeev}(2006)}]{Kharzeev:2004ey}%
  \BibitemOpen
  \bibfield  {author} {\bibinfo {author} {\bibfnamefont {Dmitri}\ \bibnamefont
  {Kharzeev}},\ }\bibfield  {title} {\enquote {\bibinfo {title} {{Parity
  violation in hot QCD: Why it can happen, and how to look for it}},}\ }\href
  {http://dx.doi.org/10.1016/j.physletb.2005.11.075} {\bibfield  {journal}
  {\bibinfo  {journal} {Phys. Lett. B}\ }\textbf {\bibinfo {volume} {633}},\
  \bibinfo {pages} {260--264} (\bibinfo {year} {2006})},\ \Eprint
  {http://arxiv.org/abs/hep-ph/0406125} {arXiv:hep-ph/0406125} \BibitemShut
  {NoStop}%
\bibitem [{\citenamefont {Kharzeev}\ and\ \citenamefont
  {Zhitnitsky}(2007)}]{Kharzeev:2007tn}%
  \BibitemOpen
  \bibfield  {author} {\bibinfo {author} {\bibfnamefont {D.}~\bibnamefont
  {Kharzeev}}\ and\ \bibinfo {author} {\bibfnamefont {A.}~\bibnamefont
  {Zhitnitsky}},\ }\bibfield  {title} {\enquote {\bibinfo {title} {{Charge
  separation induced by P-odd bubbles in QCD matter}},}\ }\href
  {http://dx.doi.org/10.1016/j.nuclphysa.2007.10.001} {\bibfield  {journal}
  {\bibinfo  {journal} {Nucl. Phys. A}\ }\textbf {\bibinfo {volume} {797}},\
  \bibinfo {pages} {67--79} (\bibinfo {year} {2007})},\ \Eprint
  {http://arxiv.org/abs/0706.1026} {arXiv:0706.1026 [hep-ph]} \BibitemShut
  {NoStop}%
\bibitem [{\citenamefont {Kharzeev}\ \emph {et~al.}(2008)\citenamefont
  {Kharzeev}, \citenamefont {McLerran},\ and\ \citenamefont
  {Warringa}}]{Kharzeev:2007jp}%
  \BibitemOpen
  \bibfield  {author} {\bibinfo {author} {\bibfnamefont {Dmitri~E.}\
  \bibnamefont {Kharzeev}}, \bibinfo {author} {\bibfnamefont {Larry~D.}\
  \bibnamefont {McLerran}}, \ and\ \bibinfo {author} {\bibfnamefont
  {Harmen~J.}\ \bibnamefont {Warringa}},\ }\bibfield  {title} {\enquote
  {\bibinfo {title} {{The Effects of topological charge change in heavy ion
  collisions: $``$Event by event P and CP violation$"$}},}\ }\href
  {http://dx.doi.org/10.1016/j.nuclphysa.2008.02.298} {\bibfield  {journal}
  {\bibinfo  {journal} {Nucl. Phys. A}\ }\textbf {\bibinfo {volume} {803}},\
  \bibinfo {pages} {227--253} (\bibinfo {year} {2008})},\ \Eprint
  {http://arxiv.org/abs/0711.0950} {arXiv:0711.0950 [hep-ph]} \BibitemShut
  {NoStop}%
\bibitem [{\citenamefont {Fukushima}\ \emph {et~al.}(2008)\citenamefont
  {Fukushima}, \citenamefont {Kharzeev},\ and\ \citenamefont
  {Warringa}}]{Fukushima:2008xe}%
  \BibitemOpen
  \bibfield  {author} {\bibinfo {author} {\bibfnamefont {Kenji}\ \bibnamefont
  {Fukushima}}, \bibinfo {author} {\bibfnamefont {Dmitri~E.}\ \bibnamefont
  {Kharzeev}}, \ and\ \bibinfo {author} {\bibfnamefont {Harmen~J.}\
  \bibnamefont {Warringa}},\ }\bibfield  {title} {\enquote {\bibinfo {title}
  {{The Chiral Magnetic Effect}},}\ }\href
  {http://dx.doi.org/10.1103/PhysRevD.78.074033} {\bibfield  {journal}
  {\bibinfo  {journal} {Phys. Rev. D}\ }\textbf {\bibinfo {volume} {78}},\
  \bibinfo {pages} {074033} (\bibinfo {year} {2008})},\ \Eprint
  {http://arxiv.org/abs/0808.3382} {arXiv:0808.3382 [hep-ph]} \BibitemShut
  {NoStop}%
\bibitem [{\citenamefont {Vilenkin}(1980{\natexlab{a}})}]{Vilenkin:1980fu}%
  \BibitemOpen
  \bibfield  {author} {\bibinfo {author} {\bibfnamefont {A.}~\bibnamefont
  {Vilenkin}},\ }\bibfield  {title} {\enquote {\bibinfo {title} {{Equilibrium
  parity-violating current in a magnetic field}},}\ }\href
  {http://dx.doi.org/10.1103/PhysRevD.22.3080} {\bibfield  {journal} {\bibinfo
  {journal} {Phys. Rev. D}\ }\textbf {\bibinfo {volume} {22}},\ \bibinfo
  {pages} {3080} (\bibinfo {year} {1980}{\natexlab{a}})}\BibitemShut {NoStop}%
\bibitem [{\citenamefont {Nielsen}\ and\ \citenamefont
  {Ninomiya}(1983)}]{Nielsen:1983rb}%
  \BibitemOpen
  \bibfield  {author} {\bibinfo {author} {\bibfnamefont {Holger~Bech}\
  \bibnamefont {Nielsen}}\ and\ \bibinfo {author} {\bibfnamefont {Masao}\
  \bibnamefont {Ninomiya}},\ }\bibfield  {title} {\enquote {\bibinfo {title}
  {{The Adler-Bell-Jackiw anomaly and Weyl fermions in a crystal}},}\ }\href
  {\doibase 10.1016/0370-2693(83)91529-0} {\bibfield  {journal} {\bibinfo
  {journal} {Phys. Lett. B}\ }\textbf {\bibinfo {volume} {130}},\ \bibinfo
  {pages} {389--396} (\bibinfo {year} {1983})}\BibitemShut {NoStop}%
\bibitem [{\citenamefont {Son}\ and\ \citenamefont
  {Zhitnitsky}(2004)}]{Son:2004tq}%
  \BibitemOpen
  \bibfield  {author} {\bibinfo {author} {\bibfnamefont {D.~T.}\ \bibnamefont
  {Son}}\ and\ \bibinfo {author} {\bibfnamefont {Ariel~R.}\ \bibnamefont
  {Zhitnitsky}},\ }\bibfield  {title} {\enquote {\bibinfo {title} {{Quantum
  anomalies in dense matter}},}\ }\href
  {http://dx.doi.org/10.1103/PhysRevD.70.074018} {\bibfield  {journal}
  {\bibinfo  {journal} {Phys. Rev. D}\ }\textbf {\bibinfo {volume} {70}},\
  \bibinfo {pages} {074018} (\bibinfo {year} {2004})},\ \Eprint
  {http://arxiv.org/abs/hep-ph/0405216} {arXiv:hep-ph/0405216 [hep-ph]}
  \BibitemShut {NoStop}%
\bibitem [{\citenamefont {Metlitski}\ and\ \citenamefont
  {Zhitnitsky}(2005)}]{Metlitski:2005pr}%
  \BibitemOpen
  \bibfield  {author} {\bibinfo {author} {\bibfnamefont {Max~A.}\ \bibnamefont
  {Metlitski}}\ and\ \bibinfo {author} {\bibfnamefont {Ariel~R.}\ \bibnamefont
  {Zhitnitsky}},\ }\bibfield  {title} {\enquote {\bibinfo {title} {{Anomalous
  axion interactions and topological currents in dense matter}},}\ }\href
  {\doibase 10.1103/PhysRevD.72.045011} {\bibfield  {journal} {\bibinfo
  {journal} {Phys. Rev. D}\ }\textbf {\bibinfo {volume} {72}},\ \bibinfo
  {pages} {045011} (\bibinfo {year} {2005})},\ \Eprint
  {http://arxiv.org/abs/hep-ph/0505072} {arXiv:hep-ph/0505072 [hep-ph]}
  \BibitemShut {NoStop}%
\bibitem [{\citenamefont {Newman}\ and\ \citenamefont
  {Son}(2006)}]{Newman:2005as}%
  \BibitemOpen
  \bibfield  {author} {\bibinfo {author} {\bibfnamefont {G.~M.}\ \bibnamefont
  {Newman}}\ and\ \bibinfo {author} {\bibfnamefont {D.~T.}\ \bibnamefont
  {Son}},\ }\bibfield  {title} {\enquote {\bibinfo {title} {{Response of
  strongly-interacting matter to magnetic field: Some exact results}},}\ }\href
  {\doibase 10.1103/PhysRevD.73.045006} {\bibfield  {journal} {\bibinfo
  {journal} {Phys. Rev. D}\ }\textbf {\bibinfo {volume} {73}},\ \bibinfo
  {pages} {045006} (\bibinfo {year} {2006})},\ \Eprint
  {http://arxiv.org/abs/hep-ph/0510049} {arXiv:hep-ph/0510049 [hep-ph]}
  \BibitemShut {NoStop}%
\bibitem [{\citenamefont {Kharzeev}\ \emph {et~al.}(1998)\citenamefont
  {Kharzeev}, \citenamefont {Pisarski},\ and\ \citenamefont
  {Tytgat}}]{Kharzeev:1998kz}%
  \BibitemOpen
  \bibfield  {author} {\bibinfo {author} {\bibfnamefont {Dmitri}\ \bibnamefont
  {Kharzeev}}, \bibinfo {author} {\bibfnamefont {R.~D.}\ \bibnamefont
  {Pisarski}}, \ and\ \bibinfo {author} {\bibfnamefont {Michel H.~G.}\
  \bibnamefont {Tytgat}},\ }\bibfield  {title} {\enquote {\bibinfo {title}
  {{Possibility of Spontaneous Parity Violation in Hot QCD}},}\ }\href
  {http://dx.doi.org/10.1103/PhysRevLett.81.512} {\bibfield  {journal}
  {\bibinfo  {journal} {Phys. Rev. Lett.}\ }\textbf {\bibinfo {volume} {81}},\
  \bibinfo {pages} {512--515} (\bibinfo {year} {1998})},\ \Eprint
  {http://arxiv.org/abs/hep-ph/9804221} {arXiv:hep-ph/9804221 [hep-ph]}
  \BibitemShut {NoStop}%
\bibitem [{\citenamefont {Kharzeev}\ and\ \citenamefont
  {Pisarski}(2000)}]{Kharzeev:1999cz}%
  \BibitemOpen
  \bibfield  {author} {\bibinfo {author} {\bibfnamefont {Dmitri}\ \bibnamefont
  {Kharzeev}}\ and\ \bibinfo {author} {\bibfnamefont {Robert~D.}\ \bibnamefont
  {Pisarski}},\ }\bibfield  {title} {\enquote {\bibinfo {title} {{Pionic
  measures of parity and CP violation in high-energy nuclear collisions}},}\
  }\href {http://dx.doi.org/10.1103/PhysRevD.61.111901} {\bibfield  {journal}
  {\bibinfo  {journal} {Phys. Rev. D}\ }\textbf {\bibinfo {volume} {61}},\
  \bibinfo {pages} {111901} (\bibinfo {year} {2000})},\ \Eprint
  {http://arxiv.org/abs/hep-ph/9906401} {arXiv:hep-ph/9906401 [hep-ph]}
  \BibitemShut {NoStop}%
\bibitem [{\citenamefont {Kharzeev}\ \emph {et~al.}(2002)\citenamefont
  {Kharzeev}, \citenamefont {Krasnitz},\ and\ \citenamefont
  {Venugopalan}}]{Kharzeev:2001ev}%
  \BibitemOpen
  \bibfield  {author} {\bibinfo {author} {\bibfnamefont {D.}~\bibnamefont
  {Kharzeev}}, \bibinfo {author} {\bibfnamefont {A.}~\bibnamefont {Krasnitz}},
  \ and\ \bibinfo {author} {\bibfnamefont {R.}~\bibnamefont {Venugopalan}},\
  }\bibfield  {title} {\enquote {\bibinfo {title} {{Anomalous chirality
  fluctuations in the initial stage of heavy ion collisions and parity odd
  bubbles}},}\ }\href {http://dx.doi.org/10.1016/S0370-2693(02)02630-8}
  {\bibfield  {journal} {\bibinfo  {journal} {Phys. Lett. B}\ }\textbf
  {\bibinfo {volume} {545}},\ \bibinfo {pages} {298--306} (\bibinfo {year}
  {2002})},\ \Eprint {http://arxiv.org/abs/hep-ph/0109253}
  {arXiv:hep-ph/0109253 [hep-ph]} \BibitemShut {NoStop}%
\bibitem [{\citenamefont {Kondo}(1964)}]{Kondo:1964}%
  \BibitemOpen
  \bibfield  {author} {\bibinfo {author} {\bibfnamefont {J.}~\bibnamefont
  {Kondo}},\ }\bibfield  {title} {\enquote {\bibinfo {title} {{Resistance
  Minimum in Dilute Magnetic Alloys}},}\ }\href
  {http://dx.doi.org/10.1143/PTP.32.37} {\bibfield  {journal} {\bibinfo
  {journal} {Prog. Theor. Phys.}\ }\textbf {\bibinfo {volume} {32}},\ \bibinfo
  {pages} {37--49} (\bibinfo {year} {1964})}\BibitemShut {NoStop}%
\bibitem [{\citenamefont {Hewson}(1993)}]{Hewson}%
  \BibitemOpen
  \bibfield  {author} {\bibinfo {author} {\bibfnamefont {A.~C.}\ \bibnamefont
  {Hewson}},\ }\href {\doibase 10.1017/CBO9780511470752} {\emph {\bibinfo
  {title} {{The Kondo Problem to Heavy Fermions}}}}\ (\bibinfo  {publisher}
  {Cambridge University Press},\ \bibinfo {year} {1993})\BibitemShut {NoStop}%
\bibitem [{\citenamefont {Yosida}(1996)}]{Yosida}%
  \BibitemOpen
  \bibfield  {author} {\bibinfo {author} {\bibfnamefont {K.}~\bibnamefont
  {Yosida}},\ }\href {https://www.springer.com/gp/book/9783540606512#aboutBook}
  {\emph {\bibinfo {title} {{Theory of Magnetism}}}}\ (\bibinfo  {publisher}
  {Springer-Verlag Berlin Heidelberg},\ \bibinfo {year} {1996})\BibitemShut
  {NoStop}%
\bibitem [{\citenamefont {Yamada}(2004)}]{Yamada}%
  \BibitemOpen
  \bibfield  {author} {\bibinfo {author} {\bibfnamefont {K.}~\bibnamefont
  {Yamada}},\ }\href {http://dx.doi.org/10.1017/CBO9780511534904} {\emph
  {\bibinfo {title} {{Electron Correlation in Metals}}}}\ (\bibinfo
  {publisher} {Cambridge University Press},\ \bibinfo {year}
  {2004})\BibitemShut {NoStop}%
\bibitem [{\citenamefont {Coleman}(2015)}]{coleman_2015}%
  \BibitemOpen
  \bibfield  {author} {\bibinfo {author} {\bibfnamefont {Piers}\ \bibnamefont
  {Coleman}},\ }\href {\doibase 10.1017/CBO9781139020916} {\emph {\bibinfo
  {title} {Introduction to Many-Body Physics}}}\ (\bibinfo  {publisher}
  {Cambridge University Press},\ \bibinfo {year} {2015})\BibitemShut {NoStop}%
\bibitem [{\citenamefont {Yasui}\ and\ \citenamefont
  {Sudoh}(2013)}]{Yasui:2013xr}%
  \BibitemOpen
  \bibfield  {author} {\bibinfo {author} {\bibfnamefont {S.}~\bibnamefont
  {Yasui}}\ and\ \bibinfo {author} {\bibfnamefont {K.}~\bibnamefont {Sudoh}},\
  }\bibfield  {title} {\enquote {\bibinfo {title} {{Heavy-quark dynamics for
  charm and bottom flavor on the Fermi surface at zero temperature}},}\ }\href
  {http://dx.doi.org/10.1103/PhysRevC.88.015201} {\bibfield  {journal}
  {\bibinfo  {journal} {Phys. Rev. C}\ }\textbf {\bibinfo {volume} {88}},\
  \bibinfo {pages} {015201} (\bibinfo {year} {2013})},\ \Eprint
  {http://arxiv.org/abs/1301.6830} {arXiv:1301.6830 [hep-ph]} \BibitemShut
  {NoStop}%
\bibitem [{\citenamefont {Hattori}\ \emph {et~al.}(2015)\citenamefont
  {Hattori}, \citenamefont {Itakura}, \citenamefont {Ozaki},\ and\
  \citenamefont {Yasui}}]{Hattori:2015hka}%
  \BibitemOpen
  \bibfield  {author} {\bibinfo {author} {\bibfnamefont {Koichi}\ \bibnamefont
  {Hattori}}, \bibinfo {author} {\bibfnamefont {Kazunori}\ \bibnamefont
  {Itakura}}, \bibinfo {author} {\bibfnamefont {Sho}\ \bibnamefont {Ozaki}}, \
  and\ \bibinfo {author} {\bibfnamefont {Shigehiro}\ \bibnamefont {Yasui}},\
  }\bibfield  {title} {\enquote {\bibinfo {title} {{QCD Kondo effect: quark
  matter with heavy-flavor impurities}},}\ }\href
  {http://dx.doi.org/10.1103/PhysRevD.92.065003} {\bibfield  {journal}
  {\bibinfo  {journal} {Phys. Rev. D}\ }\textbf {\bibinfo {volume} {92}},\
  \bibinfo {pages} {065003} (\bibinfo {year} {2015})},\ \Eprint
  {http://arxiv.org/abs/1504.07619} {arXiv:1504.07619 [hep-ph]} \BibitemShut
  {NoStop}%
\bibitem [{\citenamefont {Ozaki}\ \emph {et~al.}(2016)\citenamefont {Ozaki},
  \citenamefont {Itakura},\ and\ \citenamefont {Kuramoto}}]{Ozaki:2015sya}%
  \BibitemOpen
  \bibfield  {author} {\bibinfo {author} {\bibfnamefont {Sho}\ \bibnamefont
  {Ozaki}}, \bibinfo {author} {\bibfnamefont {Kazunori}\ \bibnamefont
  {Itakura}}, \ and\ \bibinfo {author} {\bibfnamefont {Yoshio}\ \bibnamefont
  {Kuramoto}},\ }\bibfield  {title} {\enquote {\bibinfo {title} {{Magnetically
  induced QCD Kondo effect}},}\ }\href
  {http://dx.doi.org/10.1103/PhysRevD.94.074013} {\bibfield  {journal}
  {\bibinfo  {journal} {Phys. Rev. D}\ }\textbf {\bibinfo {volume} {94}},\
  \bibinfo {pages} {074013} (\bibinfo {year} {2016})},\ \Eprint
  {http://arxiv.org/abs/1509.06966} {arXiv:1509.06966 [hep-ph]} \BibitemShut
  {NoStop}%
\bibitem [{\citenamefont {Yasui}\ \emph {et~al.}(2019)\citenamefont {Yasui},
  \citenamefont {Suzuki},\ and\ \citenamefont {Itakura}}]{Yasui:2016svc}%
  \BibitemOpen
  \bibfield  {author} {\bibinfo {author} {\bibfnamefont {Shigehiro}\
  \bibnamefont {Yasui}}, \bibinfo {author} {\bibfnamefont {Kei}\ \bibnamefont
  {Suzuki}}, \ and\ \bibinfo {author} {\bibfnamefont {Kazunori}\ \bibnamefont
  {Itakura}},\ }\bibfield  {title} {\enquote {\bibinfo {title} {{Kondo phase
  diagram of quark matter}},}\ }\href
  {http://dx.doi.org/10.1016/j.nuclphysa.2018.12.025} {\bibfield  {journal}
  {\bibinfo  {journal} {Nucl. Phys. A}\ }\textbf {\bibinfo {volume} {983}},\
  \bibinfo {pages} {90--102} (\bibinfo {year} {2019})},\ \Eprint
  {http://arxiv.org/abs/1604.07208} {arXiv:1604.07208 [hep-ph]} \BibitemShut
  {NoStop}%
\bibitem [{\citenamefont {Yasui}(2017)}]{Yasui:2016yet}%
  \BibitemOpen
  \bibfield  {author} {\bibinfo {author} {\bibfnamefont {Shigehiro}\
  \bibnamefont {Yasui}},\ }\bibfield  {title} {\enquote {\bibinfo {title}
  {{Kondo cloud of single heavy quark in cold and dense matter}},}\ }\href
  {http://dx.doi.org/10.1016/j.physletb.2017.08.066} {\bibfield  {journal}
  {\bibinfo  {journal} {Phys. Lett. B}\ }\textbf {\bibinfo {volume} {773}},\
  \bibinfo {pages} {428--434} (\bibinfo {year} {2017})},\ \Eprint
  {http://arxiv.org/abs/1608.06450} {arXiv:1608.06450 [hep-ph]} \BibitemShut
  {NoStop}%
\bibitem [{\citenamefont {Kanazawa}\ and\ \citenamefont
  {Uchino}(2016)}]{Kanazawa:2016ihl}%
  \BibitemOpen
  \bibfield  {author} {\bibinfo {author} {\bibfnamefont {Takuya}\ \bibnamefont
  {Kanazawa}}\ and\ \bibinfo {author} {\bibfnamefont {Shun}\ \bibnamefont
  {Uchino}},\ }\bibfield  {title} {\enquote {\bibinfo {title} {{Overscreened
  Kondo effect, (color) superconductivity and Shiba states in Dirac metals and
  quark matter}},}\ }\href {http://dx.doi.org/10.1103/PhysRevD.94.114005}
  {\bibfield  {journal} {\bibinfo  {journal} {Phys. Rev. D}\ }\textbf {\bibinfo
  {volume} {94}},\ \bibinfo {pages} {114005} (\bibinfo {year} {2016})},\
  \Eprint {http://arxiv.org/abs/1609.00033} {arXiv:1609.00033
  [cond-mat.str-el]} \BibitemShut {NoStop}%
\bibitem [{\citenamefont {Kimura}\ and\ \citenamefont
  {Ozaki}(2017)}]{Kimura:2016zyv}%
  \BibitemOpen
  \bibfield  {author} {\bibinfo {author} {\bibfnamefont {Taro}\ \bibnamefont
  {Kimura}}\ and\ \bibinfo {author} {\bibfnamefont {Sho}\ \bibnamefont
  {Ozaki}},\ }\bibfield  {title} {\enquote {\bibinfo {title} {{Fermi/non-Fermi
  mixing in SU($N$) Kondo effect}},}\ }\href
  {http://dx.doi.org/10.7566/JPSJ.86.084703} {\bibfield  {journal} {\bibinfo
  {journal} {J. Phys. Soc. Jap.}\ }\textbf {\bibinfo {volume} {86}},\ \bibinfo
  {pages} {084703} (\bibinfo {year} {2017})},\ \Eprint
  {http://arxiv.org/abs/1611.07284} {arXiv:1611.07284 [cond-mat.str-el]}
  \BibitemShut {NoStop}%
\bibitem [{\citenamefont {Yasui}\ and\ \citenamefont
  {Ozaki}(2017)}]{Yasui:2017bey}%
  \BibitemOpen
  \bibfield  {author} {\bibinfo {author} {\bibfnamefont {Shigehiro}\
  \bibnamefont {Yasui}}\ and\ \bibinfo {author} {\bibfnamefont {Sho}\
  \bibnamefont {Ozaki}},\ }\bibfield  {title} {\enquote {\bibinfo {title}
  {{Transport coefficients from the QCD Kondo effect}},}\ }\href
  {http://dx.doi.org/10.1103/PhysRevD.96.114027} {\bibfield  {journal}
  {\bibinfo  {journal} {Phys. Rev. D}\ }\textbf {\bibinfo {volume} {96}},\
  \bibinfo {pages} {114027} (\bibinfo {year} {2017})},\ \Eprint
  {http://arxiv.org/abs/1710.03434} {arXiv:1710.03434 [hep-ph]} \BibitemShut
  {NoStop}%
\bibitem [{\citenamefont {Suzuki}\ \emph {et~al.}(2017)\citenamefont {Suzuki},
  \citenamefont {Yasui},\ and\ \citenamefont {Itakura}}]{Suzuki:2017gde}%
  \BibitemOpen
  \bibfield  {author} {\bibinfo {author} {\bibfnamefont {Kei}\ \bibnamefont
  {Suzuki}}, \bibinfo {author} {\bibfnamefont {Shigehiro}\ \bibnamefont
  {Yasui}}, \ and\ \bibinfo {author} {\bibfnamefont {Kazunori}\ \bibnamefont
  {Itakura}},\ }\bibfield  {title} {\enquote {\bibinfo {title} {{Interplay
  between chiral symmetry breaking and the QCD Kondo effect}},}\ }\href
  {http://dx.doi.org/10.1103/PhysRevD.96.114007} {\bibfield  {journal}
  {\bibinfo  {journal} {Phys. Rev. D}\ }\textbf {\bibinfo {volume} {96}},\
  \bibinfo {pages} {114007} (\bibinfo {year} {2017})},\ \Eprint
  {http://arxiv.org/abs/1708.06930} {arXiv:1708.06930 [hep-ph]} \BibitemShut
  {NoStop}%
\bibitem [{\citenamefont {Yasui}\ \emph {et~al.}(2017)\citenamefont {Yasui},
  \citenamefont {Suzuki},\ and\ \citenamefont {Itakura}}]{Yasui:2017izi}%
  \BibitemOpen
  \bibfield  {author} {\bibinfo {author} {\bibfnamefont {Shigehiro}\
  \bibnamefont {Yasui}}, \bibinfo {author} {\bibfnamefont {Kei}\ \bibnamefont
  {Suzuki}}, \ and\ \bibinfo {author} {\bibfnamefont {Kazunori}\ \bibnamefont
  {Itakura}},\ }\bibfield  {title} {\enquote {\bibinfo {title} {{Topology and
  stability of the Kondo phase in quark matter}},}\ }\href
  {http://dx.doi.org/10.1103/PhysRevD.96.014016} {\bibfield  {journal}
  {\bibinfo  {journal} {Phys. Rev. D}\ }\textbf {\bibinfo {volume} {96}},\
  \bibinfo {pages} {014016} (\bibinfo {year} {2017})},\ \Eprint
  {http://arxiv.org/abs/1703.04124} {arXiv:1703.04124 [hep-ph]} \BibitemShut
  {NoStop}%
\bibitem [{\citenamefont {Kimura}\ and\ \citenamefont
  {Ozaki}(2019)}]{Kimura:2018vxj}%
  \BibitemOpen
  \bibfield  {author} {\bibinfo {author} {\bibfnamefont {Taro}\ \bibnamefont
  {Kimura}}\ and\ \bibinfo {author} {\bibfnamefont {Sho}\ \bibnamefont
  {Ozaki}},\ }\bibfield  {title} {\enquote {\bibinfo {title} {{Conformal field
  theory analysis of the QCD Kondo effect}},}\ }\href
  {http://dx.doi.org/10.1103/PhysRevD.99.014040} {\bibfield  {journal}
  {\bibinfo  {journal} {Phys. Rev. D}\ }\textbf {\bibinfo {volume} {99}},\
  \bibinfo {pages} {014040} (\bibinfo {year} {2019})},\ \Eprint
  {http://arxiv.org/abs/1806.06486} {arXiv:1806.06486 [hep-ph]} \BibitemShut
  {NoStop}%
\bibitem [{\citenamefont {Fariello}\ \emph {et~al.}(2019)\citenamefont
  {Fariello}, \citenamefont {Mac\'{i}as},\ and\ \citenamefont
  {Navarra}}]{Macias:2019vbl}%
  \BibitemOpen
  \bibfield  {author} {\bibinfo {author} {\bibfnamefont {R.}~\bibnamefont
  {Fariello}}, \bibinfo {author} {\bibfnamefont {Juan~C.}\ \bibnamefont
  {Mac\'{i}as}}, \ and\ \bibinfo {author} {\bibfnamefont {F.S.}\ \bibnamefont
  {Navarra}},\ }\bibfield  {title} {\enquote {\bibinfo {title} {{The QCD Kondo
  phase in quark stars}},}\ }\href@noop {} {\  (\bibinfo {year} {2019})},\
  \Eprint {http://arxiv.org/abs/1901.01623} {arXiv:1901.01623 [nucl-th]}
  \BibitemShut {NoStop}%
\bibitem [{\citenamefont {Hattori}\ \emph {et~al.}(2019)\citenamefont
  {Hattori}, \citenamefont {Huang},\ and\ \citenamefont
  {Pisarski}}]{Hattori:2019zig}%
  \BibitemOpen
  \bibfield  {author} {\bibinfo {author} {\bibfnamefont {Koichi}\ \bibnamefont
  {Hattori}}, \bibinfo {author} {\bibfnamefont {Xu-Guang}\ \bibnamefont
  {Huang}}, \ and\ \bibinfo {author} {\bibfnamefont {Robert~D.}\ \bibnamefont
  {Pisarski}},\ }\bibfield  {title} {\enquote {\bibinfo {title} {{Emergent QCD
  Kondo effect in two-flavor color superconducting phase}},}\ }\href
  {http://dx.doi.org/10.1103/PhysRevD.99.094044} {\bibfield  {journal}
  {\bibinfo  {journal} {Phys. Rev. D}\ }\textbf {\bibinfo {volume} {99}},\
  \bibinfo {pages} {094044} (\bibinfo {year} {2019})},\ \Eprint
  {http://arxiv.org/abs/1903.10953} {arXiv:1903.10953 [hep-ph]} \BibitemShut
  {NoStop}%
\bibitem [{\citenamefont {Suenaga}\ \emph
  {et~al.}(2020{\natexlab{a}})\citenamefont {Suenaga}, \citenamefont {Suzuki},\
  and\ \citenamefont {Yasui}}]{Suenaga:2019car}%
  \BibitemOpen
  \bibfield  {author} {\bibinfo {author} {\bibfnamefont {Daiki}\ \bibnamefont
  {Suenaga}}, \bibinfo {author} {\bibfnamefont {Kei}\ \bibnamefont {Suzuki}}, \
  and\ \bibinfo {author} {\bibfnamefont {Shigehiro}\ \bibnamefont {Yasui}},\
  }\bibfield  {title} {\enquote {\bibinfo {title} {{QCD Kondo excitons}},}\
  }\href {http://dx.doi.org/10.1103/PhysRevResearch.2.023066} {\bibfield
  {journal} {\bibinfo  {journal} {Phys. Rev. Research}\ }\textbf {\bibinfo
  {volume} {2}},\ \bibinfo {pages} {023066} (\bibinfo {year}
  {2020}{\natexlab{a}})},\ \Eprint {http://arxiv.org/abs/1909.07573}
  {arXiv:1909.07573 [nucl-th]} \BibitemShut {NoStop}%
\bibitem [{\citenamefont {Suenaga}\ \emph
  {et~al.}(2020{\natexlab{b}})\citenamefont {Suenaga}, \citenamefont {Suzuki},
  \citenamefont {Araki},\ and\ \citenamefont {Yasui}}]{Suenaga:2019jqu}%
  \BibitemOpen
  \bibfield  {author} {\bibinfo {author} {\bibfnamefont {Daiki}\ \bibnamefont
  {Suenaga}}, \bibinfo {author} {\bibfnamefont {Kei}\ \bibnamefont {Suzuki}},
  \bibinfo {author} {\bibfnamefont {Yasufumi}\ \bibnamefont {Araki}}, \ and\
  \bibinfo {author} {\bibfnamefont {Shigehiro}\ \bibnamefont {Yasui}},\
  }\bibfield  {title} {\enquote {\bibinfo {title} {{Kondo effect driven by
  chirality imbalance}},}\ }\href {\doibase 10.1103/PhysRevResearch.2.023312}
  {\bibfield  {journal} {\bibinfo  {journal} {Phys. Rev. Research}\ }\textbf
  {\bibinfo {volume} {2}},\ \bibinfo {pages} {023312} (\bibinfo {year}
  {2020}{\natexlab{b}})},\ \Eprint {http://arxiv.org/abs/1912.12669}
  {arXiv:1912.12669 [hep-ph]} \BibitemShut {NoStop}%
\bibitem [{\citenamefont {Kanazawa}(2020)}]{Kanazawa:2020xje}%
  \BibitemOpen
  \bibfield  {author} {\bibinfo {author} {\bibfnamefont {Takuya}\ \bibnamefont
  {Kanazawa}},\ }\bibfield  {title} {\enquote {\bibinfo {title} {{Random matrix
  model for the QCD Kondo effect}},}\ }\href@noop {} {\  (\bibinfo {year}
  {2020})},\ \Eprint {http://arxiv.org/abs/2006.00200} {arXiv:2006.00200
  [hep-th]} \BibitemShut {NoStop}%
\bibitem [{\citenamefont {Araki}\ \emph {et~al.}()\citenamefont {Araki},
  \citenamefont {Suenaga}, \citenamefont {Suzuki},\ and\ \citenamefont
  {Yasui}}]{Araki:2020fox}%
  \BibitemOpen
  \bibfield  {author} {\bibinfo {author} {\bibfnamefont {Yasufumi}\
  \bibnamefont {Araki}}, \bibinfo {author} {\bibfnamefont {Daiki}\ \bibnamefont
  {Suenaga}}, \bibinfo {author} {\bibfnamefont {Kei}\ \bibnamefont {Suzuki}}, \
  and\ \bibinfo {author} {\bibfnamefont {Shigehiro}\ \bibnamefont {Yasui}},\
  }\bibfield  {title} {\enquote {\bibinfo {title} {{Two relativistic Kondo
  effects from two HQETs}},}\ }\href@noop {} {\ }\Eprint
  {http://arxiv.org/abs/2008.08434} {arXiv:2008.08434 [hep-ph]} \BibitemShut
  {NoStop}%
\bibitem [{\citenamefont {Araki}\ \emph {et~al.}(2020)\citenamefont {Araki},
  \citenamefont {Suenaga}, \citenamefont {Suzuki},\ and\ \citenamefont
  {Yasui}}]{Araki:2020rok}%
  \BibitemOpen
  \bibfield  {author} {\bibinfo {author} {\bibfnamefont {Yasufumi}\
  \bibnamefont {Araki}}, \bibinfo {author} {\bibfnamefont {Daiki}\ \bibnamefont
  {Suenaga}}, \bibinfo {author} {\bibfnamefont {Kei}\ \bibnamefont {Suzuki}}, \
  and\ \bibinfo {author} {\bibfnamefont {Shigehiro}\ \bibnamefont {Yasui}},\
  }\bibfield  {title} {\enquote {\bibinfo {title} {{Spin-orbital magnetic
  response of relativistic fermions with band hybridization}},}\ }\href@noop {}
  {\  (\bibinfo {year} {2020})},\ \Eprint {http://arxiv.org/abs/2011.00882}
  {arXiv:2011.00882 [cond-mat.mes-hall]} \BibitemShut {NoStop}%
\bibitem [{\citenamefont {Kimura}(2020)}]{Kimura:2020uhq}%
  \BibitemOpen
  \bibfield  {author} {\bibinfo {author} {\bibfnamefont {Taro}\ \bibnamefont
  {Kimura}},\ }\bibfield  {title} {\enquote {\bibinfo {title} {{ABCD of Kondo
  effect}},}\ }\href@noop {} {\  (\bibinfo {year} {2020})},\ \Eprint
  {http://arxiv.org/abs/2011.08301} {arXiv:2011.08301 [cond-mat.str-el]}
  \BibitemShut {NoStop}%
\bibitem [{\citenamefont {Yasui}\ and\ \citenamefont
  {Sudoh}(2017)}]{Yasui:2016hlz}%
  \BibitemOpen
  \bibfield  {author} {\bibinfo {author} {\bibfnamefont {Shigehiro}\
  \bibnamefont {Yasui}}\ and\ \bibinfo {author} {\bibfnamefont {Kazutaka}\
  \bibnamefont {Sudoh}},\ }\bibfield  {title} {\enquote {\bibinfo {title}
  {{Kondo effect of $\bar{D}_{s}$ and $\bar{D}_{s}^{\ast}$ mesons in nuclear
  matter}},}\ }\href {http://dx.doi.org/10.1103/PhysRevC.95.035204} {\bibfield
  {journal} {\bibinfo  {journal} {Phys. Rev. C}\ }\textbf {\bibinfo {volume}
  {95}},\ \bibinfo {pages} {035204} (\bibinfo {year} {2017})},\ \Eprint
  {http://arxiv.org/abs/1607.07948} {arXiv:1607.07948 [hep-ph]} \BibitemShut
  {NoStop}%
\bibitem [{\citenamefont {Yasui}(2016)}]{Yasui:2016ngy}%
  \BibitemOpen
  \bibfield  {author} {\bibinfo {author} {\bibfnamefont {Shigehiro}\
  \bibnamefont {Yasui}},\ }\bibfield  {title} {\enquote {\bibinfo {title}
  {{Kondo effect in charm and bottom nuclei}},}\ }\href
  {http://dx.doi.org/10.1103/PhysRevC.93.065204} {\bibfield  {journal}
  {\bibinfo  {journal} {Phys. Rev. C}\ }\textbf {\bibinfo {volume} {93}},\
  \bibinfo {pages} {065204} (\bibinfo {year} {2016})},\ \Eprint
  {http://arxiv.org/abs/1602.00227} {arXiv:1602.00227 [hep-ph]} \BibitemShut
  {NoStop}%
\bibitem [{\citenamefont {Yasui}\ and\ \citenamefont
  {Miyamoto}(2019)}]{Yasui:2019ogk}%
  \BibitemOpen
  \bibfield  {author} {\bibinfo {author} {\bibfnamefont {Shigehiro}\
  \bibnamefont {Yasui}}\ and\ \bibinfo {author} {\bibfnamefont {Tomokazu}\
  \bibnamefont {Miyamoto}},\ }\bibfield  {title} {\enquote {\bibinfo {title}
  {{Spin-isospin Kondo effects for $\Sigma_{c}$ and $\Sigma_{c}^{\ast}$ baryons
  and $\bar{D}$ and $\bar{D}^{\ast}$ mesons}},}\ }\href
  {http://dx.doi.org/10.1103/PhysRevC.100.045201} {\bibfield  {journal}
  {\bibinfo  {journal} {Phys. Rev. C}\ }\textbf {\bibinfo {volume} {100}},\
  \bibinfo {pages} {045201} (\bibinfo {year} {2019})},\ \Eprint
  {http://arxiv.org/abs/1905.02478} {arXiv:1905.02478 [hep-ph]} \BibitemShut
  {NoStop}%
\bibitem [{\citenamefont {Gorbar}\ \emph {et~al.}(2013)\citenamefont {Gorbar},
  \citenamefont {Miransky}, \citenamefont {Shovkovy},\ and\ \citenamefont
  {Wang}}]{Gorbar:2013upa}%
  \BibitemOpen
  \bibfield  {author} {\bibinfo {author} {\bibfnamefont {E.~V.}\ \bibnamefont
  {Gorbar}}, \bibinfo {author} {\bibfnamefont {V.~A.}\ \bibnamefont
  {Miransky}}, \bibinfo {author} {\bibfnamefont {I.~A.}\ \bibnamefont
  {Shovkovy}}, \ and\ \bibinfo {author} {\bibfnamefont {Xinyang}\ \bibnamefont
  {Wang}},\ }\bibfield  {title} {\enquote {\bibinfo {title} {{Radiative
  corrections to chiral separation effect in QED}},}\ }\href
  {http://dx.doi.org/10.1103/PhysRevD.88.025025} {\bibfield  {journal}
  {\bibinfo  {journal} {Phys. Rev. D}\ }\textbf {\bibinfo {volume} {88}},\
  \bibinfo {pages} {025025} (\bibinfo {year} {2013})},\ \Eprint
  {http://arxiv.org/abs/1304.4606} {arXiv:1304.4606 [hep-ph]} \BibitemShut
  {NoStop}%
\bibitem [{\citenamefont {Eichten}\ and\ \citenamefont
  {Hill}(1990)}]{Eichten:1989zv}%
  \BibitemOpen
  \bibfield  {author} {\bibinfo {author} {\bibfnamefont {Estia}\ \bibnamefont
  {Eichten}}\ and\ \bibinfo {author} {\bibfnamefont {Brian~Russell}\
  \bibnamefont {Hill}},\ }\bibfield  {title} {\enquote {\bibinfo {title} {{An
  Effective Field Theory for the Calculation of Matrix Elements Involving Heavy
  Quarks}},}\ }\href {\doibase 10.1016/0370-2693(90)92049-O} {\bibfield
  {journal} {\bibinfo  {journal} {Phys. Lett. B}\ }\textbf {\bibinfo {volume}
  {234}},\ \bibinfo {pages} {511--516} (\bibinfo {year} {1990})}\BibitemShut
  {NoStop}%
\bibitem [{\citenamefont {Georgi}(1990)}]{Georgi:1990um}%
  \BibitemOpen
  \bibfield  {author} {\bibinfo {author} {\bibfnamefont {Howard}\ \bibnamefont
  {Georgi}},\ }\bibfield  {title} {\enquote {\bibinfo {title} {{An effective
  field theory for heavy quarks at low energies}},}\ }\href
  {http://dx.doi.org/10.1016/0370-2693(90)91128-X} {\bibfield  {journal}
  {\bibinfo  {journal} {Phys. Lett. B}\ }\textbf {\bibinfo {volume} {240}},\
  \bibinfo {pages} {447--450} (\bibinfo {year} {1990})}\BibitemShut {NoStop}%
\bibitem [{\citenamefont {Neubert}(1994)}]{Neubert:1993mb}%
  \BibitemOpen
  \bibfield  {author} {\bibinfo {author} {\bibfnamefont {Matthias}\
  \bibnamefont {Neubert}},\ }\bibfield  {title} {\enquote {\bibinfo {title}
  {{Heavy quark symmetry}},}\ }\href
  {https://doi.org/10.1016/0370-1573(94)90091-4} {\bibfield  {journal}
  {\bibinfo  {journal} {Phys. Rept.}\ }\textbf {\bibinfo {volume} {245}},\
  \bibinfo {pages} {259--396} (\bibinfo {year} {1994})},\ \Eprint
  {http://arxiv.org/abs/hep-ph/9306320} {arXiv:hep-ph/9306320 [hep-ph]}
  \BibitemShut {NoStop}%
\bibitem [{\citenamefont {Manohar}\ and\ \citenamefont
  {Wise}(2000)}]{Manohar:2000dt}%
  \BibitemOpen
  \bibfield  {author} {\bibinfo {author} {\bibfnamefont {Aneesh~V.}\
  \bibnamefont {Manohar}}\ and\ \bibinfo {author} {\bibfnamefont {Mark~B.}\
  \bibnamefont {Wise}},\ }\href {\doibase 10.1017/CBO9780511529351} {\emph
  {\bibinfo {title} {Heavy Quark Physics}}},\ Cambridge Monographs on Particle
  Physics, Nuclear Physics and Cosmology\ (\bibinfo  {publisher} {Cambridge
  University Press},\ \bibinfo {year} {2000})\BibitemShut {NoStop}%
\bibitem [{\citenamefont {Kapusta}\ and\ \citenamefont
  {Gale}(2011)}]{Kapusta:2006pm}%
  \BibitemOpen
  \bibfield  {author} {\bibinfo {author} {\bibfnamefont {Joseph~I.}\
  \bibnamefont {Kapusta}}\ and\ \bibinfo {author} {\bibfnamefont {Charles}\
  \bibnamefont {Gale}},\ }\href {\doibase 10.1017/CBO9780511535130} {\emph
  {\bibinfo {title} {{Finite-temperature field theory: Principles and
  applications}}}},\ Cambridge Monographs on Mathematical Physics\ (\bibinfo
  {publisher} {Cambridge University Press},\ \bibinfo {year}
  {2011})\BibitemShut {NoStop}%
\bibitem [{\citenamefont {Kharzeev}\ and\ \citenamefont
  {Warringa}(2009)}]{Kharzeev:2009pj}%
  \BibitemOpen
  \bibfield  {author} {\bibinfo {author} {\bibfnamefont {Dmitri~E.}\
  \bibnamefont {Kharzeev}}\ and\ \bibinfo {author} {\bibfnamefont {Harmen~J.}\
  \bibnamefont {Warringa}},\ }\bibfield  {title} {\enquote {\bibinfo {title}
  {{Chiral Magnetic conductivity}},}\ }\href
  {http://dx.doi.org/10.1103/PhysRevD.80.034028} {\bibfield  {journal}
  {\bibinfo  {journal} {Phys. Rev. D}\ }\textbf {\bibinfo {volume} {80}},\
  \bibinfo {pages} {034028} (\bibinfo {year} {2009})},\ \Eprint
  {http://arxiv.org/abs/0907.5007} {arXiv:0907.5007 [hep-ph]} \BibitemShut
  {NoStop}%
\bibitem [{\citenamefont {Hou}\ \emph {et~al.}(2011)\citenamefont {Hou},
  \citenamefont {Liu},\ and\ \citenamefont {Ren}}]{Hou:2011ze}%
  \BibitemOpen
  \bibfield  {author} {\bibinfo {author} {\bibfnamefont {Defu}\ \bibnamefont
  {Hou}}, \bibinfo {author} {\bibfnamefont {Hui}\ \bibnamefont {Liu}}, \ and\
  \bibinfo {author} {\bibfnamefont {Hai-cang}\ \bibnamefont {Ren}},\ }\bibfield
   {title} {\enquote {\bibinfo {title} {{Some Field Theoretic Issues Regarding
  the Chiral Magnetic Effect}},}\ }\href {\doibase 10.1007/JHEP05(2011)046}
  {\bibfield  {journal} {\bibinfo  {journal} {JHEP}\ }\textbf {\bibinfo
  {volume} {05}},\ \bibinfo {pages} {046} (\bibinfo {year} {2011})},\ \Eprint
  {http://arxiv.org/abs/1103.2035} {arXiv:1103.2035 [hep-ph]} \BibitemShut
  {NoStop}%
\bibitem [{\citenamefont {Son}\ and\ \citenamefont
  {Yamamoto}(2013)}]{Son:2012zy}%
  \BibitemOpen
  \bibfield  {author} {\bibinfo {author} {\bibfnamefont {Dam~Thanh}\
  \bibnamefont {Son}}\ and\ \bibinfo {author} {\bibfnamefont {Naoki}\
  \bibnamefont {Yamamoto}},\ }\bibfield  {title} {\enquote {\bibinfo {title}
  {{Kinetic theory with Berry curvature from quantum field theories}},}\ }\href
  {\doibase 10.1103/PhysRevD.87.085016} {\bibfield  {journal} {\bibinfo
  {journal} {Phys. Rev. D}\ }\textbf {\bibinfo {volume} {87}},\ \bibinfo
  {pages} {085016} (\bibinfo {year} {2013})},\ \Eprint
  {http://arxiv.org/abs/1210.8158} {arXiv:1210.8158 [hep-th]} \BibitemShut
  {NoStop}%
\bibitem [{\citenamefont {Landsteiner}\ \emph {et~al.}(2014)\citenamefont
  {Landsteiner}, \citenamefont {Megias},\ and\ \citenamefont
  {Pena-Benitez}}]{Landsteiner:2013aba}%
  \BibitemOpen
  \bibfield  {author} {\bibinfo {author} {\bibfnamefont {Karl}\ \bibnamefont
  {Landsteiner}}, \bibinfo {author} {\bibfnamefont {Eugenio}\ \bibnamefont
  {Megias}}, \ and\ \bibinfo {author} {\bibfnamefont {Francisco}\ \bibnamefont
  {Pena-Benitez}},\ }\bibfield  {title} {\enquote {\bibinfo {title} {{Frequency
  dependence of the Chiral Vortical Effect}},}\ }\href {\doibase
  10.1103/PhysRevD.90.065026} {\bibfield  {journal} {\bibinfo  {journal} {Phys.
  Rev. D}\ }\textbf {\bibinfo {volume} {90}},\ \bibinfo {pages} {065026}
  (\bibinfo {year} {2014})},\ \Eprint {http://arxiv.org/abs/1312.1204}
  {arXiv:1312.1204 [hep-ph]} \BibitemShut {NoStop}%
\bibitem [{\citenamefont {Kharzeev}\ \emph {et~al.}(2017)\citenamefont
  {Kharzeev}, \citenamefont {Stephanov},\ and\ \citenamefont
  {Yee}}]{Kharzeev:2016sut}%
  \BibitemOpen
  \bibfield  {author} {\bibinfo {author} {\bibfnamefont {Dmitri~E.}\
  \bibnamefont {Kharzeev}}, \bibinfo {author} {\bibfnamefont {Mikhail~A.}\
  \bibnamefont {Stephanov}}, \ and\ \bibinfo {author} {\bibfnamefont {Ho-Ung}\
  \bibnamefont {Yee}},\ }\bibfield  {title} {\enquote {\bibinfo {title}
  {{Anatomy of chiral magnetic effect in and out of equilibrium}},}\ }\href
  {http://dx.doi.org/10.1103/PhysRevD.95.051901} {\bibfield  {journal}
  {\bibinfo  {journal} {Phys. Rev. D}\ }\textbf {\bibinfo {volume} {95}},\
  \bibinfo {pages} {051901} (\bibinfo {year} {2017})},\ \Eprint
  {http://arxiv.org/abs/1612.01674} {arXiv:1612.01674 [hep-ph]} \BibitemShut
  {NoStop}%
\bibitem [{\citenamefont {Alford}\ \emph {et~al.}(1998)\citenamefont {Alford},
  \citenamefont {Rajagopal},\ and\ \citenamefont {Wilczek}}]{Alford:1997zt}%
  \BibitemOpen
  \bibfield  {author} {\bibinfo {author} {\bibfnamefont {Mark~G.}\ \bibnamefont
  {Alford}}, \bibinfo {author} {\bibfnamefont {Krishna}\ \bibnamefont
  {Rajagopal}}, \ and\ \bibinfo {author} {\bibfnamefont {Frank}\ \bibnamefont
  {Wilczek}},\ }\bibfield  {title} {\enquote {\bibinfo {title} {{QCD at finite
  baryon density: Nucleon droplets and color superconductivity}},}\ }\href
  {http://dx.doi.org/10.1016/S0370-2693(98)00051-3} {\bibfield  {journal}
  {\bibinfo  {journal} {Phys. Lett. B}\ }\textbf {\bibinfo {volume} {422}},\
  \bibinfo {pages} {247--256} (\bibinfo {year} {1998})},\ \Eprint
  {http://arxiv.org/abs/hep-ph/9711395} {arXiv:hep-ph/9711395 [hep-ph]}
  \BibitemShut {NoStop}%
\bibitem [{\citenamefont {Rapp}\ \emph {et~al.}(1998)\citenamefont {Rapp},
  \citenamefont {Sch\"{a}fer}, \citenamefont {Shuryak},\ and\ \citenamefont
  {Velkovsky}}]{Rapp:1997zu}%
  \BibitemOpen
  \bibfield  {author} {\bibinfo {author} {\bibfnamefont {R.}~\bibnamefont
  {Rapp}}, \bibinfo {author} {\bibfnamefont {Thomas}\ \bibnamefont
  {Sch\"{a}fer}}, \bibinfo {author} {\bibfnamefont {Edward~V.}\ \bibnamefont
  {Shuryak}}, \ and\ \bibinfo {author} {\bibfnamefont {M.}~\bibnamefont
  {Velkovsky}},\ }\bibfield  {title} {\enquote {\bibinfo {title} {{Diquark Bose
  Condensates in High Density Matter and Instantons}},}\ }\href
  {https://doi.org/10.1103/PhysRevLett.81.53} {\bibfield  {journal} {\bibinfo
  {journal} {Phys. Rev. Lett.}\ }\textbf {\bibinfo {volume} {81}},\ \bibinfo
  {pages} {53--56} (\bibinfo {year} {1998})},\ \Eprint
  {http://arxiv.org/abs/hep-ph/9711396} {arXiv:hep-ph/9711396 [hep-ph]}
  \BibitemShut {NoStop}%
\bibitem [{\citenamefont {Alford}\ \emph {et~al.}(1999)\citenamefont {Alford},
  \citenamefont {Rajagopal},\ and\ \citenamefont {Wilczek}}]{Alford:1998mk}%
  \BibitemOpen
  \bibfield  {author} {\bibinfo {author} {\bibfnamefont {Mark~G.}\ \bibnamefont
  {Alford}}, \bibinfo {author} {\bibfnamefont {Krishna}\ \bibnamefont
  {Rajagopal}}, \ and\ \bibinfo {author} {\bibfnamefont {Frank}\ \bibnamefont
  {Wilczek}},\ }\bibfield  {title} {\enquote {\bibinfo {title} {{Color flavor
  locking and chiral symmetry breaking in high density QCD}},}\ }\href
  {http://dx.doi.org/10.1016/S0550-3213(98)00668-3} {\bibfield  {journal}
  {\bibinfo  {journal} {Nucl. Phys. B}\ }\textbf {\bibinfo {volume} {537}},\
  \bibinfo {pages} {443--458} (\bibinfo {year} {1999})},\ \Eprint
  {http://arxiv.org/abs/hep-ph/9804403} {arXiv:hep-ph/9804403 [hep-ph]}
  \BibitemShut {NoStop}%
\bibitem [{\citenamefont {Alford}\ \emph {et~al.}(2008)\citenamefont {Alford},
  \citenamefont {Schmitt}, \citenamefont {Rajagopal},\ and\ \citenamefont
  {Sch{\"a}fer}}]{Alford:2007xm}%
  \BibitemOpen
  \bibfield  {author} {\bibinfo {author} {\bibfnamefont {Mark~G.}\ \bibnamefont
  {Alford}}, \bibinfo {author} {\bibfnamefont {Andreas}\ \bibnamefont
  {Schmitt}}, \bibinfo {author} {\bibfnamefont {Krishna}\ \bibnamefont
  {Rajagopal}}, \ and\ \bibinfo {author} {\bibfnamefont {Thomas}\ \bibnamefont
  {Sch{\"a}fer}},\ }\bibfield  {title} {\enquote {\bibinfo {title} {{Color
  superconductivity in dense quark matter}},}\ }\href
  {http://dx.doi.org/10.1103/RevModPhys.80.1455} {\bibfield  {journal}
  {\bibinfo  {journal} {Rev. Mod. Phys.}\ }\textbf {\bibinfo {volume} {80}},\
  \bibinfo {pages} {1455--1515} (\bibinfo {year} {2008})},\ \Eprint
  {http://arxiv.org/abs/0709.4635} {arXiv:0709.4635 [hep-ph]} \BibitemShut
  {NoStop}%
\bibitem [{\citenamefont {Nambu}\ and\ \citenamefont
  {Jona-Lasinio}(1961)}]{Nambu:1961tp}%
  \BibitemOpen
  \bibfield  {author} {\bibinfo {author} {\bibfnamefont {Yoichiro}\
  \bibnamefont {Nambu}}\ and\ \bibinfo {author} {\bibfnamefont
  {G.}~\bibnamefont {Jona-Lasinio}},\ }\bibfield  {title} {\enquote {\bibinfo
  {title} {{Dynamical Model of Elementary Particles Based on an Analogy with
  Superconductivity. 1.}}}\ }\href {\doibase 10.1103/PhysRev.122.345}
  {\bibfield  {journal} {\bibinfo  {journal} {Phys. Rev.}\ }\textbf {\bibinfo
  {volume} {122}},\ \bibinfo {pages} {345--358} (\bibinfo {year}
  {1961})}\BibitemShut {NoStop}%
\bibitem [{\citenamefont {Bardeen}\ and\ \citenamefont
  {Hill}(1994)}]{Bardeen:1993ae}%
  \BibitemOpen
  \bibfield  {author} {\bibinfo {author} {\bibfnamefont {William~A.}\
  \bibnamefont {Bardeen}}\ and\ \bibinfo {author} {\bibfnamefont
  {Christopher~T.}\ \bibnamefont {Hill}},\ }\bibfield  {title} {\enquote
  {\bibinfo {title} {{Chiral dynamics and heavy quark symmetry in a solvable
  toy field theoretic model}},}\ }\href {\doibase 10.1103/PhysRevD.49.409}
  {\bibfield  {journal} {\bibinfo  {journal} {Phys. Rev. D}\ }\textbf {\bibinfo
  {volume} {49}},\ \bibinfo {pages} {409--425} (\bibinfo {year} {1994})},\
  \Eprint {http://arxiv.org/abs/hep-ph/9304265} {arXiv:hep-ph/9304265}
  \BibitemShut {NoStop}%
\bibitem [{\citenamefont {Nowak}\ \emph {et~al.}(1993)\citenamefont {Nowak},
  \citenamefont {Rho},\ and\ \citenamefont {Zahed}}]{Nowak:1992um}%
  \BibitemOpen
  \bibfield  {author} {\bibinfo {author} {\bibfnamefont {Maciej~A.}\
  \bibnamefont {Nowak}}, \bibinfo {author} {\bibfnamefont {Mannque}\
  \bibnamefont {Rho}}, \ and\ \bibinfo {author} {\bibfnamefont
  {I.}~\bibnamefont {Zahed}},\ }\bibfield  {title} {\enquote {\bibinfo {title}
  {{Chiral effective action with heavy quark symmetry}},}\ }\href
  {http://dx.doi.org/10.1103/PhysRevD.48.4370} {\bibfield  {journal} {\bibinfo
  {journal} {Phys. Rev. D}\ }\textbf {\bibinfo {volume} {48}},\ \bibinfo
  {pages} {4370--4374} (\bibinfo {year} {1993})},\ \Eprint
  {http://arxiv.org/abs/hep-ph/9209272} {arXiv:hep-ph/9209272} \BibitemShut
  {NoStop}%
\bibitem [{\citenamefont {Suenaga}\ \emph {et~al.}(2015)\citenamefont
  {Suenaga}, \citenamefont {He}, \citenamefont {Ma},\ and\ \citenamefont
  {Harada}}]{Suenaga:2014sga}%
  \BibitemOpen
  \bibfield  {author} {\bibinfo {author} {\bibfnamefont {Daiki}\ \bibnamefont
  {Suenaga}}, \bibinfo {author} {\bibfnamefont {Bing-Ran}\ \bibnamefont {He}},
  \bibinfo {author} {\bibfnamefont {Yong-Liang}\ \bibnamefont {Ma}}, \ and\
  \bibinfo {author} {\bibfnamefont {Masayasu}\ \bibnamefont {Harada}},\
  }\bibfield  {title} {\enquote {\bibinfo {title} {{Mass degeneracy of
  heavy-light mesons with chiral partner structure in the half-Skyrmion
  phase}},}\ }\href {http://dx.doi.org/10.1103/PhysRevD.91.036001} {\bibfield
  {journal} {\bibinfo  {journal} {Phys. Rev. D}\ }\textbf {\bibinfo {volume}
  {91}},\ \bibinfo {pages} {036001} (\bibinfo {year} {2015})},\ \Eprint
  {http://arxiv.org/abs/1412.2462} {arXiv:1412.2462 [hep-ph]} \BibitemShut
  {NoStop}%
\bibitem [{\citenamefont {Harada}\ \emph {et~al.}(2017)\citenamefont {Harada},
  \citenamefont {Ma}, \citenamefont {Suenaga},\ and\ \citenamefont
  {Takeda}}]{Harada:2016uca}%
  \BibitemOpen
  \bibfield  {author} {\bibinfo {author} {\bibfnamefont {Masayasu}\
  \bibnamefont {Harada}}, \bibinfo {author} {\bibfnamefont {Yong-Liang}\
  \bibnamefont {Ma}}, \bibinfo {author} {\bibfnamefont {Daiki}\ \bibnamefont
  {Suenaga}}, \ and\ \bibinfo {author} {\bibfnamefont {Yusuke}\ \bibnamefont
  {Takeda}},\ }\bibfield  {title} {\enquote {\bibinfo {title} {{Relation
  between the mass modification of heavy\textendash{}light mesons and the
  chiral symmetry structure in dense matter}},}\ }\href
  {http://dx.doi.org/10.1093/ptep/ptx140} {\bibfield  {journal} {\bibinfo
  {journal} {PTEP}\ }\textbf {\bibinfo {volume} {2017}},\ \bibinfo {pages}
  {113D01} (\bibinfo {year} {2017})},\ \Eprint
  {http://arxiv.org/abs/1612.03496} {arXiv:1612.03496 [hep-ph]} \BibitemShut
  {NoStop}%
\bibitem [{\citenamefont {Suenaga}\ \emph {et~al.}(2017)\citenamefont
  {Suenaga}, \citenamefont {Yasui},\ and\ \citenamefont
  {Harada}}]{Suenaga:2017deu}%
  \BibitemOpen
  \bibfield  {author} {\bibinfo {author} {\bibfnamefont {Daiki}\ \bibnamefont
  {Suenaga}}, \bibinfo {author} {\bibfnamefont {Shigehiro}\ \bibnamefont
  {Yasui}}, \ and\ \bibinfo {author} {\bibfnamefont {Masayasu}\ \bibnamefont
  {Harada}},\ }\bibfield  {title} {\enquote {\bibinfo {title} {{Spectral
  functions for $\bar{D}$ and $\bar{D}_0^*$ mesons in nuclear matter with
  partial restoration of chiral symmetry}},}\ }\href
  {http://dx.doi.org/10.1103/PhysRevC.96.015204} {\bibfield  {journal}
  {\bibinfo  {journal} {Phys. Rev. C}\ }\textbf {\bibinfo {volume} {96}},\
  \bibinfo {pages} {015204} (\bibinfo {year} {2017})},\ \Eprint
  {http://arxiv.org/abs/1703.02762} {arXiv:1703.02762 [nucl-th]} \BibitemShut
  {NoStop}%
\bibitem [{\citenamefont {Satow}\ and\ \citenamefont
  {Yee}(2014)}]{Satow:2014lva}%
  \BibitemOpen
  \bibfield  {author} {\bibinfo {author} {\bibfnamefont {Daisuke}\ \bibnamefont
  {Satow}}\ and\ \bibinfo {author} {\bibfnamefont {Ho-Ung}\ \bibnamefont
  {Yee}},\ }\bibfield  {title} {\enquote {\bibinfo {title} {{Chiral magnetic
  effect at weak coupling with relaxation dynamics}},}\ }\href
  {http://dx.doi.org/10.1103/PhysRevD.90.014027} {\bibfield  {journal}
  {\bibinfo  {journal} {Phys. Rev. D}\ }\textbf {\bibinfo {volume} {90}},\
  \bibinfo {pages} {014027} (\bibinfo {year} {2014})},\ \Eprint
  {http://arxiv.org/abs/1406.1150} {arXiv:1406.1150 [hep-ph]} \BibitemShut
  {NoStop}%
\bibitem [{\citenamefont {Schrieffer}(1999)}]{schrieffer1999theory}%
  \BibitemOpen
  \bibfield  {author} {\bibinfo {author} {\bibfnamefont {John~R.}\ \bibnamefont
  {Schrieffer}},\ }\href {https://books.google.co.jp/books?id=let7wRir74MC}
  {\emph {\bibinfo {title} {Theory Of Superconductivity}}},\ Advanced Books
  Classics\ (\bibinfo  {publisher} {Avalon Publishing},\ \bibinfo {year}
  {1999})\BibitemShut {NoStop}%
\bibitem [{\citenamefont {Cheng}\ and\ \citenamefont
  {Li}(1994)}]{cheng1994gauge}%
  \BibitemOpen
  \bibfield  {author} {\bibinfo {author} {\bibfnamefont {Ta~P.}\ \bibnamefont
  {Cheng}}\ and\ \bibinfo {author} {\bibfnamefont {Ling~F.}\ \bibnamefont
  {Li}},\ }\href {https://books.google.co.jp/books?id=NUUBEAAAQBAJ} {\emph
  {\bibinfo {title} {Gauge Theory of Elementary Particle Physics}}}\ (\bibinfo
  {publisher} {Oxford University Press},\ \bibinfo {year} {1994})\BibitemShut
  {NoStop}%
\bibitem [{\citenamefont {Nambu}(1960)}]{Nambu:1960tm}%
  \BibitemOpen
  \bibfield  {author} {\bibinfo {author} {\bibfnamefont {Yoichiro}\
  \bibnamefont {Nambu}},\ }\bibfield  {title} {\enquote {\bibinfo {title}
  {{Quasiparticles and Gauge Invariance in the Theory of Superconductivity}},}\
  }\href {\doibase 10.1103/PhysRev.117.648} {\bibfield  {journal} {\bibinfo
  {journal} {Phys. Rev.}\ }\textbf {\bibinfo {volume} {117}},\ \bibinfo {pages}
  {648--663} (\bibinfo {year} {1960})}\BibitemShut {NoStop}%
\bibitem [{\citenamefont {Tserkovnyak}\ \emph {et~al.}(2015)\citenamefont
  {Tserkovnyak}, \citenamefont {Pesin},\ and\ \citenamefont
  {Loss}}]{Tserkovnyak}%
  \BibitemOpen
  \bibfield  {author} {\bibinfo {author} {\bibfnamefont {Yaroslav}\
  \bibnamefont {Tserkovnyak}}, \bibinfo {author} {\bibfnamefont {D.~A.}\
  \bibnamefont {Pesin}}, \ and\ \bibinfo {author} {\bibfnamefont {Daniel}\
  \bibnamefont {Loss}},\ }\bibfield  {title} {\enquote {\bibinfo {title} {Spin
  and orbital magnetic response on the surface of a topological insulator},}\
  }\href {https://doi.org/10.1103/PhysRevB.91.041121} {\bibfield  {journal}
  {\bibinfo  {journal} {Phys. Rev. B}\ }\textbf {\bibinfo {volume} {91}},\
  \bibinfo {pages} {041121} (\bibinfo {year} {2015})},\ \Eprint
  {http://arxiv.org/abs/1411.2070} {arXiv:1411.2070 [cond-mat.mes-hall]}
  \BibitemShut {NoStop}%
\bibitem [{\citenamefont {Koshino}\ and\ \citenamefont
  {Hizbullah}(2016)}]{Koshino}%
  \BibitemOpen
  \bibfield  {author} {\bibinfo {author} {\bibfnamefont {Mikito}\ \bibnamefont
  {Koshino}}\ and\ \bibinfo {author} {\bibfnamefont {Intan~Fatimah}\
  \bibnamefont {Hizbullah}},\ }\bibfield  {title} {\enquote {\bibinfo {title}
  {Magnetic susceptibility in three-dimensional nodal semimetals},}\ }\href
  {http://dx.doi.org/10.1103/PhysRevB.93.045201} {\bibfield  {journal}
  {\bibinfo  {journal} {Phys. Rev. B}\ }\textbf {\bibinfo {volume} {93}},\
  \bibinfo {pages} {045201} (\bibinfo {year} {2016})},\ \Eprint
  {http://arxiv.org/abs/1510.02191} {arXiv:1510.02191 [cond-mat.mes-hall]}
  \BibitemShut {NoStop}%
\bibitem [{\citenamefont {Nakai}\ and\ \citenamefont {Nomura}(2016)}]{Nakai}%
  \BibitemOpen
  \bibfield  {author} {\bibinfo {author} {\bibfnamefont {Ryota}\ \bibnamefont
  {Nakai}}\ and\ \bibinfo {author} {\bibfnamefont {Kentaro}\ \bibnamefont
  {Nomura}},\ }\bibfield  {title} {\enquote {\bibinfo {title} {Crossed
  responses of spin and orbital magnetism in topological insulators},}\ }\href
  {https://doi.org/10.1103/PhysRevB.93.214434} {\bibfield  {journal} {\bibinfo
  {journal} {Phys. Rev. B}\ }\textbf {\bibinfo {volume} {93}},\ \bibinfo
  {pages} {214434} (\bibinfo {year} {2016})},\ \Eprint
  {http://arxiv.org/abs/1604.04991} {arXiv:1604.04991 [cond-mat.mes-hall]}
  \BibitemShut {NoStop}%
\bibitem [{\citenamefont {Suzuura}\ and\ \citenamefont {Ando}(2016)}]{Suzuura}%
  \BibitemOpen
  \bibfield  {author} {\bibinfo {author} {\bibfnamefont {Hidekatsu}\
  \bibnamefont {Suzuura}}\ and\ \bibinfo {author} {\bibfnamefont {Tsuneya}\
  \bibnamefont {Ando}},\ }\bibfield  {title} {\enquote {\bibinfo {title}
  {Theory of magnetic response in two-dimensional giant rashba system},}\
  }\href {\doibase 10.1103/PhysRevB.94.085303} {\bibfield  {journal} {\bibinfo
  {journal} {Phys. Rev. B}\ }\textbf {\bibinfo {volume} {94}},\ \bibinfo
  {pages} {085303} (\bibinfo {year} {2016})}\BibitemShut {NoStop}%
\bibitem [{\citenamefont {Ando}\ and\ \citenamefont {Suzuura}(2017)}]{Ando}%
  \BibitemOpen
  \bibfield  {author} {\bibinfo {author} {\bibfnamefont {Tsuneya}\ \bibnamefont
  {Ando}}\ and\ \bibinfo {author} {\bibfnamefont {Hidekatsu}\ \bibnamefont
  {Suzuura}},\ }\bibfield  {title} {\enquote {\bibinfo {title} {Note on formula
  of weak-field hall conductivity: Singular behavior for long-range
  scatterers},}\ }\href {\doibase 10.7566/JPSJ.86.014709} {\bibfield  {journal}
  {\bibinfo  {journal} {J. Phys. Soc. Jpn.}\ }\textbf {\bibinfo {volume}
  {86}},\ \bibinfo {pages} {014709} (\bibinfo {year} {2017})}\BibitemShut
  {NoStop}%
\bibitem [{\citenamefont {Ominato}\ \emph {et~al.}(2019)\citenamefont
  {Ominato}, \citenamefont {Tatsumi},\ and\ \citenamefont {Nomura}}]{Ominato}%
  \BibitemOpen
  \bibfield  {author} {\bibinfo {author} {\bibfnamefont {Yuya}\ \bibnamefont
  {Ominato}}, \bibinfo {author} {\bibfnamefont {Shuta}\ \bibnamefont
  {Tatsumi}}, \ and\ \bibinfo {author} {\bibfnamefont {Kentaro}\ \bibnamefont
  {Nomura}},\ }\bibfield  {title} {\enquote {\bibinfo {title} {Spin-orbit
  crossed susceptibility in topological dirac semimetals},}\ }\href
  {http://dx.doi.org/10.1103/PhysRevB.99.085205} {\bibfield  {journal}
  {\bibinfo  {journal} {Phys. Rev. B}\ }\textbf {\bibinfo {volume} {99}},\
  \bibinfo {pages} {085205} (\bibinfo {year} {2019})},\ \Eprint
  {http://arxiv.org/abs/1809.10852} {arXiv:1809.10852 [cond-mat.mes-hall]}
  \BibitemShut {NoStop}%
\bibitem [{\citenamefont {Yee}(2009)}]{Yee:2009vw}%
  \BibitemOpen
  \bibfield  {author} {\bibinfo {author} {\bibfnamefont {Ho-Ung}\ \bibnamefont
  {Yee}},\ }\bibfield  {title} {\enquote {\bibinfo {title} {{Holographic Chiral
  Magnetic Conductivity}},}\ }\href
  {http://dx.doi.org/10.1088/1126-6708/2009/11/085} {\bibfield  {journal}
  {\bibinfo  {journal} {JHEP}\ }\textbf {\bibinfo {volume} {11}},\ \bibinfo
  {pages} {085} (\bibinfo {year} {2009})},\ \Eprint
  {http://arxiv.org/abs/0908.4189} {arXiv:0908.4189 [hep-th]} \BibitemShut
  {NoStop}%
\bibitem [{\citenamefont {Vilenkin}(1979)}]{Vilenkin:1979ui}%
  \BibitemOpen
  \bibfield  {author} {\bibinfo {author} {\bibfnamefont {A.}~\bibnamefont
  {Vilenkin}},\ }\bibfield  {title} {\enquote {\bibinfo {title} {{Macroscopic
  parity-violating effects: Neutrino fluxes from rotating black holes and in
  rotating thermal radiation}},}\ }\href
  {http://dx.doi.org/10.1103/PhysRevD.20.1807} {\bibfield  {journal} {\bibinfo
  {journal} {Phys. Rev. D}\ }\textbf {\bibinfo {volume} {20}},\ \bibinfo
  {pages} {1807} (\bibinfo {year} {1979})}\BibitemShut {NoStop}%
\bibitem [{\citenamefont {Vilenkin}(1980{\natexlab{b}})}]{Vilenkin:1980zv}%
  \BibitemOpen
  \bibfield  {author} {\bibinfo {author} {\bibfnamefont {A.}~\bibnamefont
  {Vilenkin}},\ }\bibfield  {title} {\enquote {\bibinfo {title} {{Quantum field
  theory at finite temperature in a rotating system}},}\ }\href
  {http://dx.doi.org/10.1103/PhysRevD.21.2260} {\bibfield  {journal} {\bibinfo
  {journal} {Phys. Rev. D}\ }\textbf {\bibinfo {volume} {21}},\ \bibinfo
  {pages} {2260} (\bibinfo {year} {1980}{\natexlab{b}})}\BibitemShut {NoStop}%
\bibitem [{\citenamefont {Erdmenger}\ \emph {et~al.}(2009)\citenamefont
  {Erdmenger}, \citenamefont {Haack}, \citenamefont {Kaminski},\ and\
  \citenamefont {Yarom}}]{Erdmenger:2008rm}%
  \BibitemOpen
  \bibfield  {author} {\bibinfo {author} {\bibfnamefont {Johanna}\ \bibnamefont
  {Erdmenger}}, \bibinfo {author} {\bibfnamefont {Michael}\ \bibnamefont
  {Haack}}, \bibinfo {author} {\bibfnamefont {Matthias}\ \bibnamefont
  {Kaminski}}, \ and\ \bibinfo {author} {\bibfnamefont {Amos}\ \bibnamefont
  {Yarom}},\ }\bibfield  {title} {\enquote {\bibinfo {title} {{Fluid dynamics
  of R-charged black holes}},}\ }\href
  {http://dx.doi.org/10.1088/1126-6708/2009/01/055} {\bibfield  {journal}
  {\bibinfo  {journal} {JHEP}\ }\textbf {\bibinfo {volume} {01}},\ \bibinfo
  {pages} {055} (\bibinfo {year} {2009})},\ \Eprint
  {http://arxiv.org/abs/0809.2488} {arXiv:0809.2488 [hep-th]} \BibitemShut
  {NoStop}%
\bibitem [{\citenamefont {Banerjee}\ \emph {et~al.}(2011)\citenamefont
  {Banerjee}, \citenamefont {Bhattacharya}, \citenamefont {Bhattacharyya},
  \citenamefont {Dutta}, \citenamefont {Loganayagam},\ and\ \citenamefont
  {Surowka}}]{Banerjee:2008th}%
  \BibitemOpen
  \bibfield  {author} {\bibinfo {author} {\bibfnamefont {Nabamita}\
  \bibnamefont {Banerjee}}, \bibinfo {author} {\bibfnamefont {Jyotirmoy}\
  \bibnamefont {Bhattacharya}}, \bibinfo {author} {\bibfnamefont {Sayantani}\
  \bibnamefont {Bhattacharyya}}, \bibinfo {author} {\bibfnamefont {Suvankar}\
  \bibnamefont {Dutta}}, \bibinfo {author} {\bibfnamefont {R.}~\bibnamefont
  {Loganayagam}}, \ and\ \bibinfo {author} {\bibfnamefont {P.}~\bibnamefont
  {Surowka}},\ }\bibfield  {title} {\enquote {\bibinfo {title} {{Hydrodynamics
  from charged black branes}},}\ }\href
  {http://dx.doi.org/10.1007/JHEP01(2011)094} {\bibfield  {journal} {\bibinfo
  {journal} {JHEP}\ }\textbf {\bibinfo {volume} {01}},\ \bibinfo {pages} {094}
  (\bibinfo {year} {2011})},\ \Eprint {http://arxiv.org/abs/0809.2596}
  {arXiv:0809.2596 [hep-th]} \BibitemShut {NoStop}%
\bibitem [{\citenamefont {Son}\ and\ \citenamefont
  {Surowka}(2009)}]{Son:2009tf}%
  \BibitemOpen
  \bibfield  {author} {\bibinfo {author} {\bibfnamefont {Dam~T.}\ \bibnamefont
  {Son}}\ and\ \bibinfo {author} {\bibfnamefont {Piotr}\ \bibnamefont
  {Surowka}},\ }\bibfield  {title} {\enquote {\bibinfo {title} {{Hydrodynamics
  with Triangle Anomalies}},}\ }\href {\doibase 10.1103/PhysRevLett.103.191601}
  {\bibfield  {journal} {\bibinfo  {journal} {Phys. Rev. Lett.}\ }\textbf
  {\bibinfo {volume} {103}},\ \bibinfo {pages} {191601} (\bibinfo {year}
  {2009})},\ \Eprint {http://arxiv.org/abs/0906.5044} {arXiv:0906.5044
  [hep-th]} \BibitemShut {NoStop}%
\bibitem [{\citenamefont {Landsteiner}\ \emph {et~al.}(2011)\citenamefont
  {Landsteiner}, \citenamefont {Meg\'{\i}as},\ and\ \citenamefont
  {Pena-Benitez}}]{Landsteiner:2011cp}%
  \BibitemOpen
  \bibfield  {author} {\bibinfo {author} {\bibfnamefont {Karl}\ \bibnamefont
  {Landsteiner}}, \bibinfo {author} {\bibfnamefont {Eugenio}\ \bibnamefont
  {Meg\'{\i}as}}, \ and\ \bibinfo {author} {\bibfnamefont {Francisco}\
  \bibnamefont {Pena-Benitez}},\ }\bibfield  {title} {\enquote {\bibinfo
  {title} {{Gravitational Anomaly and Transport}},}\ }\href
  {http://dx.doi.org/10.1103/PhysRevLett.107.021601} {\bibfield  {journal}
  {\bibinfo  {journal} {Phys. Rev. Lett.}\ }\textbf {\bibinfo {volume} {107}},\
  \bibinfo {pages} {021601} (\bibinfo {year} {2011})},\ \Eprint
  {http://arxiv.org/abs/1103.5006} {arXiv:1103.5006 [hep-ph]} \BibitemShut
  {NoStop}%
\bibitem [{\citenamefont {Stephanov}\ and\ \citenamefont
  {Yin}(2012)}]{Stephanov:2012ki}%
  \BibitemOpen
  \bibfield  {author} {\bibinfo {author} {\bibfnamefont {M.A.}\ \bibnamefont
  {Stephanov}}\ and\ \bibinfo {author} {\bibfnamefont {Y.}~\bibnamefont
  {Yin}},\ }\bibfield  {title} {\enquote {\bibinfo {title} {{Chiral Kinetic
  Theory}},}\ }\href {\doibase 10.1103/PhysRevLett.109.162001} {\bibfield
  {journal} {\bibinfo  {journal} {Phys. Rev. Lett.}\ }\textbf {\bibinfo
  {volume} {109}},\ \bibinfo {pages} {162001} (\bibinfo {year} {2012})},\
  \Eprint {http://arxiv.org/abs/1207.0747} {arXiv:1207.0747 [hep-th]}
  \BibitemShut {NoStop}%
\bibitem [{\citenamefont {Abramchuk}\ \emph {et~al.}(2018)\citenamefont
  {Abramchuk}, \citenamefont {Khaidukov},\ and\ \citenamefont
  {Zubkov}}]{Abramchuk:2018jhd}%
  \BibitemOpen
  \bibfield  {author} {\bibinfo {author} {\bibfnamefont {Ruslan}\ \bibnamefont
  {Abramchuk}}, \bibinfo {author} {\bibfnamefont {Z.V.}\ \bibnamefont
  {Khaidukov}}, \ and\ \bibinfo {author} {\bibfnamefont {M.A.}\ \bibnamefont
  {Zubkov}},\ }\bibfield  {title} {\enquote {\bibinfo {title} {{Anatomy of the
  chiral vortical effect}},}\ }\href {\doibase 10.1103/PhysRevD.98.076013}
  {\bibfield  {journal} {\bibinfo  {journal} {Phys. Rev. D}\ }\textbf {\bibinfo
  {volume} {98}},\ \bibinfo {pages} {076013} (\bibinfo {year} {2018})},\
  \Eprint {http://arxiv.org/abs/1806.02605} {arXiv:1806.02605 [hep-ph]}
  \BibitemShut {NoStop}%
\bibitem [{\citenamefont {Guo}\ and\ \citenamefont {Lin}(2017)}]{Guo:2016dnm}%
  \BibitemOpen
  \bibfield  {author} {\bibinfo {author} {\bibfnamefont {Er-dong}\ \bibnamefont
  {Guo}}\ and\ \bibinfo {author} {\bibfnamefont {Shu}\ \bibnamefont {Lin}},\
  }\bibfield  {title} {\enquote {\bibinfo {title} {{Quark mass correction to
  chiral separation effect and pseudoscalar condensate}},}\ }\href
  {http://dx.doi.org/10.1007/JHEP01(2017)111} {\bibfield  {journal} {\bibinfo
  {journal} {JHEP}\ }\textbf {\bibinfo {volume} {01}},\ \bibinfo {pages} {111}
  (\bibinfo {year} {2017})},\ \Eprint {http://arxiv.org/abs/1610.05886}
  {arXiv:1610.05886 [hep-th]} \BibitemShut {NoStop}%
\bibitem [{\citenamefont {Flachi}\ and\ \citenamefont
  {Fukushima}(2018)}]{Flachi:2017vlp}%
  \BibitemOpen
  \bibfield  {author} {\bibinfo {author} {\bibfnamefont {Antonino}\
  \bibnamefont {Flachi}}\ and\ \bibinfo {author} {\bibfnamefont {Kenji}\
  \bibnamefont {Fukushima}},\ }\bibfield  {title} {\enquote {\bibinfo {title}
  {{Chiral vortical effect with finite rotation, temperature, and
  curvature}},}\ }\href {\doibase 10.1103/PhysRevD.98.096011} {\bibfield
  {journal} {\bibinfo  {journal} {Phys. Rev. D}\ }\textbf {\bibinfo {volume}
  {98}},\ \bibinfo {pages} {096011} (\bibinfo {year} {2018})},\ \Eprint
  {http://arxiv.org/abs/1702.04753} {arXiv:1702.04753 [hep-th]} \BibitemShut
  {NoStop}%
\bibitem [{\citenamefont {Lin}\ and\ \citenamefont {Yang}(2018)}]{Lin:2018aon}%
  \BibitemOpen
  \bibfield  {author} {\bibinfo {author} {\bibfnamefont {Shu}\ \bibnamefont
  {Lin}}\ and\ \bibinfo {author} {\bibfnamefont {Lixin}\ \bibnamefont {Yang}},\
  }\bibfield  {title} {\enquote {\bibinfo {title} {{Mass correction to chiral
  vortical effect and chiral separation effect}},}\ }\href
  {http://dx.doi.org/10.1103/PhysRevD.98.114022} {\bibfield  {journal}
  {\bibinfo  {journal} {Phys. Rev. D}\ }\textbf {\bibinfo {volume} {98}},\
  \bibinfo {pages} {114022} (\bibinfo {year} {2018})},\ \Eprint
  {http://arxiv.org/abs/1810.02979} {arXiv:1810.02979 [nucl-th]} \BibitemShut
  {NoStop}%
\bibitem [{\citenamefont {Schwinger}(1951)}]{Schwinger:1951nm}%
  \BibitemOpen
  \bibfield  {author} {\bibinfo {author} {\bibfnamefont {Julian~S.}\
  \bibnamefont {Schwinger}},\ }\bibfield  {title} {\enquote {\bibinfo {title}
  {{On gauge invariance and vacuum polarization}},}\ }\href
  {http://dx.doi.org/10.1103/PhysRev.82.664} {\bibfield  {journal} {\bibinfo
  {journal} {Phys. Rev.}\ }\textbf {\bibinfo {volume} {82}},\ \bibinfo {pages}
  {664--679} (\bibinfo {year} {1951})}\BibitemShut {NoStop}%
\bibitem [{\citenamefont {Miransky}\ and\ \citenamefont
  {Shovkovy}(2015)}]{Miransky:2015ava}%
  \BibitemOpen
  \bibfield  {author} {\bibinfo {author} {\bibfnamefont {Vladimir~A.}\
  \bibnamefont {Miransky}}\ and\ \bibinfo {author} {\bibfnamefont {Igor~A.}\
  \bibnamefont {Shovkovy}},\ }\bibfield  {title} {\enquote {\bibinfo {title}
  {{Quantum field theory in a magnetic field: From quantum chromodynamics to
  graphene and Dirac semimetals}},}\ }\href
  {https://dx.doi.org/10.1016/j.physrep.2015.02.003} {\bibfield  {journal}
  {\bibinfo  {journal} {Phys. Rept.}\ }\textbf {\bibinfo {volume} {576}},\
  \bibinfo {pages} {1--209} (\bibinfo {year} {2015})},\ \Eprint
  {http://arxiv.org/abs/1503.00732} {arXiv:1503.00732 [hep-ph]} \BibitemShut
  {NoStop}%
\end{thebibliography}%

\end{document}